\documentclass[11pt,letterpaper]{article}

\usepackage[english]{babel}

\usepackage[letterpaper,top=2cm,bottom=2cm,left=3cm,right=3cm,marginparwidth=1.75cm]{geometry}
\usepackage{amsmath}
\usepackage{mathrsfs}
\usepackage{graphicx}

\usepackage{tikz}
\usetikzlibrary{arrows.meta,positioning,calc}
\usepackage{xcolor}

\usepackage[colorlinks=true, allcolors=blue, hypertexnames=false]{hyperref}
\usepackage[capitalize,noabbrev]{cleveref}
\usepackage{xspace}
\usepackage{array}

\newcommand{\R}{\mathbb{R}}

\usepackage{amssymb}
\usepackage{mathtools}
\usepackage{amsthm}

\theoremstyle{plain}
\newtheorem{theorem}{Theorem}[section]

\newtheorem{lemma}[theorem]{Lemma}
\newtheorem{corollary}[theorem]{Corollary}
\theoremstyle{definition}
\newtheorem{definition}[theorem]{Definition}
\newtheorem{assumption}[theorem]{Assumption}
\theoremstyle{remark}

\newtheorem{claim}[theorem]{Claim}

\renewcommand{\epsilon}{\varepsilon}

\newcommand{\eat}[1]{}

\DeclareMathOperator*{\E}{\ensuremath{\mathbb{E}}}

\newcommand{\OPT}{\ensuremath{\mathrm{OPT}}}

\newcommand{\err}{\ensuremath{\mathsf{err}}}

\newcommand{\cost}{\ensuremath{\mathsf{cost}}}

\usepackage{longtable}

\usepackage{url}

\def\set#1{\left\{ #1 \right\}}
\def\floor#1{\left\lfloor #1 \right\rfloor}
\def\ceil#1{\left\lceil #1 \right\rceil}

\def\abs#1{\left| #1 \right|}
\def\norm#1{\left\| #1 \right\|}

\newcommand{\ip}[2]{\left\langle #1,#2\right\rangle}

\renewcommand{\tilde}{\widetilde}

\title{When is Uniform Sampling Optimal?\\Coresets for Distributional Clustering}
\author{Junqiu Hu\thanks{
        Email: \texttt{junqiu2025@gmail.com}
    }\\
    Nanjing University
    \and
Lingxiao Huang\thanks{
        Email: \texttt{huanglingxiao@nju.edu.cn}
    }\\
    Nanjing University
    \and
    Baoduo Xu\thanks{ 
        Email: \texttt{bdxu03@gmail.com}}\\
    Nanjing University
}
\date{}
\begin{document}
\maketitle
\begin{abstract}
Recent work~\cite{HuHuangLiXu2026TightCoresetBounds} settled the optimal worst-case strong $\varepsilon$-coreset size for Euclidean $k$-means in unrestricted dimension at $\tilde{\Theta}(\min\{k^{3/2}\varepsilon^{-2}, k\varepsilon^{-4}\})$, showing that the natural target $\tilde{O}(k\varepsilon^{-2})$ is unattainable in general. For mixtures of $k$ isotropic Gaussian components with balanced weights and bounded scale imbalance in dimension $d\ge\log k$, we show that uniform sampling yields strong $\varepsilon$-coresets of size $\tilde{O}(k\varepsilon^{-2})$, without cluster separation. This saves a factor of $\min\{\sqrt{k},\varepsilon^{-2}\}$ over the worst-case bound, up to logarithms. For population approximation by independent, equally weighted samples, the near-linear rate is optimal up to logarithmic factors over this class.

More generally, for datasets generated from an isotropic Gaussian mixture $\mu$ satisfying a mild cost-comparison assumption, uniform sampling yields strong $\varepsilon$-coresets of size $\tilde{O}((k\zeta_\mu+\pi_{\min}^{-1})\varepsilon^{-2})$, where $\pi_{\min}$ is the minimum component weight and $\zeta_\mu$ measures scale imbalance. Our proof uses a base-shift cost decomposition: one query center per component describes the base costs, and graph spanners control the sparse corrections from points switching centers. For general mixtures under cost and tail assumptions, we obtain $\tilde{O}((k\zeta_\mu+\pi_{\min}^{-1})\varepsilon^{-2}\min\{\sqrt{k},\varepsilon^{-2}\})$ samples. We also prove a matching $\tilde{\Omega}(\min\{k^{3/2}\varepsilon^{-2},k\varepsilon^{-4}\})$ lower bound for population approximation by uniform sampling from anisotropic Gaussian mixtures, even with equal weights and a shared covariance matrix.

\end{abstract}

\newpage
\tableofcontents
\newpage
\pagenumbering{arabic}

\section{Introduction}\label{sec:problem-statement}
Clustering is a fundamental task in data analysis, machine learning, and theoretical computer science~\cite{Jain2010DataClustering}. Among clustering objectives, Euclidean $k$-means is one of the most extensively studied and widely used~\cite{BloemerLammersenSchmidtSohler2016}. Given a finite dataset $X\subseteq\R^d$ and an integer $k\geq1$, the goal is to find a set $C\subseteq\R^d$ of $k$ centers minimizing the average squared distance to the nearest center. We write
\[
    \cost(p,C):=d^2(p,C)=\min_{c\in C}\|p-c\|_2^2,
    \qquad
    \cost(X,C):=\frac{1}{|X|}\sum_{p\in X}\cost(p,C).
\]

When $X$ is large, storing, communicating, or repeatedly processing the entire dataset can be prohibitively expensive. This motivates coresets: small weighted summaries that approximately preserve the objective values of the original dataset~\cite{HM04,FL11}. A weighted subset $S\subseteq X$, with nonnegative weights $w$, is a strong $\varepsilon$-coreset for $X$ if
\[
    \cost(S,C):=\sum_{p\in S}w(p)\cost(p,C)
    \in(1\pm\varepsilon)\cost(X,C)
    \qquad\text{for every }C\subseteq\R^d\text{ with }|C|=k.
\]
Because the guarantee holds uniformly over all candidate center sets, a strong coreset can replace the full dataset for $k$-means optimization, reducing storage and computation; its composability also enables standard streaming and distributed implementations~\cite{FeldmanSchmidtSohler2020}.

A central question in coreset theory is to determine the smallest possible strong coreset size for Euclidean $k$-means. A natural target was a size nearly linear in $k$ and quadratic in $\varepsilon^{-1}$, namely $\widetilde O(k\varepsilon^{-2})$.\footnote{Throughout, $\lesssim$ and $\gtrsim$ suppress only constants, while $\widetilde O$, $\widetilde\Omega$, and $\widetilde\Theta$ additionally suppress polylogarithmic factors.} Cohen-Addad et al.~\cite{CohenAddadLSS22} established an $\Omega(k\varepsilon^{-2})$ lower bound together with an upper bound of $\widetilde O(k\varepsilon^{-2}\min\{k,\varepsilon^{-2}\})$, which was subsequently improved to $\widetilde O(k\varepsilon^{-2}\min\{\sqrt{k},\varepsilon^{-2}\})$~\cite{CohenAddadLSSO22}. Huang, Li, and Wu~\cite{DBLP:conf/stoc/Huang0024} showed that the target $\widetilde O(k\varepsilon^{-2})$ is unattainable in general by proving an $\Omega(k\varepsilon^{-4})$ lower bound when $\varepsilon=\Omega(k^{-1/4})$, matching the corresponding upper-bound regime. More recently, Hu, Huang, Li, and Xu~\cite{HuHuangLiXu2026TightCoresetBounds} proved a matching $\widetilde\Omega(k^{3/2}\varepsilon^{-2})$ lower bound for all $0<\varepsilon<k^{-1/4}$. Together, these results determine the optimal worst-case coreset size, over datasets of unrestricted dimension, as
\[
    \widetilde\Theta\!\left(\min\{k^{3/2}\varepsilon^{-2},\,k\varepsilon^{-4}\}\right).
\]

These lower bounds rely on delicate worst-case constructions~\cite{DBLP:conf/stoc/Huang0024,HuHuangLiXu2026TightCoresetBounds} that plant aligned offset vectors across clusters. In practice, simple sampling on real-world datasets performs much better than the worst-case bounds predict~\cite{SchwiegelshohnSheikhOmar2022EmpiricalKMeansCoresets}, motivating structured input models that admit $\widetilde O(k\varepsilon^{-2})$ coresets. One approach is to assume cost stability: Bansal et al.~\cite{bansal2024sensitivitysampling} showed that sensitivity sampling yields $\widetilde O(k\varepsilon^{-2})$ coresets under constant cost stability, $\OPT_{k-1}(X)\ge(1+\beta)\OPT_k(X)$ for a constant $\beta>0$, where $\OPT_m(X)$ is the minimum $m$-means cost. However, requiring a constant relative cost jump when a single center is removed is restrictive---it can fail for overlapping mixtures, or even for well-separated balanced clusters once $k$ is large enough that removing one center increases the total cost by only $O(1/k)$. 

We instead study a distributional source of structure, modeling the input $X$ as an i.i.d.\ sample from an underlying $k$-component mixture distribution---starting with isotropic Gaussian mixtures~\cite{McLachlanBasford1988,BanfieldRaftery1993,FraleyRaftery2002} and extending to general mixtures under a tail condition. For isotropic Gaussians, the offset directions are independent and uniform on the sphere, conditional on the component labels and radii. This contrasts with the aligned patterns of the worst-case constructions and leads us to ask:

\begin{quote}
\textbf{Question.} \emph{If a finite dataset $X$ is generated from an underlying mixture distribution, does uniform sampling give a strong $\varepsilon$-coreset of size $\widetilde O(k\varepsilon^{-2})$ for $X$, bypassing the worst-case barrier $\widetilde\Theta(\min\{k^{3/2}\varepsilon^{-2},k\varepsilon^{-4}\})$?}
\end{quote}

\subsection{Our contributions}\label{sec:contributions}

\paragraph{Isotropic mixtures: near-linear size.}
We answer this question affirmatively for isotropic Gaussian mixtures with balanced weights and bounded scale imbalance in dimension $d\ge\log k$, without cluster separation. More generally, the bound holds under the comparison condition~\eqref{eq:comparison_condition}. The parameters $\zeta_\mu$ and $\pi_{\min}$ capture scale and weight imbalance: $\pi_{\min}$ is the minimum component weight, and $\zeta_\mu$ is the largest component mean-squared radius divided by their mixture-weighted average (Section~\ref{sec:preliminaries}).

For datasets generated from balanced mixtures with $\pi_{\min}=\Omega(1/k)$ and $\zeta_\mu=O(1)$, uniform sampling yields strong $\varepsilon$-coresets of size $\widetilde O(k\varepsilon^{-2})$ under the same comparison condition. Equal-weight components with a common variance are a special case, with $\pi_{\min}=1/k$ and $\zeta_\mu=1$. This removes a factor $\sqrt{k}$ for $\varepsilon\le k^{-1/4}$ and $\varepsilon^{-2}$ for $\varepsilon\ge k^{-1/4}$ from the worst-case size, up to logarithms. The sampling rule needs no recovery of component labels, means, or covariances.\footnote{The logarithmic factors in our sample bounds depend on $1/(\varepsilon\pi_{\min})$, not on $d$. Constants may depend on the fixed comparison and tail parameters; logarithmic powers may depend on the fixed tail exponent.} We prove the result by first establishing a sampling guarantee for the population cost $\cost(\mu,C):=\E_{p\sim\mu}\cost(p,C)$ and then transferring it to the generated dataset.

\begin{theorem}[Coreset size for isotropic Gaussian mixtures]\label{thm:main_k_gaussian}
Let $\mu$ be an isotropic Gaussian mixture satisfying the comparison condition~\eqref{eq:comparison_condition}. For every $\varepsilon\in(0,1)$, an i.i.d.\ sample $X\sim\mu^N$ of size
\[
    N=\widetilde O\!\left((k\zeta_\mu+\pi_{\min}^{-1})\varepsilon^{-2}\right)
\]
forms a strong $\varepsilon$-coreset for $\mu$ with probability at least $0.9$.
\end{theorem}

\begin{corollary}[Uniform sampling]\label{cor:finite-generated-dataset}
Under the assumptions of Theorem~\ref{thm:main_k_gaussian}, for every $\varepsilon\in(0,1)$, let $S$ consist of
\[
    m=\widetilde O\!\left((k\zeta_\mu+\pi_{\min}^{-1})\varepsilon^{-2}\right)
\]
observations sampled uniformly without replacement from the indices of $X\sim\mu^N$, where $N\ge m$. With weight $1/m$ per observation, $S$ is a strong $\varepsilon$-coreset of $X$ with probability at least $0.9$.
\end{corollary}

The probability is over dataset generation and sampling. This finite-data reduction is proved in Appendix~\ref{appendix:finite-generated-dataset}.

Appendices~\ref{appendix:pi-min-necessity} and~\ref{appendix:zeta-necessity} establish the necessity of the dependence on $\pi_{\min}$ and $\zeta_\mu$ for population approximation by independent, equally weighted samples. In particular, the balanced near-linear rate is optimal up to logarithmic factors over the isotropic mixture class.

\paragraph{General mixtures: worst-case size by uniform sampling.}
We extend this guarantee to general mixtures under the cost and tail assumptions in Section~\ref{sec:mixture-assumptions}. In the balanced regime, the bound matches the worst-case size; the gain here is the uniform sampling rule rather than a smaller bound:

\begin{theorem}[Coreset size for general mixtures]\label{thm:vanilla_bound}
Let $\mu$ be a mixture with $V_\mu>0$, satisfying the comparison condition~\eqref{eq:comparison_condition} and a tail condition (Assumption~\ref{ass:general-tail}). For every $\varepsilon\in(0,1)$, an i.i.d.\ sample $X\sim\mu^N$ of size
\[
    N=\widetilde O\!\left((k\zeta_\mu+\pi_{\min}^{-1})\varepsilon^{-2}\min\{\sqrt{k},\varepsilon^{-2}\}\right)
\]
forms an $\varepsilon$-coreset for $\mu$ with probability at least $0.9$.
\end{theorem}

The same reduction gives finite-data coresets of this size. Appendices~\ref{appendix:comparison-necessity} and~\ref{appendix:tail-necessity} justify the cost-comparison and tail assumptions.

\paragraph{Anisotropic mixtures: worst-case lower bound.}
In contrast, for anisotropic Gaussian mixtures, the required sample size can rise to the worst-case scale, even with equal component weights and a shared covariance matrix. The following lower bound for uniform sampling is proved in Section~\ref{sec:lower-bounds}.

\begin{theorem}[Lower bound for uniform sampling]
\label{thm:gaussian-sampling-lower-bound}
For every sufficiently large $k$ and every $\varepsilon\in(0,1)$, there
is an equal-weight Gaussian mixture $\mu=k^{-1}\sum_{j=1}^k\mathcal N(a_j,\Sigma)$
in some dimension, with $\Sigma\succeq0$ and
\[
 \mathsf{tr}\Sigma=V_\mu=1=O(\cost(\mu,C))
 \quad\text{for every set $C$ of at most $k$ centers},
\]
such that, for every
\begin{equation}
 1\le N\le c\,
 \frac{\min\{k^{3/2}\varepsilon^{-2},\,k\varepsilon^{-4}\}}
      {\log^8(k/\varepsilon)},
 \label{eq:sample-range}
\end{equation}
an independent sample $X\sim\mu^N$ satisfies, with probability at least $0.9$,
\[
 |\cost(X,C)-\cost(\mu,C)|>\varepsilon\cost(\mu,C)
 \quad\text{for some set $C$ of at most $k$ centers}.
\]
Here $c>0$ is universal, and $\mu$ depends only on $k,\varepsilon$.
\end{theorem}

\subsection{Related work}

\paragraph{Coresets for structured instances.}
Bansal et al.~\cite{bansal2024sensitivitysampling} show that sensitivity sampling yields strong $\varepsilon$-coresets of size $\widetilde O(k\varepsilon^{-2})$ for datasets satisfying constant cost stability. Their result exploits a cost-stability gap in the finite dataset, whereas ours uses distributional assumptions on the generating components and uniform sampling, without requiring such a gap.

\paragraph{Uniform sampling for clustering coresets.}
Braverman et al.~\cite{braverman2022powerUniformSampling} reduce coreset construction to ring instances and apply uniform sampling within these instances; our construction samples uniformly from the entire dataset without computing such a partition.
Huang, Jiang, and Lou~\cite{huang2023poweruniformsampling} obtain $(1+\varepsilon)$-approximation guarantees for $k$-median via uniform sampling under balancedness.
Carmel and Krauthgamer~\cite{carmel2025stablecoresets} introduce \emph{stable coresets}, which approximately preserve the relative ordering of candidate costs rather than every cost individually, and construct them by uniform sampling for $1$-median in $\ell_1$ and related metrics.

Bachem et al.~\cite{bachem2017uniform} obtain kurtosis-dependent uniform deviation bounds for $k$-means over all center sets, with dimension-dependent sample complexity and an additive error term proportional to the global variance.
In concurrent and independent work, Carmel and Krauthgamer~\cite{carmel2026uniform} obtain dimension-free relative-error guarantees over $O(1)$-approximate candidate solutions under cluster balancedness, sufficiently strong $\rho$-separation, and bounded kurtosis within each optimal cluster.
In contrast, for datasets generated by isotropic Gaussian mixtures in dimension $d\ge\log k$, we obtain a \emph{strong} coreset guarantee across \emph{all} candidate center sets $C\in(\R^d)^k$ without requiring cluster separation.

\paragraph{Statistical learning and distributional coresets.}
Related work studies Gaussian-mixture density estimation in total variation~\cite{ashtiani2020nearoptimal}, coresets for likelihood-based Gaussian-mixture fitting~\cite{lucic2018traininggaussianmixturemodels,feldman2019coresets}, and excess-risk bounds for PAC $k$-means learning~\cite{cohenaddad2025almostoptimalpac}.
Our task instead uses mixture assumptions on the input distribution to justify uniform sampling that preserves ordinary $k$-means costs within relative error for every candidate center set, first for the population and then for the generated finite dataset, without learning the mixture.
Alishahi and Phillips~\cite{alishahi2024nodimensional} also study dimension-free sampling coresets for distributional inputs, but for regularized classification losses.
Work on stochastic noise~\cite{huang2025coresetsStochasticNoise} targets unobserved clean data rather than the observed dataset itself.

\section{Preliminaries}\label{sec:preliminaries}

\subsection{Mixture model and parameters}\label{sec:model}

Fix integers $k,d\ge1$. We write $[k]=\{1,\ldots,k\}$ and use $d(x,y)=\|x-y\|_2$ for Euclidean distance. We consider a mixture $\mu=\sum_{j=1}^k\pi_j\mu_j$ on $\mathbb R^d$, where each $\pi_j>0$, $\sum_{j=1}^k\pi_j=1$, and each component has finite second moment. Its minimum mixture weight is $\pi_{\min}:=\min_{j\in[k]}\pi_j$.

For each component $\mu_j$, define its mean $a_j$ and root-mean-square radius $r_j\ge0$ by
\[
    a_j:=\mathbb E_{p\sim\mu_j}p,
    \qquad r_j^2:=\mathbb E_{p\sim\mu_j}d^2(p,a_j).
\]
Let $A=\{a_1,\ldots,a_k\}$. The latent population cost, which assigns each point to its generating mean, is
\[
    V_\mu:=\sum_{j\in[k]}\pi_jr_j^2.
\]
For $V_\mu>0$, define the scale imbalance
\[
    \zeta_\mu:=\frac{\max_{j\in[k]}r_j^2}{V_\mu}.
\]
For isotropic Gaussian mixtures, $\mu_j=\mathcal N(a_j,\sigma_j I_d)$ with per-coordinate variance $\sigma_j>0$, so
\[
    r_j^2=d\sigma_j,
    \qquad \zeta_\mu=\frac{\max_j\sigma_j}{\sum_j\pi_j\sigma_j}.
\]

\subsection{Samples, costs, and population coresets}\label{sec:population-coresets}

Let $X=(p_1,\ldots,p_N)\sim\mu^N$ be a sample of $N\ge1$ independent observations. Each observation is generated by first drawing a component index $j$ with probability $\pi_j$, and then drawing a point from $\mu_j$. Let $X_j\subseteq X$ denote the submultiset of observations generated from component $\mu_j$, and set $a(p)=a_j$ for each observation $p\in X_j$. The latent labels are used only in the analysis; uniform sampling does not need them.

A center set $C\in(\mathbb R^d)^k$ is a $k$-tuple of centers. For $x\in\mathbb R^d$, write $d(x,C)=\min_{c\in C}d(x,c)$ and let $c(x)\in\arg\min_{c\in C}d(x,c)$. The pointwise cost is $\cost(p,C)=d^2(p,C)$, and the population and normalized empirical costs are
\[
    \cost(\mu,C):=\mathbb E_{p\sim\mu}d^2(p,C),
    \qquad \cost(X,C):=\frac1N\sum_{p\in X}d^2(p,C).
\]
For every fixed query $C$, $\mathbb E_X\cost(X,C)=\cost(\mu,C)$.

We also write
\[
    V_X:=\frac1N\sum_{p\in X}d^2(p,a(p))
    =\frac1N\sum_{j\in[k]}\sum_{p\in X_j}d^2(p,a_j)
\]
for the empirical within-component variance (assigning each observation $p\in X_j$ to its generating mean $a_j$ rather than its nearest mean in $A$).

\begin{definition}[$\varepsilon$-coreset for a distribution]\label{def:population-coreset}
Let $\mu$ be a distribution over $\mathbb R^d$ with finite second moment and let $\varepsilon\in(0,1)$. A finite sample $X$, with weight $1/|X|$ per observation, is an $\varepsilon$-coreset for $\mu$ with respect to $k$-means if, for every center set $C\in(\mathbb R^d)^k$,
\[
    (1-\varepsilon)\cost(\mu,C)
    \le \cost(X,C)
    \le (1+\varepsilon)\cost(\mu,C).
\]
\end{definition}

\paragraph{Restricted costs.}
For a subset $J\subseteq[k]$, define
\[
    \pi(J):=\sum_{j\in J}\pi_j,
    \qquad \mu_J:=\sum_{j\in J}\pi_j\mu_j,
    \qquad X_J:=\bigcup_{j\in J}X_j.
\]
Here $\mu_J$ is an unnormalized subprobability measure of mass $\pi(J)$, and both restricted costs retain the global normalization:
\[
    \cost(\mu_J,C):=\sum_{j\in J}\pi_j\mathbb E_{p\sim\mu_j}d^2(p,C),
    \qquad
    \cost(X_J,C):=\frac1N\sum_{p\in X_J}d^2(p,C),
\]
so that $\mathbb E_X\cost(X_J,C)=\cost(\mu_J,C)$.

\subsection{Structural assumptions}\label{sec:mixture-assumptions}

\paragraph{Comparison condition.}
We say that $\mu$ satisfies the comparison condition if
\begin{equation}\tag{CC}\label{eq:comparison_condition}
    V_\mu\lesssim\cost(\mu,C)
    \qquad\text{for every }C\in(\mathbb R^d)^k.
\end{equation}
For isotropic Gaussian mixtures, Corollary~\ref{cor:mixture_quantization_lower_bound} gives $\cost(\mu,C)\ge k^{-2/d}V_\mu$, so~\eqref{eq:comparison_condition} holds automatically whenever $d\ge\log k$. In smaller dimensions, several centers can fit one component at a cost far below its within-component variance; a counterexample without this condition is given in Appendix~\ref{appendix:comparison-necessity}.

The triangle inequality and~\eqref{eq:comparison_condition} give the component-mean distance bound
\[
    \sum_{j\in[k]}\pi_jd^2(a_j,C)
    \le2\cost(\mu,C)+2V_\mu
    \lesssim\cost(\mu,C).
\]

For general mixtures, we additionally require uniform control of large offsets relative to each component radius, which permits averaging over the radius shells in the proof.

\begin{assumption}[Tail assumption]\label{ass:general-tail}
For a fixed exponent $q>0$, every component with $r_j>0$ satisfies, for every $u\ge1$,
\begin{equation}\label{eq:general-tail}
\Pr_{p\sim\mu_j}[d(p,a_j)>ur_j]\le2\exp(-\Omega(u^q)).
\end{equation}
The positive constant implicit in the exponent is uniform over $u$, the components, and the mixtures under consideration.
\end{assumption}
The choices $q=2$ and $q=1$ give radial subgaussian and subexponential tails, respectively. Theorem~\ref{thm:tail-lower-bound} in Appendix~\ref{appendix:tail-necessity} shows the need for quantitative tail control.

\section{Technical overview}\label{sec:technical-overview}

\subsection{Isotropic Gaussian mixtures: covers and spanner counting}

\paragraph{Guiding obstruction from the worst case.}
The worst-case lower bounds~\cite{DBLP:conf/stoc/Huang0024,HuHuangLiXu2026TightCoresetBounds} hinge on reusing aligned offset patterns across clusters. The starting point of our analysis is that an isotropic Gaussian mixture behaves quite differently: conditioned on the latent labels and radii, the normalized offsets $(p-a(p))/\|p-a(p)\|_2$ are drawn independently and uniformly from $S^{d-1}$. In high dimensions, these unit offsets are pairwise almost orthogonal rather than aligned across clusters. The candidate centers may depend on the entire sample, so concentration for a fixed query or direction does not give a simultaneous guarantee over all queries. We use this directional randomness to bound the covering numbers of the realized cost vectors uniformly over all candidate center sets $C$. The target is a cover of logarithmic size $\widetilde O(k\alpha^{-2})$ at accuracy scale $\alpha$; the following reduction explains how it yields the sample bound.

\paragraph{Symmetrization, chaining, and covering numbers.}
Following the chaining framework for clustering coresets~\cite{CohenAddadLSS22,DBLP:conf/stoc/Huang0024,bansal2024sensitivitysampling}, symmetrization reduces the problem to bounding $\mathbb E_X\mathbb E_g\sup_C|Y_C|$, where
\[
    Y_C:=\sum_{p\in X}\frac{g_p d^2(p,C)}{N\cost(\mu,C)}.
\]
Here the $g_p$ are independent standard Gaussians, independent of $X$, which can now be fixed when constructing covers.

Classifying components and grouping points by offset radius and query distance reduce the covering problem to blocks of comparable geometry. We seek covers $\mathcal V(\alpha)$ with pointwise error $\alpha\err(p,C)$, where
\[
    \err(p,C):=d(p,C)d(p,a(p))+d^2(p,a(p)).
\]
The Gaussian maximal inequality pairs increment variance with logarithmic cover size. Suppressing the logarithmic and radius-shell factors accounted for below, the key cancellation is
\[
    \sqrt{\frac{\zeta_\mu\alpha^2}{N}\cdot k\alpha^{-2}}
    =\sqrt{\frac{k\zeta_\mu}{N}}.
\]
The component-mean contribution is treated separately. Thus the geometric target is a cover of logarithmic size $\widetilde O(k\alpha^{-2})$; we explain its construction first, then return to the classification and shell estimates.

\paragraph{The main technical barrier.}
To see where existing covering arguments hit a wall, consider the idealized setting where $\mu$ is an equal-weight mixture of uniform distributions on $k$ spheres of common radius $R$ centered at $a_1,\ldots,a_k$, and every component lies at comparable distance $d(a_j,C)+R=\Theta(tR)$ ($t\ge1$) from the query $C$. Then every point satisfies
\[
    \|p-a_j\|_2=R,\qquad \err(p,C)=\Theta(tR^2),
\]
and our goal is to construct a net of the realized cost vectors $(\cost(p,C))_{p\in X}$ with coordinate error $O(\alpha tR^2)$ and logarithmic size $\widetilde O(k\alpha^{-2})$. Directly discretizing the $k$ centers in $\R^d$ gives a logarithmic net size of $\widetilde O(kd)$, which is prohibitive in high dimensions. The standard remedy is to first project the points into a low-dimensional space via dimension reduction and then discretize the projected centers.

However, neither multiplicative nor additive terminal embeddings alone achieve $\widetilde O(k\alpha^{-2})$ here:
\begin{itemize}
    \item \emph{Multiplicative terminal embeddings}~\cite{narayanan2019optimal} preserve all point-to-center distances up to $(1\pm\alpha)$, which incurs squared-distance error $O(\alpha t^2R^2)$---a factor of $t$ larger than our target $\Theta(\alpha tR^2)$. Tightening the distortion to $O(\alpha/t)$ inflates the embedding dimension to $\widetilde O(t^2\alpha^{-2})$, degrading the logarithmic net size to $\widetilde O(kt^2\alpha^{-2})$.
    \item \emph{Additive terminal embeddings}~\cite{DBLP:conf/stoc/Huang0024} guarantee additive distance error $O(\alpha R)$ in $\widetilde O(\alpha^{-2})$ dimensions for terminals inside a single ball of radius $R$, which yields the desired $O(\alpha tR^2)$ squared-distance error at distance $\Theta(tR)$. Because the component means $a_1,\ldots,a_k$ can be arbitrarily far apart, however, we must apply a separate additive terminal embedding to each component ball $B(a_j,R)$. Discretizing all $k$ query centers inside each of the $k$ component embeddings then introduces $k^2$ center--component pairs, degrading the logarithmic net size to $\widetilde O(k^2\alpha^{-2})$.
\end{itemize}

\paragraph{Base-shift decomposition: one center per component.}
Instead of representing every query center inside every component's embedding, we retain only $c(a_j)$, the center nearest to the generating mean $a_j$. This defines a base cost for every point, while a nonnegative shift corrects for points whose nearest center differs from that of their mean. On the fixed distance block, write
\[
    b_p^C:=d^2(p,c(a(p))),\qquad
    s_p^C:=d^2(p,c(a(p)))-d^2(p,C),
\]
so the true cost vector satisfies $v^C=b^C-s^C$ exactly. To cover $b^C$, a componentwise additive terminal embedding needs to represent just one center per component. Discretizing its image in a ball of radius $O(tR)$ at mesh $O(\alpha R)$ gives total logarithmic net size $\widetilde O(k\alpha^{-2})$. Crucially, because we only need a net for the realized cost vectors $b^C\in\R^N$ rather than a globally consistent center set in $\R^d$, each component $\mu_j$ can discretize its own nearest center $c(a_j)$ independently inside its own low-dimensional terminal embedding, avoiding the $k^2$ cross-term blowup.

It remains to cover the shift vector. The triangle inequality gives $0\le s_p^C\le O(tR^2)$. At scale $\alpha$, only the coordinates of the current block in
\[
    S_\alpha(C):=\{p:s_p^C\ge\alpha R^2t/16\}
\]
need to be recorded; the other shift entries can be replaced by zero. If $|S_\alpha(C)|=\widetilde O(k\alpha^{-2})$ uniformly over $C$, we can give a net to the shift vector with logarithmic size $\widetilde O(k\alpha^{-2})$. Together with the base cover, this gives the desired cost-vector cover. The remaining geometric task is therefore to prove that large center-switch corrections are sparse.

\paragraph{First counting attempt: pairwise directions.}
For $p\in S_\alpha(C)\cap X_j$, the fact that $c(a_j)$ is nearest to $a_j$ implies
\[
    2\langle p-a_j,c(p)-c(a_j)\rangle=s_p^C+d^2(a_j,c(p))-d^2(a_j,c(a_j))\ge\alpha R^2t/16.
\]
Since $\|c(p)-c(a_j)\|_2=O(tR)$, the normalized offset $x_p:=(p-a(p))/\|p-a(p)\|_2$ has projection $\Omega(\alpha)$ onto its normalized center-switch direction. Our first attempt requires all sampled unit offsets to be pairwise almost orthogonal:
\[
    |\langle x_p,x_q\rangle|\le\frac1N
    \qquad\text{for all distinct sample coordinates }p,q.
\]
Their Gram matrix is then positive definite with operator norm less than $2$, so any unit direction has projection $\Omega(\alpha)$ onto at most $O(\alpha^{-2})$ offsets. Applying this bound to each ordered center pair gives $|S_\alpha(C)|=O(k^2\alpha^{-2})$. Spherical concentration and a union bound guarantee the good event with inverse-polynomial failure probability when $d\gtrsim N^2\log(eN)$. Thus there are two bottlenecks: counting center pairs incurs a $k^2$ factor rather than $k$, and global near-orthogonality already forces $d\ge N$, leaving in particular the intermediate range $\log(eN)\lesssim d<N$ untreated by this argument.

\paragraph{Second counting attempt: the center subspace.}
To address the first bottleneck, observe that the center-difference directions are not independent choices: they all belong to
\[
    L_C:=\operatorname{span}\{c_i-c_1:i\in[k]\},\qquad \dim(L_C)\le k-1.
\]
With $P_{L_C}$ denoting the orthogonal projection onto this subspace, the projection inequality gives
\[
    \alpha^2|S_\alpha(C)|
    \lesssim\sum_{p\in S_\alpha(C)}\|P_{L_C}x_p\|_2^2
    \le(k-1)\left\|\sum_{p\in X}x_px_p^\top\right\|_{\mathrm{op}}.
\]
Under the same near-orthogonality event, the operator norm is $O(1)$, giving $|S_\alpha(C)|=O(k\alpha^{-2})$ and resolving the first bottleneck. More generally, this weaker full-sample spectral bound suffices. However, the matrix has trace $N$, so its operator norm is at least $N/d$. The second bottleneck therefore remains: even the spectral formulation requires $d\gtrsim N$, leaving the intermediate-dimensional range unresolved.

\paragraph{Final argument: spanner-based signed counting.}
To overcome the dimension bottleneck while retaining nearly linear dependence on $k$, we replace full-sample spectral control by signed-subset concentration and organize the center differences through a sparse spanner. For isotropic Gaussians, the normalized offsets $x_p$ are independent and uniform on $S^{d-1}$, even conditional on the latent labels and radii. Product-sphere concentration and a union bound give the good event $\mathcal E_s$ of Lemma~\ref{lem:signed-subset-sum-good-event}: with high probability, simultaneously for every nonempty subset $T$ of sample coordinates and every choice of signs $\tau_p\in\{-1,1\}$,
\[
    \left\|\sum_{p\in T}\tau_p x_p\right\|_2
    \lesssim\sqrt{|T|}+|T|\sqrt{\frac{\log(eN/|T|)}d}.
\]
This event holds in every dimension and controls subsets and signs selected after observing the sample. It replaces the global $N/d$ obstruction by a loss depending on the selected subset.

To avoid reintroducing the $k^2$ bottleneck, form the directed shift graph on the $k$ query centers, with an edge for each ordered center pair used by a shifted point. Its underlying undirected graph has a spanner with $O(k)$ edges and paths of length $O(\log k)$ between the endpoints of every original edge~\cite{althofer1993sparse}. Every spanner edge is itself a shift-graph edge, so its center difference has norm $O(tR)$. A spanning tree can require $\Theta(k)$ hops between such endpoints; the spanner keeps the path multiplicity logarithmic.

Each center-switch difference is a signed telescoping sum along its spanner path. Summing the projection inequalities and regrouping by spanner edges produces only $O(k)$ signed subset sums, with each shifted point appearing in $O(\log k)$ of them. Applying $\mathcal E_s$ and Cauchy--Schwarz gives
\[
    \alpha|S_\alpha(C)|\lesssim\sqrt{k\log k|S_\alpha(C)|}
    +|S_\alpha(C)|\log k\sqrt{\frac{\log(eN)}d}.
\]
Let $\alpha_\star$ be a large enough universal multiple of $\log k\sqrt{\log(eN)/d}$. For $\alpha\ge\alpha_\star$, the second term can be absorbed into the left-hand side, yielding
\[
    |S_\alpha(C)|=O\!\left(k\alpha^{-2}\log k\right).
\]
The dimension requirement is now scale-dependent, with only logarithmic dependence on $N$. For $\alpha<\alpha_\star$, we have $d=O(\alpha^{-2}\log^2 k\log(eN))$, so direct discretization already gives logarithmic net size $\widetilde O(k\alpha^{-2})$. If $\alpha_\star\ge1$, the direct cover suffices at every scale. Together with the base-vector cover, these constructions resolve both bottlenecks: nearly linear dependence on $k$, with no intermediate-dimensional gap.

\paragraph{From the geometric cover to the sample bound.}
Standard cluster classifications, as in Bansal et al.~\cite[Section~2.2]{bansal2024sensitivitysampling}, separate far/close and tiny/huge components, reducing the main task to huge close components. Inner points with $d(p,a_j)\le\varepsilon^2r_j$ are controlled through component-mean costs. Under an upper-radius bound $d(p,a_j)\le2^\ell r_j$, the remaining outer points admit dyadic blocks of comparable offset radii and mean-to-center distances.

On a high-probability good event, the resulting covers $\mathcal V(\alpha)$ have pointwise error $\alpha\err(p,C)$. The increment variance is $O(2^{4\ell}\zeta_\mu\alpha^2/N)$, and our covering bound is $\log|\mathcal V(\alpha)|=\widetilde O(k\alpha^{-2}(1+\ell)^3)$. Their $\alpha^2$ and $\alpha^{-2}$ factors cancel in the Gaussian maximal inequality. Summing over logarithmically many chaining scales gives
\[
    \widetilde O\!\left(2^{2\ell}(1+\ell)^{3/2}\sqrt{\frac{k\zeta_\mu}{N}}\right).
\]

Finally, we average over shells indexed by the smallest $\ell\ge1$ bounding all normalized radii by $2^\ell$. Gaussian radial concentration gives shell probability at most $\min\{1,N\exp(-\Omega(d4^\ell))\}$ for $\ell\ge2$, so the shell factors cost only polylogarithms. Together with the component-mean contribution $\widetilde O((N\pi_{\min})^{-1/2})$, this yields Theorem~\ref{thm:main_k_gaussian}.

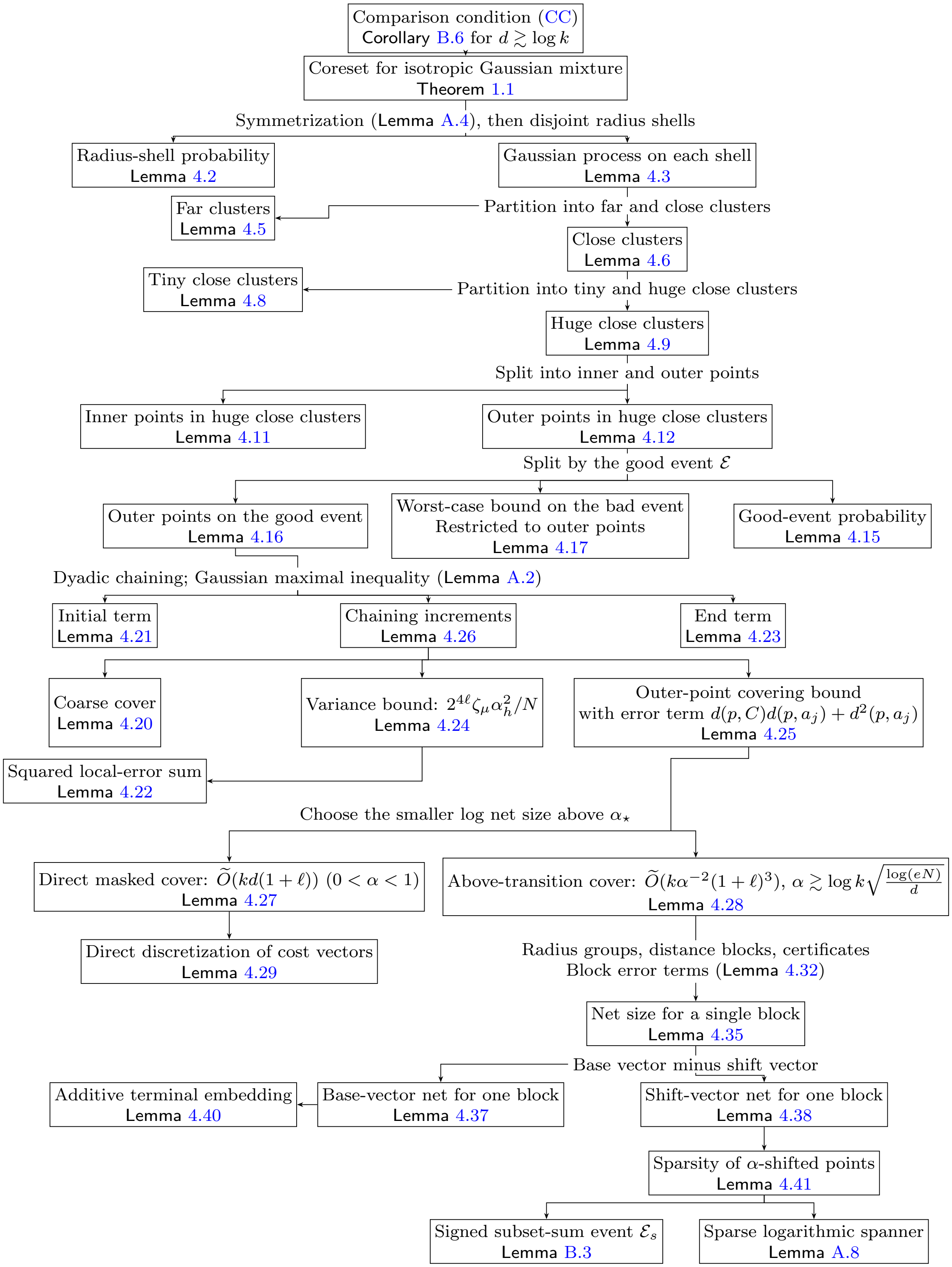
\begin{figure}[p]
\centering
\begingroup
\hypersetup{linkcolor=blue,citecolor=blue}
\begin{tikzpicture}[
  x=1cm,y=1cm,
  font=\normalfont\rmfamily\fontsize{8}{9.5}\selectfont,
  iso result/.style={draw=black,line width=.35pt,rectangle,align=center,
    inner xsep=2pt,inner ysep=2.5pt},
  iso step/.style={align=center,inner xsep=2pt,inner ysep=1pt},
  iso line/.style={line width=.35pt},
  iso arrow/.style={-{Stealth[length=1mm,width=.75mm]},line width=.35pt}
]
\node[iso result] (iso-assumptions) at (0,0)
  {Comparison condition~\eqref{eq:comparison_condition}\\
  \textsf{Corollary}~\ref{cor:mixture_quantization_lower_bound} for $d\gtrsim\log k$};
\node[iso result] (iso-main) at (0,-.8)
  {Coreset for isotropic Gaussian mixture\\\textsf{Theorem}~\ref{thm:main_k_gaussian}};
\draw[iso arrow] (iso-assumptions.south) -- (iso-main.north);
\node[iso step] (iso-symm) at (0,-1.5)
  {Symmetrization (\textsf{Lemma}~\ref{lemma:symm}), then disjoint radius shells};
\draw[iso line] (iso-main.south) -- (iso-symm.north);
\node[iso result] (iso-shellprob) at (-4.7,-2.2)
  {Radius-shell probability\\\textsf{Lemma}~\ref{lem:radius-shell-prob}};
\node[iso result] (iso-shell) at (2.6,-2.2)
  {Gaussian process on each shell\\\textsf{Lemma}~\ref{lemma:main_lemma_conditioned_on_shell}};
\draw[iso arrow] (iso-symm.south) -- ++(0,-.05) -| (iso-shellprob.north);
\draw[iso arrow] (iso-symm.south) -- ++(0,-.05) -| (iso-shell.north);
\node[iso step] (iso-farclose) at (2.6,-2.85)
  {Partition into far and close clusters};
\draw[iso line] (iso-shell.south) -- (iso-farclose.north);
\node[iso result] (iso-far) at (-3.9,-3.05)
  {Far clusters\\\textsf{Lemma}~\ref{lemma:far_clusters}};
\node[iso result] (iso-close) at (2.6,-3.55)
  {Close clusters\\\textsf{Lemma}~\ref{lemma:close_clusters}};
\draw[iso arrow] (iso-farclose.west) -- (-2.2,-2.85) |- (iso-far.east);
\draw[iso arrow] (iso-farclose.south) -- (iso-close.north);
\node[iso step] (iso-tinyhuge) at (2.6,-4.2)
  {Partition into tiny and huge close clusters};
\draw[iso line] (iso-close.south) -- (iso-tinyhuge.north);
\node[iso result] (iso-tiny) at (-3.9,-4.2)
  {Tiny close clusters\\\textsf{Lemma}~\ref{lemma:tiny_clusters}};
\node[iso result] (iso-huge) at (2.6,-4.9)
  {Huge close clusters\\\textsf{Lemma}~\ref{lemma:huge_clusters}};
\draw[iso arrow] (iso-tinyhuge.west) -- (iso-tiny.east);
\draw[iso arrow] (iso-tinyhuge.south) -- (iso-huge.north);
\node[iso step] (iso-inout) at (2.6,-5.55)
  {Split into inner and outer points};
\draw[iso line] (iso-huge.south) -- (iso-inout.north);
\node[iso result] (iso-inner) at (-3.9,-6.4)
  {Inner points in huge close clusters\\\textsf{Lemma}~\ref{lemma:inner_truncation}};
\node[iso result] (iso-outer) at (2.6,-6.4)
  {Outer points in huge close clusters\\\textsf{Lemma}~\ref{lemma:outer_points}};
\draw[iso arrow] (iso-inout.south) -- ++(0,-.1) -| (iso-inner.north);
\draw[iso arrow] (iso-inout.south) -- ++(0,-.1) -| (iso-outer.north);
\node[iso step] (iso-eventsplit) at (2.6,-7)
  {Split by the good event $\mathcal E$};
\draw[iso line] (iso-outer.south) -- (iso-eventsplit.north);
\node[iso result] (iso-good) at (-3.7,-8)
  {Outer points on the good event\\\textsf{Lemma}~\ref{lemma:bound_conditioned_on_good_event}};
\node[iso result] (iso-bad) at (1.2,-8)
  {Worst-case bound on the bad event\\Restricted to outer points\\\textsf{Lemma}~\ref{lemma:worst_case_bound}};
\node[iso result] (iso-goodprob) at (5.9,-8)
  {Good-event probability\\\textsf{Lemma}~\ref{lemma:prob-good-event}};
\draw[iso arrow] (iso-eventsplit.south) -- ++(0,-.1) -| (iso-good.north);
\draw[iso arrow] (iso-eventsplit.south) -- ++(0,-.1) -| (iso-bad.north);
\draw[iso arrow] (iso-eventsplit.south) -- ++(0,-.1) -| (iso-goodprob.north);
\node[iso step] (iso-chaining) at (-2.7,-8.85)
  {Dyadic chaining; Gaussian maximal inequality (\textsf{Lemma}~\ref{lemma:gaussian_maximal_ineq})};
\draw[iso line] (iso-good.south) -- ++(0,-.12) -| (iso-chaining.north);
\node[iso result] (iso-init) at (-5.8,-9.6)
  {Initial term\\\textsf{Lemma}~\ref{lemma:init_term_in_chaining}};
\node[iso result] (iso-increment) at (-.6,-9.6)
  {Chaining increments\\\textsf{Lemma}~\ref{lemma:main_term_in_chaining}};
\node[iso result] (iso-end) at (4.3,-9.6)
  {End term\\\textsf{Lemma}~\ref{lemma:end_term_in_chaining}};
\draw[iso arrow] (iso-chaining.south) -- ++(0,-.08) -| (iso-init.north);
\draw[iso arrow] (iso-chaining.south) -- ++(0,-.08) -| (iso-increment.north);
\draw[iso arrow] (iso-chaining.south) -- ++(0,-.08) -| (iso-end.north);
\node[iso result,minimum height=1.1cm] (iso-variance) at (-.7,-11)
  {Variance bound: $2^{4\ell}\zeta_\mu\alpha_h^2/N$\\\textsf{Lemma}~\ref{lemma:variance_bound}};
\node[iso result] (iso-errorsum) at (-5.8,-12.1)
  {Squared local-error sum\\\textsf{Lemma}~\ref{lemma:error-sum-high-dimensional}};
\node[iso result,minimum height=1.1cm] (iso-net) at (4.55,-11)
  {Outer-point covering bound\\ with error term $d(p,C)d(p,a_j)+d^2(p,a_j)$\\\textsf{Lemma}~\ref{lemma:net_size_bound}};
\node[iso result,minimum height=1.1cm] (iso-coarse) at (-5.8,-11)
  {Coarse cover\\\textsf{Lemma}~\ref{lemma:net_size_in_coarse_level}};
\coordinate (iso-incrementfork) at (-.6,-10.15);
\draw[iso arrow] (iso-increment.south) -- (iso-incrementfork) -| (iso-variance.north);
\draw[iso arrow] (iso-increment.south) -- (iso-incrementfork) -| (iso-net.north);
\draw[iso arrow] (iso-increment.south) -- (iso-incrementfork) -| (iso-coarse.north);
\draw[iso arrow] (iso-variance.south) |- (iso-errorsum.east);
\node[iso step] (iso-logsizes) at (0,-12.65)
  {Choose the smaller log net size above $\alpha_\star$};
\draw[iso line] (iso-net.south) -- ++(0,-.18) -| (3.3,-12.9);
\node[iso result] (iso-direct) at (-3.8,-13.8)
  {Direct masked cover: $\widetilde O(kd(1+\ell))$ ($0<\alpha<1$)\\\textsf{Lemma}~\ref{lemma:direct_masked_cover}};
\node[iso result] (iso-baseshift) at (3.7,-13.8)
  {Above-transition cover: $\widetilde O(k\alpha^{-2}(1+\ell)^3)$, $\alpha\gtrsim \log k\sqrt{\frac{\log(eN)}{d}}$\\
  \textsf{Lemma}~\ref{lemma:base_shift_cover}};
\draw[iso arrow] (3.3,-12.9) -| (iso-direct.north);
\draw[iso arrow] (3.3,-12.9) -| (iso-baseshift.north);
\node[iso result] (iso-discretization) at (-3.8,-15)
  {Direct discretization of cost vectors\\\textsf{Lemma}~\ref{lemma:general_alpha_net_size}};
\draw[iso arrow] (iso-direct.south) -- (iso-discretization.north);
\node[iso step] (iso-groups) at (3.7,-15)
  {Radius groups, distance blocks, certificates\\Block error terms (\textsf{Lemma}~\ref{prop:scale_properties})};
\draw[iso line] (iso-baseshift.south) -- (iso-groups.north);
\node[iso result] (iso-block) at (3.7,-16)
  {Net size for a single block\\\textsf{Lemma}~\ref{lemma:fixed-block-net}};
\draw[iso arrow] (iso-groups.south) -- (iso-block.north);
\node[iso step] (iso-subtract) at (3.7,-16.65)
  {Base vector minus shift vector};
\draw[iso line] (iso-block.south) -- (iso-subtract.north);
\node[iso result] (iso-base) at (-.4,-17.3)
  {Base-vector net for one block\\\textsf{Lemma}~\ref{lemma:base_vec_net}};
\node[iso result] (iso-shift) at (4.8,-17.3)
  {Shift-vector net for one block\\\textsf{Lemma}~\ref{lemma:shift_vec_net}};
\draw[iso arrow] (iso-subtract.west) -- ++(-.1,0) -| (iso-base.north);
\draw[iso arrow] (iso-subtract.south) -- ++(0,-.04) -| (iso-shift.north);
\node[iso result] (iso-embedding) at (-4.7,-17.3)
  {Additive terminal embedding\\\textsf{Lemma}~\ref{lem:additive_terminal_embedding}};
\draw[iso arrow] (iso-base.west) -- (iso-embedding.east);
\node[iso result] (iso-sparsity) at (4.8,-18.4)
  {Sparsity of $\alpha$-shifted points\\\textsf{Lemma}~\ref{lemma:alpha_shifted_points_num}};
\draw[iso arrow] (iso-shift.south) -- (iso-sparsity.north);
\node[iso result] (iso-signed) at (1.3,-19.5)
  {Signed subset-sum event $\mathcal E_s$\\\textsf{Lemma}~\ref{lem:signed-subset-sum-good-event}};
\node[iso result] (iso-spanner) at (5.6,-19.5)
  {Sparse logarithmic spanner\\\textsf{Lemma}~\ref{lem:sparse-log-spanner}};
\draw[iso arrow] (iso-sparsity.south) -- ++(0,-.12) -| (iso-signed.north);
\draw[iso arrow] (iso-sparsity.south) -- ++(0,-.12) -| (iso-spanner.north);
\end{tikzpicture}
\endgroup
\caption{Structural diagram of the proof of the isotropic Gaussian mixture upper bound (Theorem~\ref{thm:main_k_gaussian}).}
\label{fig:proof-tree-isotropic}
\end{figure}

The proof dependencies are summarized in Figure~\ref{fig:proof-tree-isotropic}.

\subsection{Extension to general mixtures}

For the more general mixtures in Section~\ref{sec:general-mixtures}, a tail condition still permits the shell argument. Without directional concentration, however, the sparse-shift construction is unavailable. We instead adapt two deterministic covering constructions, together with their corresponding variance estimates. Relative-error coverings based on multiplicative terminal embeddings~\cite{CohenAddadLSS22} retain entropy $\widetilde O(k\alpha^{-2})$, but their larger approximation error incurs an additional $\varepsilon^{-2}$ variance factor. The distance-sensitive construction of Huang, Li, and Wu~\cite{DBLP:conf/stoc/Huang0024} combines multiplicative and componentwise additive terminal embeddings, yielding blockwise logarithmic net size proportional to $k\alpha^{-2}\min\{t^2,1+k/t^2\}$ and hence bounded by $\widetilde O(k^{3/2}\alpha^{-2})$. Taking the smaller of the sample bounds from these two branches yields the factor $\min\{\sqrt{k},\varepsilon^{-2}\}$ in Theorem~\ref{thm:vanilla_bound}.

\begin{figure}[!htb]
\centering
\hypersetup{linkcolor=blue,citecolor=blue}
\begin{tikzpicture}[
 x=1cm,y=1cm,
 font=\normalfont\rmfamily\fontsize{8.2}{9.4}\selectfont\hyphenpenalty10000\exhyphenpenalty10000,
 gen result/.style={draw=black,line width=.35pt,rectangle,align=center,inner xsep=2pt,inner ysep=2pt},
 gen step/.style={align=center,inner xsep=2pt,inner ysep=2pt},
 gen arrow/.style={-{Stealth[length=1mm,width=.75mm]},line width=.35pt},
 gen line/.style={line width=.35pt}
]
\node[gen result] (gen-assumptions) at (0,0)
 {General mixtures with $V_\mu>0$\\
  Comparison condition~\eqref{eq:comparison_condition}; tail assumption: Assumption~\ref{ass:general-tail}};
\node[gen result] (gen-root) at (0,-1.05)
 {Coreset size for general mixtures\\
  \textsf{Theorem}~\ref{thm:vanilla_bound}};
\draw[gen arrow] (gen-assumptions.south) -- (gen-root.north);
\node[gen step] (gen-shell) at (0,-2.1)
 {Symmetrization (\textsf{Lemma}~\ref{lemma:symm}); sum over radius shells (\textsf{Lemma}~\ref{lem:general-shells}); Markov's inequality};
\draw[gen line] (gen-root.south) -- (gen-shell.north);
\node[gen result] (gen-countfail) at (-4.9,-3.35)
 {Unconditional count-failure bound on $\overline{\mathcal E_c}$\\
  \textsf{Lemma}~\ref{lem:general-shells}};
\node[gen result] (gen-reduce) at (2.3,-3.35)
 {Bound contributions of far and tiny close clusters\\
  and inner points; retain outer points on $\mathcal E_c\cap\mathscr E_\ell$\\
  \textsf{Lemma}~\ref{lem:general-reductions}};
\draw[gen arrow] (gen-shell.south) -- (0,-2.7) -| (gen-countfail.north);
\draw[gen arrow] (gen-shell.south) -- (0,-2.7) -| (gen-reduce.north);
\node[gen step] (gen-chain) at (2.3,-4.65)
 {Dyadic chaining on $X_j^{\mathrm{out}}$, $j\in J_{\mathrm{huge}}(C)$\\
  Coarse cover: \textsf{Lemma}~\ref{lemma:net_size_in_coarse_level}};
\draw[gen line] (gen-reduce.south) -- (gen-chain.north);
\node[gen result] (gen-endpoints) at (-4.9,-4.65)
 {Initial and end terms\\
  \textsf{Lemma}~\ref{lem:general-endpoints}\\
  End: branch square-sum bounds};
\draw[gen arrow] (gen-chain.west) -- (gen-endpoints.east);
\node[gen step] (gen-maximal) at (0,-5.8)
 {Increment bound~\eqref{eq:general-finite-maximum}\\
  Gaussian maximal inequality: \textsf{Lemma}~\ref{lemma:gaussian_maximal_ineq}};
\draw[gen line] (gen-chain.south) -- (2.3,-5.25) -| (gen-maximal.north);
\node[gen result,text width=6.5cm] (gen-rel) at (-3.8,-7.8)
 {Relative-error branch (Section~\ref{sec:general-relative})\\
  error term $d^2(p,C)+d^2(p,a_j)$ \\
  \textsf{Lemma}~\ref{lem:general-relative-main}\\
  $N=\widetilde O(k\zeta_\mu\varepsilon^{-4}+\pi_{\min}^{-1}\varepsilon^{-2})$};
\node[gen result,text width=7.3cm] (gen-dist) at (3.7,-7.8)
 {Distance-sensitive branch (Section~\ref{sec:general-distance})\\
  \mbox{error term $\bigl(d(p,C)d(p,a_j)+d^2(p,a_j)\bigr)\sqrt{1+\frac{\cost(\mu,C)}{k\pi_{\min}R^2}}$}\\
  \textsf{Lemma}~\ref{lem:general-distance-main}\\
  $N=\widetilde O((k\zeta_\mu+\pi_{\min}^{-1})\sqrt{k}\,\varepsilon^{-2})$};
\draw[gen arrow] (gen-maximal.south) -- (0,-6.45) -| (gen-rel.north);
\draw[gen arrow] (gen-maximal.south) -- (0,-6.45) -| (gen-dist.north);
\node[gen result,text width=3.15cm,minimum height=1.45cm] (gen-relvar) at (-5.55,-10.6)
 {Variance bound $2^{4\ell}\alpha_h^2\varepsilon^{-2}\zeta_\mu/N$\\
  \textsf{Lemma}~\ref{lem:general-relative-variance}\\
  };
\node[gen result,text width=3.15cm,minimum height=1.45cm] (gen-relcover) at (-2.05,-10.6)
 {Logarithmic net size\\
  $\widetilde O\!\left(k\alpha^{-2}(1+\ell)\right)$\\
  \textsf{Lemma}~\ref{lem:general-relative-cover}\\
  };
\node[gen result,text width=3.55cm,minimum height=1.45cm] (gen-distvar) at (1.8,-10.6)
 {Variance bound\\
  $2^{4\ell}\alpha_h^2\left(\zeta_\mu+k^{-1}\pi_{\min}^{-1}\right)/N$\\
  \textsf{Lemma}~\ref{lem:general-distance-variance}\\
  };
\node[gen result,text width=3.15cm,minimum height=1.45cm] (gen-distcover) at (5.6,-10.6)
 {Logarithmic net size\\
 $\widetilde O\!\left(k^{3/2}\alpha^{-2}(1+\ell)^3\right)$\\
  \textsf{Lemma}~\ref{lem:general-distance-cover}\\
  };
\draw[gen arrow] (gen-rel.south) -- (-3.8,-9.25) -| (gen-relvar.north);
\draw[gen arrow] (gen-rel.south) -- (-3.8,-9.25) -| (gen-relcover.north);
\draw[gen arrow] (gen-dist.south) -- (3.7,-9.25) -| (gen-distvar.north);
\draw[gen arrow] (gen-dist.south) -- (3.7,-9.25) -| (gen-distcover.north);
\node[gen result,text width=6.5cm] (gen-relconstruct) at (-3.8,-12.6)
 {Multiplicative terminal embedding~\cite{narayanan2019optimal}\\
  Direct discretization\\
  \textsf{Lemma}~\ref{lemma:general_alpha_net_size}\\
  Union over active component subsets};
\draw[gen arrow] (gen-relcover.south) -- ++(0,-.25) -| (gen-relconstruct.north);
\end{tikzpicture}
\caption{Structural diagram of the proof of the general-mixture upper bound (Theorem~\ref{thm:vanilla_bound}). The two branches are alternatives. Both covers approximate the active outer costs and are zero on all other coordinates.}
\label{fig:proof-tree-general}
\end{figure}
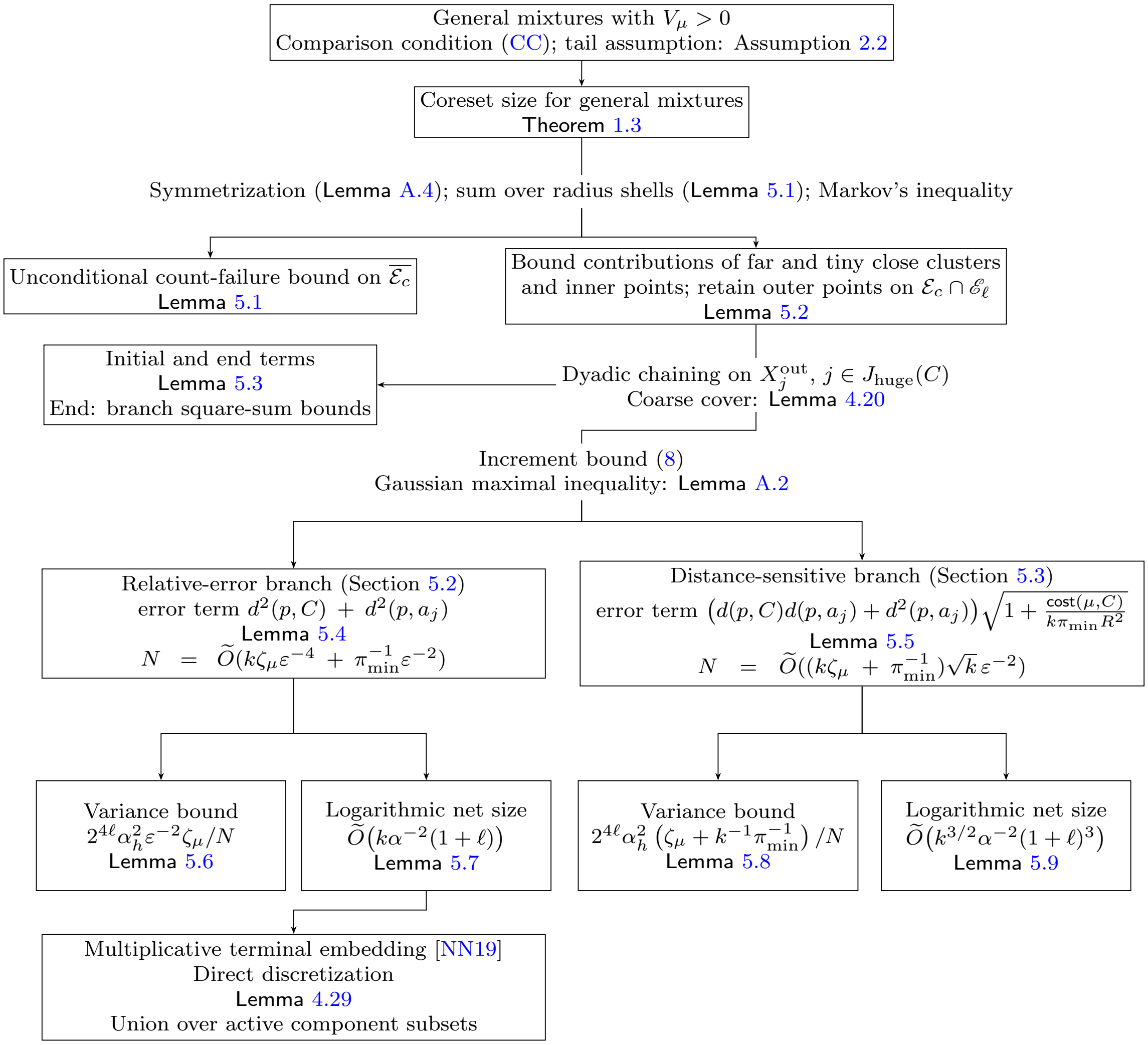

The proof dependencies are summarized in Figure~\ref{fig:proof-tree-general}.

\subsection{Anisotropic lower bound via Gaussian orthant cells}
The construction in Section~\ref{sec:lower-bounds} places all random variation in a common $r$-dimensional subspace, using covariance $\Sigma=r^{-1}I_r\oplus0$ with $2^r=\Theta(k)$. The offsets remain independent, but their shared subspace lets the same query centers affect many components. Auxiliary coordinates with zero variance distinguish $\Theta(\sqrt{k})$ active components, while the Gaussian signs partition each of them into $\Theta(k)$ orthants. A norm-adjustment coordinate keeps each query center's squared norm fixed, so changing a component control affects distances only in that component. Fixed even-orthant centers serve as alternatives to adjustable odd-orthant centers, localizing each change to its own orthant. The odd orthants of the active components give $\Theta(k^{3/2})$ cells whose costs can be adjusted independently using at most $k$ centers. Because the cost gap can shrink near orthant boundaries, we bound its first, second, and fourth moments; the resulting loss is only logarithmic in $k$.

Opposite choices of the adjustable center coordinates cancel the common costs and accumulate the absolute empirical errors across these cells, following the shared-center and opposite-query approach of Hu et al.~\cite{HuHuangLiXu2026TightCoresetBounds}. Thus the common low-dimensional support recreates the coordinated geometry behind the discrete worst-case obstruction, despite each component being Gaussian. Empty cells and frequency fluctuations yield the fine-accuracy bounds, while separated copies and a coarse-accuracy construction cover the remaining parameter range. Exact copy separation and a Gaussian small-ball estimate preserve the ordinary nearest-center objective and establish the comparison condition for every query.

\section{Proof of Theorem~\ref{thm:main_k_gaussian}: coreset upper bounds for isotropic Gaussian mixtures}\label{sec:upper-bounds}

Section~\ref{sec:upper-bounds} establishes our population coreset upper bounds for isotropic Gaussian mixtures (Theorem~\ref{thm:main_k_gaussian}). The new geometric ingredient is the single-block base-shift argument. Readers can start with its local setup in Section~\ref{sec:proof_of_fixed_block_net}, then read the base-net and shift-sparsity proofs; the preceding reductions explain how the sampled points form these blocks.

\subsection{Reduction to a localized Gaussian process}\label{sec:proof_main_theorem}

Throughout this section, fix $\varepsilon\in(0,1)$ and let $X\sim\mu^N$, where $N$ is sufficiently large according to the sample-size bound in Theorem~\ref{thm:main_k_gaussian}. The hidden polylogarithmic factor is chosen sufficiently large to absorb all logarithmic losses appearing below. We use the comparison condition in the form
\[
    V_\mu\lesssim\cost(\mu,C),
    \quad\sum_{j=1}^k\pi_jd^2(a_j,C)\lesssim\cost(\mu,C),
    \qquad\text{for every }C\in(\mathbb R^d)^k, 
\]
with one fixed universal comparison constant.

\paragraph{Symmetrization argument.}
Apply Lemma~\ref{lemma:symm} to the normalized costs and their negatives. With independent $g_p\sim\mathcal N(0,1)$, this gives
\[
    \mathbb E_X\sup_C
    \frac{|\cost(X,C)-\cost(\mu,C)|}{\cost(\mu,C)}
    \lesssim\mathbb E_X\mathbb E_g\sup_C|Y_C|,
    \qquad
    Y_C:=\sum_{p\in X}\frac{g_p d^2(p,C)}{N\cost(\mu,C)}.
\]
Symmetrization precedes every shell conditioning and replacement by a component-mean cost. All subsequent restrictions use this original sample and its Gaussian signs.

\subsubsection{Decomposition into radius shells}

Because Gaussian offsets are unbounded, we partition the sample space in Definition~\ref{def:goodevent_E_ell_minus_E_ell-1} into disjoint dyadic shells indexed by the largest normalized offset radius.

\begin{definition}[Radius events and shells]\label{def:goodevent_E_ell_minus_E_ell-1}
Set $\mathcal E_0:=\varnothing$. For every integer $\ell\ge1$, define
\[
    \mathcal E_\ell:=\{d(p,a_j)\le2^\ell r_j
    \text{ for every }j\in[k]\text{ and }p\in X_j\},
    \qquad
    \mathscr E_\ell:=\mathcal E_\ell\setminus\mathcal E_{\ell-1}.
\]
\end{definition}
These events are increasing, and the shells are disjoint and cover the sample space up to a null set. In particular, $\mathscr E_1=\mathcal E_1$. There is no pointwise lower-radius requirement. For $\ell\ge2$, at least one normalized radius exceeds $2^{\ell-1}$.

Using this lower-radius condition for higher shells, the next lemma bounds the probability of each radius shell via Gaussian radial concentration.

\begin{lemma}[Probability of the radius shell]\label{lem:radius-shell-prob}
For every $\ell\ge2$,
\[
    \Pr[\mathscr E_\ell]\le\min\{1,N\exp(-d4^{\ell-1}/8)\}.
\]
For the first shell, $\Pr[\mathscr E_1]\le1$.
\end{lemma}
The proof is in Appendix~\ref{appendix:radius-shell-prob}. To pair with this shell tail bound, the next lemma controls the symmetrized Gaussian process conditioned on each radius shell.

\begin{lemma}[Gaussian process bound on a radius shell]\label{lemma:main_lemma_conditioned_on_shell}
Under the standing assumptions, for every $d\ge1$ and $\ell\ge1$,
\[
    \mathbb E_{X\mid\mathscr E_\ell}\mathbb E_g\sup_C|Y_C|
    \lesssim 2^{4\ell}\ell\varepsilon.
\]
\end{lemma}

\begin{proof}[Proof of Theorem~\ref{thm:main_k_gaussian}]
Lemma~\ref{lem:radius-shell-prob} implies the explicit shell-moment bound
\begin{equation}\label{eq:shell_moment}\tag{M}
    \sum_{\ell\ge1}\Pr[\mathscr E_\ell]\,2^{4\ell}\ell
    \lesssim \left(1+\frac{\log(eN)}{d}\right)^2
        \left(1+\log\left(1+\frac{\log(eN)}{d}\right)\right)=\widetilde{O}(1).
\end{equation}
The first comparison is up to a universal constant in every dimension. Only the final $\widetilde{O}(1)$ suppresses a log factor. To prove the first comparison, split the sum at
\[
    m:=\left\lceil\log_4\left(64\left(1+\frac{\log(eN)}d\right)\right)\right\rceil.
\]
The sum up to $m$ is $\lesssim16^m m$. For $\ell=m+i$, $i\ge1$, the shell probability is at most $\exp(-4^i)$, so the remaining sum is
\[
    16^m\sum_{i\ge1}16^i(m+i)e^{-4^i}\lesssim16^m m.
\]
Substituting the value of $m$ proves \eqref{eq:shell_moment}.

For target accuracy $\varepsilon$, apply all intermediate estimates, including the count and chaining bounds, at a sufficiently small universal multiple of $\varepsilon/[\log(eN)]^3$. The explicit bound in~\eqref{eq:shell_moment} then makes the shell-averaged error a sufficiently small multiple of $\varepsilon$.

Set $L=\log(e/(\varepsilon\pi_{\min}))$ and take
\[
N=\left\lceil C(k\zeta_\mu+\pi_{\min}^{-1})\varepsilon^{-2}L^B\right\rceil
\]
for sufficiently large fixed constants $B,C$. Since $k,\zeta_\mu\le\pi_{\min}^{-1}$, the prefactor is at most $2\pi_{\min}^{-2}\varepsilon^{-2}$, and hence $\log(eN)=O(L)$ for fixed $B,C$. The accuracy replacement, the chain length, and the entropy bound in Lemma~\ref{lemma:net_size_bound} therefore cost only fixed powers of $L$. Choosing $B,C$ large enough absorbs these losses and ensures the component-count hypothesis.

Combining symmetrization with the shell lemma at this smaller accuracy and \eqref{eq:shell_moment} gives
\[
    \mathbb E_X\sup_C
    \frac{|\cost(X,C)-\cost(\mu,C)|}{\cost(\mu,C)}
    \le\varepsilon/10,
\]
after choosing the universal constant in the accuracy replacement sufficiently small. Markov's inequality gives failure probability at most $1/10$.
\end{proof}

\subsubsection{Proof of Lemma~\ref{lemma:main_lemma_conditioned_on_shell}: far and close clusters}

To prove Lemma~\ref{lemma:main_lemma_conditioned_on_shell} on a fixed shell, we separate components whose offsets are small relative to their mean-to-center distance. For these far components, point costs can be approximated by component-mean costs. The remaining close components require the geometric cover.

\begin{definition}[Far and close clusters]
For a fixed shell $\ell$, define
\[
    J_{\mathrm{far}}(C):=\{j\in[k]:d(a_j,C)>\varepsilon^{-1}2^\ell r_j\},
    \qquad
    J_{\mathrm{close}}(C):=[k]\setminus J_{\mathrm{far}}(C).
\]
\end{definition}
The corresponding restrictions of the Gaussian process satisfy
\[
    \sup_C|Y_C|
    \le\sup_C\left|\sum_{p\in X_{J_{\mathrm{far}}(C)}}
    \frac{g_p d^2(p,C)}{N\cost(\mu,C)}\right|
    +\sup_C\left|\sum_{p\in X_{J_{\mathrm{close}}(C)}}
    \frac{g_p d^2(p,C)}{N\cost(\mu,C)}\right|.
\]

We bound the two terms on the right-hand side in turn. The next two lemmas control the far and close-component contribution. 

\begin{lemma}[Contribution of far clusters]\label{lemma:far_clusters}
If $N\pi_{\min}\gtrsim\varepsilon^{-2}\log(2k)$, then, for every $\ell\ge1$,
\[
    \mathbb E_{X\mid\mathscr E_\ell}\mathbb E_g\sup_C
    \left|\sum_{p\in X_{J_{\mathrm{far}}(C)}}
    \frac{g_p d^2(p,C)}{N\cost(\mu,C)}\right|\lesssim\varepsilon.
\]
\end{lemma}

\begin{lemma}[Contribution of close clusters]\label{lemma:close_clusters}
Under the standing assumptions, for every $\ell\ge1$,
\[
    \mathbb E_{X\mid\mathscr E_\ell}\mathbb E_g\sup_C
    \left|\sum_{p\in X_{J_{\mathrm{close}}(C)}}
    \frac{g_p d^2(p,C)}{N\cost(\mu,C)}\right|
    \lesssim2^{4\ell}\ell\varepsilon.
\]
\end{lemma}

Lemma~\ref{lemma:far_clusters} is proved by approximating point costs with component-mean costs; the proof is deferred to Appendix~\ref{appendix:far_clusters}. And the proof of Lemma~\ref{lemma:close_clusters} is given in Section~\ref{sec:proof_of_close_clusters}. Now we can give the proof of Lemma~\ref{lemma:main_lemma_conditioned_on_shell}.

\begin{proof}[Proof of Lemma~\ref{lemma:main_lemma_conditioned_on_shell}]
The far and close sets partition the components. Apply the triangle inequality and Lemmas~\ref{lemma:far_clusters} and~\ref{lemma:close_clusters} to obtain
\[
    \mathbb E_{X\mid\mathscr E_\ell}\mathbb E_g\sup_C|Y_C|
    \lesssim\varepsilon+2^{4\ell}\ell\varepsilon\lesssim2^{4\ell}\ell\varepsilon.
\]
\end{proof}

\subsubsection{Proof of Lemma~\ref{lemma:close_clusters}: tiny and huge close clusters}\label{sec:proof_of_close_clusters}

To prove Lemma \ref{lemma:close_clusters}, we subdivide them at a polynomial radius threshold, separating tiny close components from huge close components with a controlled radius ratio, since close components can have arbitrarily small radii relative to the largest component. We first give the definitions of tiny and huge clusters.

\begin{definition}[Tiny and huge clusters]
Define
\[
    J_{\mathrm{tiny}}(C):=\left\{j\in J_{\mathrm{close}}(C):
    r_j<k^{-1}\varepsilon^{3/2}\zeta_\mu^{-1/2}\max_{i\in[k]}r_i\right\},
    \qquad
    J_{\mathrm{huge}}(C):=J_{\mathrm{close}}(C)\setminus J_{\mathrm{tiny}}(C).
\]
\end{definition}

Under \eqref{eq:comparison_condition}, the next lemma bounds the normalized Gaussian-process contribution of the tiny close components on each fixed shell.

\begin{lemma}[Contribution of tiny close clusters]\label{lemma:tiny_clusters}
For every deterministic sample in $\mathscr E_\ell$, under the comparison condition,
\[
    \mathbb E_g\sup_C\left|\sum_{p\in X_{J_{\mathrm{tiny}}(C)}}
    \frac{g_p d^2(p,C)}{N\cost(\mu,C)}\right|
    \lesssim2^{2\ell}\varepsilon/k^2.
\]
\end{lemma}

The proof is in Appendix~\ref{appendix:tiny_clusters}. With the tiny close components controlled, proving Lemma~\ref{lemma:close_clusters} reduces to establishing the following shell-conditioned bound for the remaining huge close components. The proof of the following lemma is in Section \ref{sec:proof_of_huge_clusters}.

\begin{lemma}[Contribution of huge close clusters]\label{lemma:huge_clusters}
Under the standing assumptions, for every $\ell\ge1$,
\[
    \mathbb E_{X\mid\mathscr E_\ell}\mathbb E_g\sup_C
    \left|\sum_{p\in X_{J_{\mathrm{huge}}(C)}}
    \frac{g_p d^2(p,C)}{N\cost(\mu,C)}\right|
    \lesssim2^{4\ell}\ell\varepsilon.
\]
\end{lemma}
Now we give the proof of Lemma~\ref{lemma:close_clusters}.
\begin{proof}[Proof of Lemma~\ref{lemma:close_clusters}]
Partition the close-cluster sum into the tiny and huge components. Apply the triangle inequality and Lemmas~\ref{lemma:tiny_clusters} and~\ref{lemma:huge_clusters}.
\end{proof}

\subsubsection{Proof of Lemma~\ref{lemma:huge_clusters}: inner and outer points}\label{sec:proof_of_huge_clusters}

Within each huge close component, we split the sample points at the fixed radius cutoff $\varepsilon^2r_j$, separating inner points near the mean from outer points.

\begin{definition}[Inner and outer points]
For each component $j\in[k]$, define
\[
    X_j^{\mathrm{in}}:=\{p\in X_j:d(p,a_j)\le\varepsilon^2r_j\},\qquad X_j^{\mathrm{out}}:=X_j\setminus X_j^{\mathrm{in}}.
\]
We call the points in $X_j^{\mathrm{in}}$ \emph{inner} and those in $X_j^{\mathrm{out}}$ \emph{outer}. On the radius shell $\mathscr E_\ell$, every outer point satisfies
\[
    \varepsilon^2r_j<d(p,a_j)\le2^\ell r_j.
\]
The split is fixed by $X$ and $\varepsilon$ and does not depend on $\ell$, $C$, or the Gaussian signs. For each query $C$, we use this split only within the huge close components $j\in J_{\mathrm{huge}}(C)$.
\end{definition}

We bound the inner and outer parts of the huge close components separately. The next lemma controls the inner-point contribution by approximating point costs with component-mean costs.

\begin{lemma}[Contribution of inner points]\label{lemma:inner_truncation}
Under the standing assumptions, for every $\ell\ge1$,
\[
    \mathbb E_{X\mid\mathscr E_\ell}\mathbb E_g\sup_C\left|\sum_{j\in J_{\mathrm{huge}}(C)}\sum_{p\in X_j^{\mathrm{in}}}\frac{g_p d^2(p,C)}{N\cost(\mu,C)}\right|\lesssim\varepsilon.
\]
\end{lemma}

The proof is in Appendix~\ref{appendix:inner-truncation}. It remains to bound the outer points of the huge close components conditioned only on the shell, as stated next.

\begin{lemma}[Contribution of outer points]\label{lemma:outer_points}
Under the standing assumptions, for every $\ell\ge1$,
\[
    \mathbb E_{X\mid\mathscr E_\ell}\mathbb E_g\sup_C
    \left|\sum_{j\in J_{\mathrm{huge}}(C)}\sum_{p\in X_j^{\mathrm{out}}}
    \frac{g_p d^2(p,C)}{N\cost(\mu,C)}\right|
    \lesssim2^{4\ell}\ell\varepsilon.
\]
\end{lemma}
The proof is in Section~\ref{sec:proof_of_outer_points}. Now we are ready to prove Lemma~\ref{lemma:huge_clusters}.

\begin{proof}[Proof of Lemma~\ref{lemma:huge_clusters}]
Split each huge close component into its inner and outer points. The triangle inequality and Lemmas~\ref{lemma:inner_truncation} and~\ref{lemma:outer_points} give
\begin{align*}
    &\mathbb E_{X\mid\mathscr E_\ell}\mathbb E_g\sup_C\left|\sum_{p\in X_{J_{\mathrm{huge}}(C)}}
    \frac{g_p d^2(p,C)}{N\cost(\mu,C)}\right|\\
    &\qquad\le\mathbb E_{X\mid\mathscr E_\ell}\mathbb E_g\sup_C\left|\sum_{j\in J_{\mathrm{huge}}(C)}\sum_{p\in X_j^{\mathrm{in}}}
    \frac{g_p d^2(p,C)}{N\cost(\mu,C)}\right|\\
    &\qquad\quad+\mathbb E_{X\mid\mathscr E_\ell}\mathbb E_g\sup_C\left|\sum_{j\in J_{\mathrm{huge}}(C)}\sum_{p\in X_j^{\mathrm{out}}}
    \frac{g_p d^2(p,C)}{N\cost(\mu,C)}\right|\\
    &\qquad\lesssim\varepsilon+2^{4\ell}\ell\varepsilon
    \lesssim2^{4\ell}\ell\varepsilon.
\end{align*}
\end{proof}

\subsubsection{Proof of Lemma~\ref{lemma:outer_points}: good events}\label{sec:proof_of_outer_points}

To prove Lemma~\ref{lemma:outer_points}, we use two good events. Component-count control bounds the local-error sums and hence the increment variances, while signed-subset control bounds large center-switch corrections and the covering entropy. Their intersection holds with high probability.

\begin{definition}[Signed subset-sum good event $\mathcal E_s$]\label{def:signed-subset-sum-good-event}
    Fix a sufficiently large universal constant $C_2$. For disjoint subsets $T_+,T_-$ of the sample coordinates, write $m=|T_+|+|T_-|$. The event $\mathcal E_s$ requires
    \[
    \left\|\sum_{p\in T_+}\frac{p-a(p)}{\|p-a(p)\|_2}
    -\sum_{p\in T_-}\frac{p-a(p)}{\|p-a(p)\|_2}\right\|_2
    \le C_2\left(\sqrt m+m\sqrt{\frac{\log(eN/m)}{d}}\right).
    \]
    For empty subsets the right-hand side is zero. A zero-radius point has probability zero; its normalized direction may be assigned any unit vector. Subsets always retain sample-coordinate multiplicities.
\end{definition}

\begin{definition}[Good event $\mathcal E$]\label{def:good-event_E}
Set $\mathcal E:=\mathcal E_c\cap\mathcal E_s$, where
\[
    \mathcal E_c:=\left\{ \frac{1}{2}N\pi_j\le|X_j|\le\frac32N\pi_j
    \text{ for every }j\in[k] \right\}.
\]
\end{definition}
Since $\mathcal E_s$ holds for every pair of disjoint subsets of the original sample, its bound also holds for the outer subsets. On $\mathcal E_c$, these subsets satisfy $|X_j^{\mathrm{out}}|\le|X_j|\le\frac32 N\pi_j$.

Because Lemma~\ref{lemma:outer_points} averages over the entire shell $\mathscr E_\ell$, we split that expectation according to whether $\mathcal E$ holds and first bound the unconditional failure probability $\Pr[\overline{\mathcal E}]$.

\begin{lemma}[Probability of the good event $\mathcal E$]\label{lemma:prob-good-event}
If $N\ge2$ and $N\pi_{\min}\gtrsim\log(kN)$ with a large enough implicit constant, then $C_2$ can be chosen universally so that
\[
    \Pr\left[\overline{\mathcal E}\right]\le N^{-2}.
\]
These hypotheses hold for the prescribed sample size.
\end{lemma}
The proof is in Appendix~\ref{appendix:prob-good-event}. For the contribution on $\mathcal E$, it suffices to bound the outer-point process for each fixed sample in $\mathcal E\cap\mathscr E_\ell$.

\begin{lemma}[Contribution of outer points on the good event $\mathcal E$]\label{lemma:bound_conditioned_on_good_event}
Under the standing assumptions, every deterministic $X\in\mathcal E\cap\mathscr E_\ell$ satisfies
\[
    \mathbb E_g\sup_C\left|\sum_{j\in J_{\mathrm{huge}}(C)}\sum_{p\in X_j^{\mathrm{out}}}
    \frac{g_p d^2(p,C)}{N\cost(\mu,C)}\right|
    \lesssim2^{4\ell}\ell\varepsilon.
\]
\end{lemma}
The proof is in Section~\ref{sec:chaining}. On the complementary set $\overline{\mathcal E}\cap\mathscr E_\ell$, where the count and directional bounds need not hold, we pair the small probability of $\overline{\mathcal E}$ with a coarse deterministic bound valid for every sample in $\mathscr E_\ell$.

\begin{lemma}[Worst-case bound on a shell]\label{lemma:worst_case_bound}
Under the comparison condition, every deterministic sample in $\mathscr E_\ell$ satisfies
\[
    \mathbb E_g\sup_C\left|\sum_{p\in X_{J_{\mathrm{huge}}(C)}}
    \frac{g_p d^2(p,C)}{N\cost(\mu,C)}\right|
    \lesssim\zeta_\mu2^{2\ell}\varepsilon^{-2}.
\]
\end{lemma}
The proof is in Appendix~\ref{appendix:worst_case_bound}. Now we give the proof of Lemma~\ref{lemma:outer_points}.

\begin{proof}[Proof of Lemma~\ref{lemma:outer_points}]
The latent labels, original directions, and normalized radii of isotropic Gaussian samples are mutually independent. The events $\mathcal E_c$, $\mathcal E_s$, and $\mathscr E_\ell$ depend respectively on these three independent collections. Therefore Lemma~\ref{lemma:prob-good-event} gives $\Pr[\overline{\mathcal E}\mid\mathscr E_\ell]\le N^{-2}$. The pointwise cost bound in the proof of Lemma~\ref{lemma:worst_case_bound} also applies to outer points: taking absolute values bounds their sum by the same cost envelope times $N^{-1}\sum_{p\in X}|g_p|$. Thus the same worst-case estimate holds for the outer restriction. Splitting according to $\mathcal E$, applying Lemma~\ref{lemma:bound_conditioned_on_good_event} on $\mathcal E$, and using this worst-case estimate on its complement gives
\[
    \mathbb E_{X\mid\mathscr E_\ell}\mathbb E_g\sup_C
    \left|\sum_{j\in J_{\mathrm{huge}}(C)}\sum_{p\in X_j^{\mathrm{out}}}
    \frac{g_p d^2(p,C)}{N\cost(\mu,C)}\right|
    \lesssim2^{4\ell}\ell\varepsilon+\zeta_\mu2^{2\ell}\varepsilon^{-2}N^{-2}
    \lesssim2^{4\ell}\ell\varepsilon.
\]
The last inequality follows from $N\gtrsim\zeta_\mu\varepsilon^{-2}$, which implies $\zeta_\mu\varepsilon^{-2}N^{-2}\lesssim\varepsilon^2\le\varepsilon$.
\end{proof}

\subsection{Proof of Lemma~\ref{lemma:bound_conditioned_on_good_event}: chaining from a covering bound}\label{sec:chaining}

We work with the outer subsets $X_j^{\mathrm{out}}$ of the full component samples $X_j$. On $\mathcal E_c$, the outer subsets satisfy $|X_j^{\mathrm{out}}|\le|X_j|\le3N\pi_j/2$. 
Throughout this section, $X$ is fixed in $\mathcal E\cap\mathscr E_\ell$, under the standing assumptions of Section~\ref{sec:proof_main_theorem}. 

\subsubsection{Outer points cost vectors, covers and chaining decomposition}

\begin{definition}[Cost vector]\label{def:cost_vector}
Define the cost vector and the family of all such vectors by
\[
    v_p^C:=d^2(p,C),\qquad
    V:=\{v^C:C\subset\mathbb R^d,\ |C|\le k\}.
\]
All vectors have the original $N$ sample coordinates, including repeated coordinates. In the following chain, only outer points of huge close components contribute.
\end{definition}

\begin{definition}[Local error term and $\alpha$-cover on outer points]\label{def:error_term_and_alpha_cover}
For $p\in X_j^{\mathrm{out}}$, use the local error term
\[
    \err(p,C):=d(p,C)d(p,a_j)+d^2(p,a_j).
\]
A finite family $\mathcal V(\alpha)\subset\mathbb R^N$ is a cover of the required coordinates if every query $C$ has a representative $v^{C,\alpha}$ satisfying
\begin{equation}\label{eq:outer-cover-error}
    |v_p^C-v_p^{C,\alpha}|\le\alpha\err(p,C)
    \quad\text{when }j\in J_{\mathrm{huge}}(C),\ p\in X_j^{\mathrm{out}},
\end{equation}
and $v_p^{C,\alpha}=0$ on all other coordinates. Representatives are chosen independently of the Gaussian signs.
\end{definition}
Both cover constructions in Section~\ref{sec:net-size-bound} use this same error term. On outer points, $\err(p,C)\ge d^2(p,a_j)>\varepsilon^4r_j^2$. The original cost vector remains unmasked; its representatives are padded with zeros outside the coordinates used in the chain.

Let $h_{\min}=0$ and $h_{\max}:=\lceil\log_2(\sqrt{\zeta_\mu}/\varepsilon)\rceil$. At the coarse scale, define the representative on the original sample coordinates by
\begin{equation}\label{eq:initial-representative}
    v_p^{C,1}:=
    \begin{cases}
    \displaystyle\varepsilon^4r_j^2
    \left\lfloor\frac{d^2(a_j,C)}{\varepsilon^4r_j^2}\right\rfloor,
    &j\in J_{\mathrm{huge}}(C),\ p\in X_j^{\mathrm{out}},\\
    0,&\text{otherwise}.
    \end{cases}
\end{equation}
For $1\le h\le h_{\max}+1$, choose a representative $v^{C,2^{-h}}\in\mathcal V(2^{-h})$ satisfying the cover inequalities for $C$. On each outer point of a huge close component, use
\begin{equation}\label{eq:chaining-telescoping}
    v_p^C=v_p^{C,1}
    +\sum_{h=0}^{h_{\max}}(v_p^{C,2^{-(h+1)}}-v_p^{C,2^{-h}})
    +(v_p^C-v_p^{C,2^{-(h_{\max}+1)}}).
\end{equation}
Define the chaining terms by
\begin{align*}
    Y_{C,\mathrm{Init}}&:=\sum_{p\in X}
    \frac{g_p v_p^{C,1}}{N\cost(\mu,C)},\\
    Y_{C,h}&:=\sum_{p\in X}
    \frac{g_p(v_p^{C,2^{-(h+1)}}-v_p^{C,2^{-h}})}{N\cost(\mu,C)},\\
    Y_{C,\mathrm{End}}&:=\sum_{j\in J_{\mathrm{huge}}(C)}
    \sum_{p\in X_j^{\mathrm{out}}}
    \frac{g_p(v_p^C-v_p^{C,2^{-(h_{\max}+1)}})}{N\cost(\mu,C)}.
\end{align*}
Their sum is the outer contribution.

To control $Y_{C,\mathrm{Init}}$, the next lemma verifies that the coarse representatives form a constant-accuracy cover on outer points with bounded logarithmic size.

\begin{lemma}[Coarse-scale cover and net size bound]\label{lemma:net_size_in_coarse_level}
The representatives $v^{C,1}$ form a cover with constant accuracy on outer points, denoted $\mathcal V(1)$, with
\[
    \log|\mathcal V(1)|\lesssim k\bigl(\ell+\log(1/\varepsilon)\bigr).
\]
\end{lemma}
\begin{proof}
For active components, $0\le d^2(a_j,C)\le\varepsilon^{-2}2^{2\ell}r_j^2$. Hence each component has at most $2+\varepsilon^{-6}2^{2\ell}$ rounded states, including the inactive state. Also,
\[
    |d^2(p,C)-d^2(a_j,C)|
    \le2d(p,C)d(p,a_j)+d^2(p,a_j)\le2\err(p,C).
\]
The rounding error is at most $\varepsilon^4r_j^2\le\err(p,C)$ on outer points. Taking logarithms of the state count gives
\[
    \log|\mathcal V(1)|\le k\log(2+\varepsilon^{-6}2^{2\ell})\lesssim k\bigl(\ell+\log(1/\varepsilon)\bigr).
\]
\end{proof}
The first increment uses this coarse cover with the same universal constant in its local-error estimate.
\subsubsection{Bounds for the initial, increment, and end terms}

\begin{lemma}[Initial term]\label{lemma:init_term_in_chaining}\label{appendix:init_in_chaining}
Under the stated good-event hypotheses,
\[
    \mathbb E_g\sup_C|Y_{C,\mathrm{Init}}|
    \lesssim\sqrt{\frac{\log(2k)}{N\pi_{\min}}}\lesssim\varepsilon.
\]
\end{lemma}
\begin{proof}
The rounded value in~\eqref{eq:initial-representative} on each active component lies between zero and $d^2(a_j,C)$. Grouping by component and using \eqref{eq:comparison_condition} gives
\begin{align*}
    |Y_{C,\mathrm{Init}}|
    &\le\frac{1}{N\cost(\mu,C)}
    \sum_{j\in J_{\mathrm{huge}}(C)}d^2(a_j,C)
    \left|\sum_{p\in X_j^{\mathrm{out}}}g_p\right|\\
    &\lesssim\max_j\frac{
    \left|\sum_{p\in X_j^{\mathrm{out}}}g_p\right|}{N\pi_j}.
\end{align*}
Since $X_j^{\mathrm{out}}$ is fixed independently of $C$ and the Gaussian signs, each normalized sum has variance $|X_j^{\mathrm{out}}|/(N\pi_j)^2\le 3/(2N\pi_j)$. The Gaussian maximal inequality over the $k$ components gives the claim.
\end{proof}

To control the resulting chaining increments and end term, the next lemma bounds the normalized sum of squared local errors.

\begin{lemma}[Error-sum bound]
\label{lemma:error-sum-high-dimensional}
Under the standing comparison condition, for every $d\ge1$, $\ell\ge1$, and deterministic sample $X\in\mathcal E\cap\mathscr E_\ell$, the outer subsets satisfy, for every $C\in(\mathbb R^d)^k$,
\begin{equation*}\tag{ES}\label{eq:error_sum}
    \sum_{j\in J_{\mathrm{huge}}(C)}\sum_{p\in X_j^{\mathrm{out}}}\frac{\err(p,C)^2}{(N\cdot\cost(\mu,C))^2}\lesssim \frac{2^{4\ell}\zeta_\mu}{N}.
\end{equation*}
\end{lemma}
The proof is in Appendix~\ref{sec:variance-bound}. Combining Lemma~\ref{lemma:error-sum-high-dimensional} with the covering error at the finest scale yields the end-term bound in the next lemma.

\begin{lemma}[End term]\label{lemma:end_term_in_chaining}\label{appendix:end_in_chaining}
Under the stated good-event hypotheses,
\[
    \mathbb E_g\sup_C|Y_{C,\mathrm{End}}|\lesssim2^{2\ell}\varepsilon.
\]
\end{lemma}
\begin{proof}
By Cauchy--Schwarz on the original $N$ coordinates, the covering inequality~\eqref{eq:outer-cover-error} at the finest scale, and Lemma~\ref{lemma:error-sum-high-dimensional},
\begin{align*}
    \mathbb E_g\sup_C|Y_{C,\mathrm{End}}|
    &\le\mathbb E_g\left(\sum_{p\in X}g_p^2\right)^{1/2}
    \sup_C\left(
    \sum_{j\in J_{\mathrm{huge}}(C)}
    \sum_{p\in X_j^{\mathrm{out}}}
    \frac{(v_p^C-v_p^{C,2^{-(h_{\max}+1)}})^2}{N^2\cost(\mu,C)^2}
    \right)^{1/2}\\
    &\lesssim\sqrt N\,2^{2\ell}2^{-h_{\max}}\sqrt{\zeta_\mu/N}
    \le2^{2\ell}\varepsilon.
\end{align*}
\end{proof}
To control the intermediate chaining increments, the next lemma combines the pointwise covering error between adjacent scales with Lemma~\ref{lemma:error-sum-high-dimensional} to bound the increment variance.

\begin{lemma}[Variance bound]\label{lemma:variance_bound}
For $0\le h\le h_{\max}$ and $\alpha_h:=2^{-h}$,
\[
    \sup_C\operatorname{Var}_g(Y_{C,h})\lesssim\frac{2^{4\ell}\zeta_\mu\alpha_h^2}{N}.
\]
\end{lemma}
The proof is in Appendix~\ref{sec:variance-bound}. To pair with this variance bound in the Gaussian maximal inequality, the next lemma bounds the logarithmic size of the outer-point covers.

\begin{lemma}[Net size bound]\label{lemma:net_size_bound}
For every $\alpha\in(0,1)$, the outer costs admit a cover satisfying
\[
    \log|\mathcal V(\alpha)|=\widetilde O\!\left(\min\left\{kd(1+\ell),\,k\alpha^{-2}(1+\ell)^3\right\}\right).
\]
\end{lemma}
The proof is in Section~\ref{sec:net-size-bound}. Combining the variance bound of Lemma~\ref{lemma:variance_bound} with the covering bounds of Lemmas~\ref{lemma:net_size_in_coarse_level} and~\ref{lemma:net_size_bound} now controls each chaining increment.

\begin{lemma}[Chaining increment]\label{lemma:main_term_in_chaining}
For every $0\le h\le h_{\max}$,
\[
    \mathbb E_g\sup_C|Y_{C,h}|
    =\widetilde O\!\left(2^{2\ell}\ell^{3/2}\sqrt{\frac{k\zeta_\mu}{N}}\right).
\]
\end{lemma}
\begin{proof}
For each realized pair of representatives
\[
    (v,v')\in\mathcal V(2^{-h})\times\mathcal V(2^{-(h+1)}),
\]
consider the centered Gaussian variable
\[
    \left(\sup_{\substack{C:\ v^{C,2^{-h}}=v\\
    v^{C,2^{-(h+1)}}=v'}}\frac1{N\cost(\mu,C)}\right)
    \sum_{p\in X}g_p(v'_p-v_p).
\]
Ignore pairs with no realizing query. The comparison condition makes the displayed supremum finite. Because the numerator is fixed for a given pair, this variable has variance at most $\sup_C\operatorname{Var}_g(Y_{C,h})$. Its absolute value dominates the increment for every query realizing that pair. The finite Gaussian maximal inequality therefore gives
\[
    \mathbb E_g\sup_C|Y_{C,h}|
    \lesssim\sqrt{\sup_C\operatorname{Var}_g(Y_{C,h})
    \log\!\left(2|\mathcal V(2^{-h})|
    |\mathcal V(2^{-(h+1)})|\right)}.
\]

Lemma~\ref{lemma:net_size_bound}, with the coarse bound when $h=0$, gives
\[
    \log\left(2|\mathcal V(2^{-h})||\mathcal V(2^{-(h+1)})|\right)=\widetilde O(k\,2^{2h}\ell^3).
\]
Since $h\le h_{\max}$, all logarithms of the reciprocal covering scale are $O(1+\log(\sqrt{\zeta_\mu}/\varepsilon))$. Combining the entropy and variance bounds cancels the factors $2^{2h}$ and $2^{-2h}$.
\end{proof}

Now we are ready to prove Lemma~\ref{lemma:bound_conditioned_on_good_event}.
\begin{proof}[Proof of Lemma~\ref{lemma:bound_conditioned_on_good_event}]
The outer contribution is the sum of the initial, increment, and end terms in the telescoping identity~\eqref{eq:chaining-telescoping}. Applying Lemmas~\ref{lemma:init_term_in_chaining}, ~\ref{lemma:end_term_in_chaining}, and~\ref{lemma:main_term_in_chaining} gives
\[
    \mathbb E_g\sup_C\left|\sum_{j\in J_{\mathrm{huge}}(C)}\sum_{p\in X_j^{\mathrm{out}}}\frac{g_p d^2(p,C)}{N\cost(\mu,C)}\right|
    \le O(2^{2\ell}\varepsilon)+\widetilde O\!\left((h_{\max}+1)2^{2\ell}\ell^{3/2}\sqrt{\frac{k\zeta_\mu}{N}}\right).
\]
There are $h_{\max}+1=O(1+\log(\sqrt{\zeta_\mu}/\varepsilon))$ increments. At the prescribed sample size, the second term is $O(2^{2\ell}\ell^{3/2}\varepsilon)$, so the entire display is $O(2^{4\ell}\ell\varepsilon)$.

\end{proof}

\subsection{Proof of Lemma~\ref{lemma:net_size_bound}: isotropic covering by base vectors and sparse shifts}\label{sec:isotropic-covering}
\label{sec:net-size-bound}

As in Section~\ref{sec:chaining}, we cover the outer subsets $X_j^{\mathrm{out}}$; $X$, $N$, and all good events refer to the original sample. Write $\mathcal E^{(\ell)}:=\mathcal E\cap\mathscr E_\ell$. Throughout, $\mathsf M(C)=\bigcup_{j\in J_{\mathrm{huge}}(C)}X_j^{\mathrm{out}}$, and all representatives are zero on the other original coordinates. Thus the masked cover $\mathcal V_{\mathsf M}(\alpha)$ below is the outer-point cover $\mathcal V(\alpha)$ of Section~\ref{sec:chaining}.

The proof of Lemma~\ref{lemma:net_size_bound} has two ingredients. The first is a direct masked discretization, which works for every scale \(\alpha\in(0,1)\) and gives the dimension-dependent cover bound in the next lemma.

\begin{lemma}[Direct masked cover]\label{lemma:direct_masked_cover}
Fix $\ell\ge1$, and let $X$ satisfy $\mathcal E^{(\ell)}$. For every $\alpha\in(0,1)$, there exists a masked $\alpha$-cover of $X$ with respect to $\mathsf M$ such that
\[
    \log|\mathcal V_{\mathsf M}(\alpha)|\lesssim k\log k+kd\bigl(\ell+\log((\varepsilon\alpha)^{-1})\bigr).
\]
\end{lemma}

The second construction applies above the transition scale $\alpha_\star\asymp \log k\sqrt{(\log(eN))/d}$ utilizing a base-shift decomposition argument to establish a dimension-free covering bound.

\begin{lemma}[$\alpha$-cover above the transition scale]
\label{lemma:base_shift_cover}
Fix $\ell\ge1$, and let $X$ satisfy $\mathcal E^{(\ell)}$. For every $\alpha\in(0,1)$ with $\alpha\ge\alpha_\star$, there exists a masked $\alpha$-cover of $X$ with respect to $\mathsf M$ such that
\[
    \log|\mathcal V_{\mathsf M}(\alpha)|=\widetilde O\!\left(k\alpha^{-2}(1+\ell)^3\right).
\]
\end{lemma}

We now combine the two bounds and give the proof of Lemma \ref{lemma:net_size_bound}. If $\alpha\ge\alpha_\star$, both constructions are available, and we choose one with smaller cardinality. If $\alpha<\alpha_\star$, use the direct cover and
\[
    d\lesssim\alpha^{-2}\log(eN)\log^2 k.
\]
The direct bound is then also at most the bound in Lemma~\ref{lemma:base_shift_cover}. This includes $\alpha_\star\ge1$, when the direct cover is used at every scale. Consequently, for every $\alpha\in(0,1)$,
\[
    \log|\mathcal V_{\mathsf M}(\alpha)|=\widetilde O\!\left(\min\left\{kd(1+\ell),\,k\alpha^{-2}(1+\ell)^3\right\}\right).
\]
This proves Lemma~\ref{lemma:net_size_bound}.

\subsubsection{Proof of Lemma~\ref{lemma:direct_masked_cover}: direct masked cover}
To prove Lemma~\ref{lemma:direct_masked_cover}, we first bound the cost-vector cover size for a fixed collection of bounded component balls under prescribed componentwise additive tolerances by discretizing candidate centers on Euclidean nets.

\begin{lemma}[Direct discretization of cost vectors; Lemma B.7 in \cite{DBLP:conf/stoc/Huang0024}]\label{lemma:general_alpha_net_size}
    Let $d\ge1$. Let $X\subset \mathbb{R}^d$ be an indexed dataset consisting of $X_j\subset B(a_j,\delta_j),a_j\in\mathbb{R}^d,\delta_j>0,j\in [k'],1\le k'\le k$. Let $\alpha\in (0,1)$, $u\ge 1$ and $\xi_1,\cdots,\xi_{k'}>0$. There exists a $V\subset \mathbb{R}^{|X|}$ with 
    $$
    \log |V|
    \lesssim
    k\log\left(
        1+
        \sum_{j=1}^{k'}
        \left(
            1+\frac{u^2\delta_j^2}{\alpha\xi_j}
        \right)^d
    \right)\lesssim k(\log k+d\log (1+u^2\delta_{\max}^2\alpha^{-1}\xi_{\min}^{-1}))
    $$
    such that the following holds:
    for any $C\in (\mathbb{R}^d)^k$ satisfying $d(a_j,C)\le u\delta_j,\forall j\in [k']$, there exists $v\in V$ such that, for every $j\in [k']$ and every $p\in X_j$,  
    $$
    \abs{d^2(p,C)-v_p}\le \alpha \xi_j,
    $$
    where $\delta_{\max}=\max_{j\in [k']}\delta_j$, $\xi_{\min}=\min_{j\in [k']}\xi_j$.
\end{lemma}

\begin{proof}
For indices with $\alpha\xi_j\ge4u^2\delta_j^2$, set $v_p=0$ on $X_j$: every admissible query satisfies $d^2(p,C)\le4u^2\delta_j^2\le\alpha\xi_j$. If all indices satisfy this inequality, the zero vector suffices.

For each remaining index $j$, fix a Euclidean net of $B(a_j,3u\delta_j)$ with mesh $\alpha\xi_j/(16u\delta_j)$ and at most $(1+96u^2\delta_j^2/(\alpha\xi_j))^d$ points. Given $C$, discard centers outside all these balls. For each remaining center $c$, choose, among balls containing $c$, an index $i$ minimizing $\xi_i/\delta_i$, and record a net point $q$ within $\alpha\xi_i/(16u\delta_i)$ of $c$, together with $i$.
For a coordinate in a remaining component, define $v_p$ as the minimum of $(d(p,q)+\alpha\xi_i/(16u\delta_i))^2$ over the recorded pairs. By the triangle inequality, each such quantity is at least $d^2(p,c)$ for its original center, so $v_p\ge d^2(p,C)$.

For $p\in X_j$ in a remaining component, its nearest center $c(p)$ lies in $B(a_j,3u\delta_j)$ and is therefore recorded. The chosen mesh for this center is at most $\alpha\xi_j/(16u\delta_j)$. Since $d(p,C)\le2u\delta_j$ and $\alpha\xi_j<4u^2\delta_j^2$, its contribution gives
\[
    d^2(p,C)\le v_p\le\left(d(p,C)+\frac{\alpha\xi_j}{8u\delta_j}\right)^2\le d^2(p,C)+\alpha\xi_j.
\]
There are at most $k$ recorded pairs. Taking all nonempty lists of at most $k$ such pairs, padded by an unused state, gives a finite family independent of $C$ with
\[
    \log|V|\le k\log\left(1+\sum_{j=1}^{k'}\left(1+\frac{96u^2\delta_j^2}{\alpha\xi_j}\right)^d\right)\lesssim k\log\left(1+\sum_{j=1}^{k'}\left(1+\frac{u^2\delta_j^2}{\alpha\xi_j}\right)^d\right).
\]
The second bound in the statement follows by taking the largest ratio inside the sum. No lower bound on $d$ beyond $d\ge1$ is used.
\end{proof}

Now we prove Lemma \ref{lemma:direct_masked_cover}.

\begin{proof}[Proof of Lemma \ref{lemma:direct_masked_cover}]
    For every subset \(J\subseteq[k]\), apply
Lemma~\ref{lemma:general_alpha_net_size} to \(\bigcup_{j\in J}X_j^{\mathrm{out}}\), with
\[
    \delta_j:=2^\ell r_j,\qquad
    u:=\epsilon^{-1},\qquad
    \xi_j:=\varepsilon^4r_j^2,
    \qquad j\in J.
\]
By the event \(\mathcal E^{(\ell)}\), we have
\(X_j^{\mathrm{out}}\subset B(a_j,2^\ell r_j)\). Hence these choices are valid. Empty component subsets are omitted; if $\bigcup_{j\in J}X_j^{\mathrm{out}}=\emptyset$, take the singleton zero vector.

For each \(J\subseteq[k]\), Lemma~\ref{lemma:general_alpha_net_size}
gives a family \(\mathcal U_J(\alpha)\subset\mathbb R^{\bigcup_{j\in J}X_j^{\mathrm{out}}}\) such that,
for every center set \(C\) satisfying $d(a_j,C)\le \epsilon^{-1}2^\ell r_j,j\in J$,
there exists \(u\in\mathcal U_J(\alpha)\) with
\[
    |d^2(p,C)-u_p|
    \le
    \alpha\,\varepsilon^4r_j^2,
    \forall p\in X_j^{\mathrm{out}},\ j\in J.
\]
Moreover,
\[
\begin{aligned}
    \log|\mathcal U_J(\alpha)|
    &\lesssim
    k\log\left(
        1+\sum_{j\in J}
         k\left(1+2^{2\ell}\alpha^{-1}\epsilon^{-6}\right)^d
    \right)                                                     \\
    &\lesssim
    k\left[
        \log k+
        d\log\left(1+2^{2\ell}\alpha^{-1}\epsilon^{-6}\right)
    \right]\\
    &\lesssim k\log k+kd(\ell+\log((\varepsilon\alpha)^{-1}))
\end{aligned}
\]
For \(u\in\mathcal U_J(\alpha)\), define its zero-padding
\(\widehat{u}^J\in\mathbb R^N\) by
\[
    \widehat{u}_p^J
    :=
    \begin{cases}
        u_p, & p\in \bigcup_{j\in J}X_j^{\mathrm{out}},\\
        0,   & p\notin \bigcup_{j\in J}X_j^{\mathrm{out}}.
    \end{cases}
\]
Set $\mathcal V_{\mathsf M}(\alpha):=
    \{\widehat{u}^J:J\subseteq[k],\ u\in\mathcal U_J(\alpha)\}$.
Then
\[
\begin{aligned}
    \log|\mathcal V_{\mathsf M}(\alpha)|
    &\le
    k\log 2+\max_{J\subseteq[k]}\log|\mathcal U_J(\alpha)|  \\
    &\lesssim
    k\log k+kd\left(\ell+\log((\varepsilon\alpha)^{-1})\right).
\end{aligned}
\]
We now verify the covering property. Fix \(C\) and let
\(J=J_{\mathrm{huge}}(C)\). By the definition of the huge cluster mask, for every
\(j\in J\), $d(a_j,C)\le \epsilon^{-1}2^\ell r_j$.
Therefore there exists \(u\in\mathcal U_J(\alpha)\) such that, for every
\(p\in X_j^{\mathrm{out}}\), \(j\in J\), $|d^2(p,C)-u_p|
    \le
    \alpha\,\varepsilon^4r_j^2$.
On the other hand, by the deterministic outer cutoff, $\err(p,C)
    =
    d(p,C)d(p,a_j)+d^2(p,a_j)
    \ge
    d^2(p,a_j)
    >
    \varepsilon^4r_j^2$.
Thus
\[
    |d^2(p,C)-u_p|
    \le
    \alpha\,\err(p,C),
    \qquad p\in \bigcup_{j\in J}X_j^{\mathrm{out}}.
\]
For \(p\notin \bigcup_{j\in J}X_j^{\mathrm{out}}\), both the masked cost vector and
\(\widehat{u}^J_p\) are zero. Hence
\(\mathcal V_{\mathsf M}(\alpha)\) is a valid masked
\(\alpha\)-cover for $X$ with respect to $\mathsf{M}$.
\end{proof}

\subsubsection{Proof of Lemma~\ref{lemma:base_shift_cover}: $\alpha$-cover above the transition scale}
\label{subsec:ring-product-reduction}
\label{subsec:high-dimensional-certificates}

To prove Lemma~\ref{lemma:base_shift_cover}, we partition the outer points into dyadic radius groups and distance-class blocks so that the global cover reduces to the single-block net bound of Lemma~\ref{lemma:fixed-block-net}.

\paragraph{Radius groups.}

\begin{definition}[Dyadic radius groups]
\label{def:group-structure}

For every integer $0\le m\le m_{\max}$, where $m_{\max}:=\ceil{\log_2(2^\ell k\varepsilon^{-7/2}\zeta_\mu^{1/2})}$, define the group $G(m)$ by 
\[
G(m):=\left\{
p\in X_j^{\mathrm{out}}:j\in[k],\
\begin{array}{l}
r_j\ge k^{-1}\varepsilon^{3/2}\zeta_\mu^{-1/2}
\max_{i\in[k]}r_i,\\
2^{-m-1}2^\ell\max_{i\in[k]}r_i
<d(p,a_j)\le
2^{-m}2^\ell\max_{i\in[k]}r_i
\end{array}
\right\}.
\]
Call $G(m)\cap \left(\bigcup_{j\in J_{\mathrm{huge}}(C)}X_j^{\mathrm{out}}\right)$ a huge group, and let $\mathcal{G}=\left\{ G(m): 0 \le m\le m_{\max} \right\}$ be the collection of all groups. 

\end{definition}

Each group collects rings with comparable radii. Tiny components do not meet these groups by their scale restriction. Every outer point in a huge component satisfies $d(p,a_j)>k^{-1}\varepsilon^{7/2}\zeta_\mu^{-1/2}\max_{i\in[k]}r_i$. Thus the disjoint sets $X_j^{\mathrm{out}}\cap G(m)$, $j\in J_{\mathrm{huge}}(C)$ and $0\le m\le m_{\max}$, partition $\bigcup_{j\in J_{\mathrm{huge}}(C)}X_j^{\mathrm{out}}$. The last group may be empty.
Let $R_m:=2^{-m}\cdot2^{\ell}\max_{i\in[k]}r_i$. Then for any $C\in (\mathbb{R}^d)^k$, each $j\in J_{\mathrm{huge}}(C)$ and $p\in X_j^{\mathrm{out}}\cap G(m)$, we have $R_m/2<d(p,a_j)\le R_m$.

\paragraph{Distance blocks and certificates.}

\leavevmode
We next refine each radius group into dyadic mean-to-center distance blocks in Definition~\ref{def:certificates-and-blocks} and index the resulting partition by a discrete certificate.

\begin{definition}[Distance blocks of a radius group]
\label{def:certificates-and-blocks}
For any center set $C\in (\mathbb{R}^d)^k$, let $G:=G(m)\in \mathcal{G}$ for integer $0\le m\le m_{\max}$ be a group. For an integer $1\le\beta\le \beta_{\max}:=m+\ceil{\log_2(\varepsilon^{-1})}$, we define 
$$
\mathsf{B}(G,C,\beta):=\left\{ p\in X_j^{\mathrm{out}}\cap G: j\in J_{\mathrm{huge}}(C), 2^{\beta-1}R_m<d(a_j,C)\le 2^{\beta}R_m \right\}.
$$
Also, define
$$
\begin{aligned}
    \mathsf{B}(G,C,0)&:=\left\{ p\in X_j^{\mathrm{out}}\cap G:j\in J_{\mathrm{huge}}(C), d(a_j,C)\le R_m \right\},\\
    \mathsf{B}(G,C,-1)&:=G\setminus \left(\bigcup_{j\in J_{\mathrm{huge}}(C)}X_j^{\mathrm{out}}\right),
\end{aligned}
$$
then $\mathsf{B}(G,C,\beta),\beta\in\left\{-1,0,1,\cdots,\beta_{\max}\right\}$ form a partition of $G$.
\end{definition}
For a huge close component, $d(a_j,C)\le\varepsilon^{-1}2^\ell r_j\le\varepsilon^{-1}2^mR_m$, which justifies the endpoint $\beta_{\max}$. Finite blocks partition $G\cap\mathsf M(C)$; the block with index $-1$ is inactive. Any block may be empty.

Combining the group radius and distance-class bounds, the next lemma shows that the error term is uniform up to constants on each active block.

\begin{lemma}[Distance scales and local error terms on a block]\label{prop:scale_properties}
    Let $t:=2^{\beta}$, $R:=R_m$, for center set $C\in (\mathbb{R}^d)^k$ and $j\in [k]$, we have following properties:
    \begin{enumerate}
        \item If $p\in X_j^{\mathrm{out}}\cap \mathsf{B}(G,C,0)$, $d(p,C)\lesssim d(p,a_j)$, and $\mathsf{err}(p,C)\asymp R^2$;
        \item If $p\in X_j^{\mathrm{out}}\cap \mathsf{B}(G,C,\beta)$ for some integer $1\le \beta\le \beta_{\max}$, then $d(a_j,C)-d(p,a_j)\le d(p,C)\le d(a_j,C)+d(p,a_j)$, and $R^2t/4<\err(p,C)\le R^2t+2R^2$, so $\err(p,C)\asymp R^2t$.
    \end{enumerate}
\end{lemma}
\begin{proof}
For $\beta=0$, $d(p,C)\le R+d(p,a_j)\le3d(p,a_j)$, and $\err(p,C)\ge d^2(p,a_j)>R^2/4$. For $\beta\ge1$, the lower bound follows from
\[
    \err(p,C)=d(p,a_j)\bigl(d(p,C)+d(p,a_j)\bigr)\ge d(p,a_j)d(a_j,C)>R^2t/4.
\]
The upper bounds follow from $d(p,C)\le d(a_j,C)+R\le Rt+R$. In particular, $\err(p,C)\ge R^2t/4$ in both cases, including $\beta=1$.
\end{proof}

\begin{definition}[Certificate]
For each $C\in (\mathbb{R}^d)^k$, denote the certificate $\phi^{(C,m)}\in\{-1,0,1,\ldots,\beta_{\max}\}^{k}$
defined as follows: for each $j\in [k]$,
\begin{enumerate}
    \item If $X_j^{\mathrm{out}}\cap G(m)=\emptyset$ or $j\notin J_{\mathrm{huge}}(C)$, then $\phi_j^{(C,m)}=-1$; 
    \item If $j\in J_{\mathrm{huge}}(C)$ and $X_j^{\mathrm{out}}\cap G\ne\emptyset$, then there exist exactly one $\beta$ such that $X_j^{\mathrm{out}}\cap G\subseteq \mathsf{B}(G,C,\beta)$. Then we set $\phi_j^{(C,m)}=\beta$.

\end{enumerate}
Let $\Phi_m :=\left\{\phi^{(C,m)}:C\in(\mathbb R^d)^k\right\}$ be the collection of all possible $m$-th level certificates.
\end{definition}
Since each coordinate has at most $\beta_{\max}+2$ choices, we have $|\Phi_m|\le(\beta_{\max}+2)^k$. For a group $G:=G(m)\in \mathcal{G}$, fix a certificate $\phi\in\Phi_m$ and a scale
$\beta\in\{0,1,\ldots,\beta_{\max}\}$, define $X_{m,\phi,\beta}=\left\{ p\in X_j^{\mathrm{out}}\cap G(m):j\in [k], \phi_j=\beta \right\}$, then 
for each $C$ with $\phi^{(C,m)}=\phi$, the finite blocks $X_{m,\phi,\beta}$ partition $G\cap\mathsf M(C)$ and equal $\mathsf B(G,C,\beta)$. The inactive value $-1$ produces zero coordinates; it does not require $X_j^{\mathrm{out}}\cap G$ to be empty.

Fixing the certificate $\phi$ fixes the sample coordinates in each block $X_{m,\phi,\beta}$: as $C$ varies over queries with this certificate, the costs change but the blocks do not. We can therefore construct one cover per block and take their product.

\begin{definition}[Block $\alpha$-cover]\label{def:cover_for_block}
    Fix $m,\phi,\beta$, let $$\mathcal{C}_{m,\phi}:=\left\{ C\in (\mathbb{R}^d)^k:\phi^{(C,m)}=\phi \right\},$$
    and $\alpha\in(0,1]$. A finite set $\mathcal{V}_{m,\phi,\beta}(\alpha)\subset \mathbb{R}^{|X_{m,\phi,\beta}|}$ is called an $\alpha$-cover for $X_{m,\phi,\beta}$, if for every $C\in\mathcal{C}_{m,\phi}$, there exists $u\in \mathcal{V}_{m,\phi,\beta}(\alpha)$, such that for any $p\in X_{m,\phi,\beta}$, $\abs{v_p^C-u_p}\le \alpha\err(p,C)$, where $R:= 2^{-m}2^{\ell}\max_{i\in[k]}r_i$, $t:=2^{\beta}$.
\end{definition}

Using $\mathsf{err}(p,C)\asymp R^2t$ from Lemma~\ref{prop:scale_properties}, the next lemma bounds the cover size for a single fixed-certificate block above the transition scale.

\begin{lemma}[Net size for a single block]
\label{lemma:fixed-block-net}

For an original sample in \(\mathcal E^{(\ell)}\), fix \(m\in\{0,\ldots,m_{\max}\}\), a certificate \(\phi\in\Phi_m\), a scale \(\beta\in\{0,1,\ldots,\beta_{\max}\}\), and a chaining scale \(\alpha\in(0,1]\) satisfying \(\alpha\ge\alpha_\star\). Suppose \(X_{m,\phi,\beta}\ne\emptyset\). Then there exists a cost-vector \(\alpha\)-net \(\mathcal V_{m,\phi,\beta}(\alpha)\) for \(X_{m,\phi,\beta}\), such that
\[
    \log|\mathcal V_{m,\phi,\beta}(\alpha)|=\widetilde O\!\left(k\alpha^{-2}(1+\beta)\right).
\]
Their coordinate errors are at most $\alpha R^2t/8\le\alpha\err(p,C)$. The block inherits the original signed subset-sum property with the original $N$; no fresh sampling or block count event is required.
\end{lemma}

The proof of Lemma~\ref{lemma:fixed-block-net} is given in Section \ref{sec:proof_of_fixed_block_net} by decomposing the cost vector into a base vector and a sparse shift vector. We now give the proof of Lemma~\ref{lemma:base_shift_cover}.

\begin{proof}[Proof of Lemma~\ref{lemma:base_shift_cover}]

We construct a masked $\alpha$-cover for $X$ with respect to $\mathsf{M}$, i.e., $\mathcal{V}_{\mathsf{M}}(\alpha)\subset \mathbb{R}^N$, through the following steps, where $\mathsf{M}(C)=\bigcup_{j\in J_{\mathrm{huge}}(C)}X_j^{\mathrm{out}}, \forall C\in (\mathbb{R}^d)^k$. Throughout the construction, whenever a vector is defined only on a subset of
\(X\), we identify it with its zero extension to \(\mathbb R^N\).

\begin{enumerate}
    \item Fix \(0\le m\le m_{\max}\), \(\phi\in\Phi_m\), and
    \(0\le\beta\le\beta_{\max}\). By Lemma~\ref{lemma:fixed-block-net},
    there exists an \(\alpha\)-cover
    \(\mathcal V_{m,\phi,\beta}(\alpha)\subseteq
    \mathbb R^{X_{m,\phi,\beta}}\) for
    \[
        \left\{
        (v_p^C)_{p\in X_{m,\phi,\beta}}
        :
        C\in(\mathbb R^d)^k,\ \phi^{(C,m)}=\phi
        \right\}.
    \]
    We view each element of \(\mathcal V_{m,\phi,\beta}(\alpha)\) as a vector
    in \(\mathbb R^N\) by padding zeros outside \(X_{m,\phi,\beta}\). For an empty block, use the singleton zero vector. The stronger bound in the lemma gives error at most $\alpha\err(p,C)$.

    \item For fixed \(m\) and \(\phi\), define the group-level cover
    \[
        \mathcal V_{m,\phi}(\alpha)
        :=
        \left\{
        \sum_{\beta=0}^{\beta_{\max}}u_\beta
        :
        u_\beta\in\mathcal V_{m,\phi,\beta}(\alpha)
        \right\}
        \subseteq\mathbb R^N .
    \]
    Since, for every \(C\) with \(\phi^{(C,m)}=\phi\), the sets
    \(X_{m,\phi,\beta}\), \(0\le\beta\le\beta_{\max}\), form the huge part of the partition of \(G(m)\), the set
    \(\mathcal V_{m,\phi}(\alpha)\) is a masked \(\alpha\)-cover for $X$ with respect to $G(m)\cap \left(\bigcup_{j\in J_{\mathrm{huge}}(C)}X_j^{\mathrm{out}}\right)$, and the set of masked cost vector with mask mapping $G(m)\cap \left(\bigcup_{j\in J_{\mathrm{huge}}(C)}X_j^{\mathrm{out}}\right)$ for $C\in \mathcal{C}_{m,\phi}$ is
    \[
        \left\{
        \left(
        \mathbf{1}[p\in G(m)\cap \left(\bigcup_{j\in J_{\mathrm{huge}}(C)}X_j^{\mathrm{out}}\right)]\cdot v_p^C
        \right)_{p\in X}
        :
        C\in(\mathbb R^d)^k,\ \phi^{(C,m)}=\phi
        \right\}.
    \]
    Moreover,
    \[
        \log|\mathcal V_{m,\phi}(\alpha)|
        \le
        \sum_{\beta=0}^{\beta_{\max}}
        \log|\mathcal V_{m,\phi,\beta}(\alpha)|.
    \]

    \item For fixed \(m\), remove the conditioning on the certificate by setting
    \[
        \mathcal V_m(\alpha)
        :=
        \bigcup_{\phi\in\Phi_m}\mathcal V_{m,\phi}(\alpha).
    \]
    Then \(\mathcal V_m(\alpha)\) is a masked \(\alpha\)-cover of $X$ with respect to $G(m)\cap \left(\bigcup_{j\in J_{\mathrm{huge}}(C)}X_j^{\mathrm{out}}\right)$, and the set of masked cost vector with mask $G(m)\cap \left(\bigcup_{j\in J_{\mathrm{huge}}(C)}X_j^{\mathrm{out}}\right)$ for $C\in (\mathbb{R}^d)^k$ is
    \[
        \left\{
        \left(
        \mathbf 1_{\{p\in G(m)\cap \left(\bigcup_{j\in J_{\mathrm{huge}}(C)}X_j^{\mathrm{out}}\right)\}}v_p^C
        \right)_{p\in X}
        :
        C\in(\mathbb R^d)^k
        \right\}.
    \]
    Its logarithmic size satisfies
    \[
    \begin{aligned}
        \log|\mathcal V_m(\alpha)|
        &\le
        \log|\Phi_m|
        +
        \max_{\phi\in\Phi_m}
        \sum_{\beta=0}^{\beta_{\max}}
        \log|\mathcal V_{m,\phi,\beta}(\alpha)|  \\
        & =\widetilde O\!\left(k\alpha^{-2}(1+\beta_{\max})^2\right).
    \end{aligned}
    \]

    \item Finally, construct the masked $\alpha$-cover $\mathcal{V}_{\mathsf{M}}(\alpha)$ by taking the product over groups:
    \[
        \mathcal V_{\mathsf{M}}(\alpha)
        :=
        \left\{
        \sum_{m=0}^{m_{\max}}u_m
        :
        u_m\in\mathcal V_m(\alpha)
        \right\}
        \subseteq\mathbb R^N .
    \]
\end{enumerate}

For each query, choose its certificates in all groups and the corresponding block representatives. Every active outer coordinate occurs in exactly one block, and every other coordinate is zero. Thus errors do not accumulate over blocks, and no division of $\alpha$ by the number of groups is needed. The size bound follows from the preceding estimates:
\[
\begin{aligned}
    \log|\mathcal V_{\mathsf{M}}(\alpha)|
    &\le
    \sum_{m=0}^{m_{\max}}\log|\mathcal V_m(\alpha)|  \\
    &=\widetilde O\!\left(k\alpha^{-2}\sum_{m=0}^{m_{\max}}(1+m+\lceil\log_2(\varepsilon^{-1})\rceil)^2\right)\\
    &=\widetilde O\!\left(k\alpha^{-2}(1+\ell)^3\right).
\end{aligned}
\]

Here $m_{\max}=O(\ell+\log(k\zeta_\mu/\varepsilon))$ and $\beta_{\max}=m+\ceil{\log_2(\varepsilon^{-1})}$. The group count, distance-class count, and base-net factor $1+\log t=1+\beta\log2$ together give the cubic factor. This proves Lemma~\ref{lemma:base_shift_cover}.
\end{proof}

\subsubsection{Single-block geometry: base vectors and sparse shifts}\label{sec:proof_of_fixed_block_net}

We first state the geometric setup independently of the preceding classifications. Let $X$ be a fixed indexed set of at most $N$ points, with $N\ge2$, partitioned into at most $k$ nonempty component subsets indexed by $J$, with reference means $a_j$. Write $a(p)=a_j$ for a point in component $j$. Fix $R>0$, $t\ge1$, and a family $\mathcal C$ of queries with at most $k$ centers. For each query, let $c(x)$ be a nearest center to $x$, with ties broken consistently. Assume that, for every component point $p$ and every $C\in\mathcal C$,
\[
R/2<d(p,a_j)\le R,\qquad d(a_j,C)\le Rt,
\]
and that $d(a_j,C)>Rt/2$ when $t>1$. With
\[
\err(p,C):=d(p,a(p))d(p,C)+d^2(p,a(p)),
\]
these scale bounds imply $\err(p,C)\ge R^2t/4$.

For the shift argument, assume that the normalized offsets $x_p=(p-a(p))/\|p-a(p)\|_2$ satisfy the signed subset-sum bound: for every nonempty $T\subseteq X$ and all signs $\tau_p\in\{-1,1\}$,
\[
\left\|\sum_{p\in T}\tau_p x_p\right\|_2
\le C_2\left(\sqrt{|T|}+|T|\sqrt{\frac{\log(eN/|T|)}d}\right),
\]
where $C_2$ is a fixed universal constant. This is the property denoted by $\mathcal E_s$; isotropic sampling supplies it through Lemma~\ref{lem:signed-subset-sum-good-event}. Set $\alpha_\star=K\log k\sqrt{\log(eN)/d}$ for a sufficiently large universal constant $K$. Our goal, for $\alpha\in(0,1]$ with $\alpha\ge\alpha_\star$, is to cover the vectors $(d^2(p,C))_{p\in X}$ with coordinate error at most $\alpha\err(p,C)$ and logarithmic size $\widetilde O(k\alpha^{-2}(1+\log t))$.

For the application to Lemma~\ref{lemma:fixed-block-net}, fix the block $X=X_{m,\phi,\beta}$, let $J$ be its set of nonempty component indices, and write $\mathcal C=\mathcal C_{m,\phi}$. The component subsets in this block are $X_j^{\mathrm{out}}\cap X_{m,\phi,\beta}$, and the original $N$ is unchanged. With $R=2^{-m}2^\ell \max_{i\in[k]}r_i$ and $t=2^\beta$, every $p\in X_j^{\mathrm{out}}\cap X_{m,\phi,\beta}$ satisfies $R/2<d(p,a_j)\le R$ and $d(a_j,C)\le Rt$. Consequently,
\[
    d(p,C)\le d(p,c(a_j))\le R(t+1),\qquad d^2(p,c(a_j))\le4R^2t^2.
\]
The triangle inequality also gives $d(p,c(a_j))-d(p,C)\le2d(p,a_j)$, hence
\[
    0\le d^2(p,c(a_j))-d^2(p,C)\le4d(p,a_j)\bigl(d(a_j,C)+d(p,a_j)\bigr)\le8R^2t.
\]
These bounds include $\beta=0$, where $t=1$ and $d(a_j,C)$ may vanish. Lemma~\ref{prop:scale_properties} gives $\err(p,C)\ge R^2t/4$ in every block.

To avoid discretizing all query centers inside every component, we decompose the block cost vector into a componentwise base vector minus a nonnegative shift vector recording the cost reduction from switching to a closer center.

\begin{definition}[Base vector, shift vector, and their nets]
Let $X=X_{m,\phi,\beta}$ be the fixed block with nonempty component indices $J\subseteq[k]$.
For $p\in X$, define 
\[
    b_p^C:=d^2(p,c(a(p))),\qquad s_p^C:=d^2(p,c(a(p)))-d^2(p,C)
\]
Then $b^C,s^C$ are called the \textit{base vector} and \textit{shift vector}, respectively. Writing $v^C=(d^2(p,C))_{p\in X}$, we have $v^C=b^C-s^C$ in $\mathbb R^{|X|}$.

The finite sets
$\mathcal B(\alpha),\mathcal S(\alpha)\subseteq\mathbb R^{|X|}$ are called $\alpha$-cover for the base vector and shift vector
on $X_{m,\phi,\beta}$ over $\mathcal C$ if for every $C\in\mathcal C$, there exists
$b^{C,\alpha}\in\mathcal B(\alpha)$ and $s^{C,\alpha}\in \mathcal S(\alpha)$ such that
\[
    |b_p^C-b_p^{C,\alpha}|
    \le \alpha R^2t/16,\quad|s_p^C-s_p^{C,\alpha}|
    \le \alpha R^2t/16
    \qquad\text{for every }p\in X.
\]
\end{definition}

The next two lemmas bound nets for the base and shift vectors separately. The base cover uses one embedded query center per component; the shift cover uses the directional condition to bound the number of corrections above the target tolerance.

\begin{lemma}[Base-vector net for one block]\label{lemma:base_vec_net}
For the fixed deterministic block above and every $\alpha\in(0,1]$, the base vectors $\{b^C:C\in\mathcal C\}$ admit an $\alpha$-net $\mathcal B(\alpha)$ satisfying
\[
    \log|\mathcal B(\alpha)|\lesssim k\alpha^{-2}\log(eN)\log(et/\alpha).
\]
No component-count or signed subset-sum event is needed for this construction.
\end{lemma}

\begin{lemma}[Shift-vector net for one block]
\label{lemma:shift_vec_net}
Assume that the fixed block inherits the signed subset-sum property $\mathcal E_s$ of the original sample, with the original $N$. For every $\alpha\in(0,1]$ satisfying $\alpha\ge\alpha_\star$, the shift vectors $\{s^C:C\in\mathcal C\}$ admit an $\alpha$-net $\mathcal S(\alpha)$ satisfying
\[
    \log|\mathcal S(\alpha)|\lesssim k\alpha^{-2}\log k\log(eN/\alpha).
\]
\end{lemma}

The proof of Lemma \ref{lemma:base_vec_net} is in Section~\ref{subsec:base-vector-net}, and the proof of Lemma \ref{lemma:shift_vec_net} is in Section~\ref{subsec:shift-vector-net}. Now we are ready to prove Lemma \ref{lemma:fixed-block-net}.

\begin{proof}[Proof of Lemma~\ref{lemma:fixed-block-net}]

If $X=\emptyset$, the claim is trivial. Otherwise Lemmas~\ref{lemma:base_vec_net} and~\ref{lemma:shift_vec_net} give finite families $\mathcal B(\alpha),\mathcal S(\alpha)$ such that, for every $C\in\mathcal C$, there are $b\in\mathcal B(\alpha)$ and $s\in\mathcal S(\alpha)$ satisfying, for every $p\in X$,
\[
    |b^C_p-b_p|\le\alpha R^2t/16,\qquad |s^C_p-s_p|\le\alpha R^2t/16.
\]
Set $v=b-s$. Since \(d^2(p,C)=b^C_p-s^C_p\), for every \(p\in X\),
\[
    |d^2(p,C)-v_p|=|(b^C_p-b_p)-(s^C_p-s_p)|\le\alpha R^2t/8\le\alpha\err(p,C).
\]
Thus \(\mathcal V_{m,\phi,\beta}(\alpha):=
    \{b-s:\ b\in\mathcal B(\alpha),\ s\in\mathcal S(\alpha)\}\) is an \(\alpha\)-net for the cost vectors on the fixed block \(X_{m,\phi,\beta}\). Finally,
\[
    \begin{aligned}
    \log|\mathcal V_{m,\phi,\beta}(\alpha)|
    &\le \log|\mathcal B(\alpha)|+\log|\mathcal S(\alpha)|\\
    &\lesssim k\alpha^{-2}\left[\log(eN)\bigl(1+\beta+\log(1/\alpha)\bigr)+\log k\log(eN/\alpha)\right].
    \end{aligned}
\]
This proves the lemma.
\end{proof} 

\subsubsection{Proof of Lemma~\ref{lemma:base_vec_net}: bounding the base vector net}
\label{subsec:base-vector-net}

We first introduce the notion of $\alpha$-terminal embedding:
\begin{definition}[Additive terminal embedding, Definition 3.1 in \cite{DBLP:conf/stoc/Huang0024}]\label{def:additive_terminal_embedding}
    Let  $\alpha \in (0,1)$, $r > 0$,  and  $X \subset \mathbb{R}^d$  be a collection of  $n$  points within a ball  $B(0,r)$  with  $0^d \in X$.  A mapping   $f :\mathbb{R}^d \rightarrow \mathbb{R}^m$  is called an  $\alpha$-additive terminal embedding of  X  if for any  $p \in X$  and  $q \in \mathbb{R}^d$, \, $|d(p,q) - d(f(p),f(q))| \leq \alpha \cdot r$.
\end{definition}
The next lemma supplies the required additive terminal embedding with target dimension $\alpha^{-2}$ times logarithmic in the terminal count.

\begin{lemma}[Additive terminal embedding, Theorem 3.2 in \cite{DBLP:conf/stoc/Huang0024}]\label{lem:additive_terminal_embedding}
     Let  $\alpha \in (0,1)$, $r > 0$,  and  $X \subset \mathbb{R}^d$  be a collection of  n  points within a ball  $B(0,r)$  with  $0^d \in X$.  There exists an  $\alpha$-additive terminal embedding with a target dimension  $O(\alpha^{-2} \log(en))$.
\end{lemma}

Now we give the proof of Lemma \ref{lemma:base_vec_net}.

\begin{proof}[Proof of Lemma~\ref{lemma:base_vec_net}]
Recall that, on this fixed block, \(d(p,a_j)\le R\) for every \(p\in X_j^{\mathrm{out}}\cap X_{m,\phi,\beta}\), \(d(a_j,C)+R=\Theta(tR)\), and the target base-vector error is $\alpha R^2t/16$.

For each \(j\in J\), apply Lemma~\ref{lem:additive_terminal_embedding} to the terminal set \(\{p-a_j:p\in X_j^{\mathrm{out}}\cap X_{m,\phi,\beta}\}\cup\{0\}\subset B(0,R)\). This gives a map \(g_j:\mathbb R^d\to\mathbb R^{d(\alpha)}\), where
\[
    d(\alpha)=O(\alpha^{-2}\log(eN)),
\]
such that, writing \(f_j(x):=g_j(x-a_j)\), for every \(p\in X_j^{\mathrm{out}}\cap X_{m,\phi,\beta}\) and every \(q\in\mathbb R^d\),
\[
    \abs{d(p,q)-d(f_j(p),f_j(q))}\le \alpha R.
\]
Replacing \(\alpha\) by a sufficiently small constant multiple of \(\alpha\) in the construction makes the total error below at most $\alpha R^2t/16$. The terminal set has at most $N+1$ points; no count condition on the block is used.

For each \(j\in J\), let \(\mathcal Q_j(\alpha)\subset\mathbb R^{d(\alpha)}\) be an \(\alpha R\)-net of the ball \(B(f_j(a_j),O(tR))\). This ball contains \(f_j(c(a_j))\) for every \(C\in\mathcal C\), since
\[
    \norm{f_j(c(a_j))-f_j(a_j)}_2\le d(c(a_j),a_j)+\alpha R\le O(tR).
\]
For every \(q=(q_j)_{j\in J}\in\prod_{j\in J}\mathcal Q_j(\alpha)\), define \(b^q\in\mathbb R^{\abs{X}}\) by
\[
    b^q_p:=d^2(f_j(p),q_j)\qquad\text{for }p\in X_j^{\mathrm{out}}\cap X_{m,\phi,\beta}.
\]
Set
\[
    \mathcal B(\alpha):=\left\{b^q:q\in\prod_{j\in J}\mathcal Q_j(\alpha)\right\}.
\]

Fix \(C\in\mathcal C\). For every \(j\in J\), choose \(q_j(C)\in\mathcal Q_j(\alpha)\) such that
\[
    \norm{q_j(C)-f_j(c(a_j))}_2\le \alpha R.
\]
Let \(b^{C,\alpha}:=b^{q(C)}\in\mathcal B(\alpha)\). For \(p\in X_j^{\mathrm{out}}\cap X_{m,\phi,\beta}\), we have
\[
    d(p,c(a_j))\le d(p,a_j)+d(a_j,C)\le O(tR),
\]
and hence \(d(f_j(p),f_j(c(a_j)))\le O(tR)\). Therefore,
\[
    \abs{d^2(p,c(a_j))-d^2(f_j(p),f_j(c(a_j)))}\le \alpha R\cdot O(tR)\le O(\alpha tR^2).
\]
Moreover, since \(\norm{q_j(C)-f_j(c(a_j))}_2\le \alpha R\) and \(d(f_j(p),q_j(C))\le O(tR)\),
\[
    \abs{d^2(f_j(p),f_j(c(a_j)))-d^2(f_j(p),q_j(C))}\le \alpha R\cdot O(tR)\le O(\alpha tR^2).
\]
Combining the last two estimates gives, for every \(p\in X\),
\[
    \abs{b^C_p-b^{C,\alpha}_p}\le O(\alpha tR^2)\le\alpha R^2t/16,
\]
after the constant rescaling of \(\alpha\). Thus \(\mathcal B(\alpha)\) is an \(\alpha\)-net for the base vectors.

It remains to bound the size. Since \(\mathcal Q_j(\alpha)\) is an \(\alpha R\)-net of a \(d(\alpha)\)-dimensional ball of radius \(O(tR)\),
\[
    \log\abs{\mathcal Q_j(\alpha)}\le d(\alpha)\log\left(\frac{O(tR)}{\alpha R}\right)=O\left(\alpha^{-2}\log(eN)\cdot\log\frac{et}{\alpha}\right).
\]
Taking the product over at most \(k\) indices,
\[
    \log\abs{\mathcal B(\alpha)}\le \sum_{j\in J}\log\abs{\mathcal Q_j(\alpha)}\le O\left(k\alpha^{-2}\log(eN)\cdot\log\frac{et}{\alpha}\right).
\]
This proves the lemma.
\end{proof}

\subsubsection{Proof of Lemma~\ref{lemma:shift_vec_net}: bounding the shift-vector net}
\label{subsec:shift-vector-net}

To obtain a cost-vector net from the base net, we record only center-switch corrections above the coordinate tolerance. Uniformly over the query $C$, at most $O(k\alpha^{-2}\log k)$ such corrections are needed; rounding their values gives the shift net. Define the $\alpha$-shift set 
\[
S_\alpha(C):=\{p\in X_j^{\mathrm{out}}\cap X_{m,\phi,\beta}:j\in J,d^2(p,c(a_j))-d^2(p,C)\ge\alpha R^2t/16\}
\]

Coordinates outside this shift set have shifts below the target tolerance and can be zeroed out, while the next lemma bounds the support size of the remaining coordinates.

\begin{lemma}[Sparsity of \(\alpha\)-shifted points]
\label{lemma:alpha_shifted_points_num}

Under the inherited event $\mathcal E_s$, let \(\alpha\in(0,1]\) satisfy \(\alpha\ge\alpha_\star\). Then, for every \(C\in\mathcal C\),
\[
    |S_\alpha(C)|\lesssim k\alpha^{-2}\log k.
\]
\end{lemma}

The proof of Lemma~\ref{lemma:alpha_shifted_points_num} is deferred to Section \ref{sec:alpha_shifted_points_num}. We now prove Lemma~\ref{lemma:shift_vec_net} by encoding the active support and rounding its shift entries.

\begin{proof}[ Proof of Lemma~\ref{lemma:shift_vec_net}] We use Lemma~\ref{lemma:alpha_shifted_points_num} and $0\le s_p^C\le8R^2t$. 
Now for any $C\in\mathcal{C}$, $j\in J$ and $p\in X_j^{\mathrm{out}}\cap X_{m,\phi,\beta}$, let
$$
s_p^{C,\alpha}=\begin{cases}
    \floor{\frac{16s_p^C}{\alpha R^2t}}\cdot\frac{\alpha R^2t}{16},&p\in S_{\alpha}(C)\\
    0,&\text{otherwise}
\end{cases}
$$
and $\mathcal S(\alpha)=\set{s^{C,\alpha}:C\in\mathcal{C}}$. Then we have that $\forall p\in X$, $\abs{s_p^{C,\alpha}-s_p^C}\le\alpha R^2t/16$. Each nonzero coordinate has at most $1+128/\alpha$ possible values. For a sufficiently large universal constant $c>0$, counting all supports up to the sparsity bound gives
\[
    \log|\mathcal S(\alpha)|\le\log\left(\sum_{i=0}^{\min\{|X|,\lceil ck\alpha^{-2}\log k\rceil\}}\binom{|X|}{i}(1+128/\alpha)^i\right)\lesssim k\alpha^{-2}\log k\log(eN/\alpha).
\]

\end{proof}

\subsubsection{Proof of Lemma~\ref{lemma:alpha_shifted_points_num}: sparsity of $\alpha$-shifted points}\label{sec:alpha_shifted_points_num}

We prove Lemma~\ref{lemma:alpha_shifted_points_num} using the inherited event in Definition~\ref{def:signed-subset-sum-good-event}.

\begin{proof}[Proof of Lemma~\ref{lemma:alpha_shifted_points_num}]
The coefficient in the definition of $\mathcal E_s$ is now fixed; the implicit constants below are universal.

Fix $C\in\mathcal C$ and label its distinct center locations $c_1,\ldots,c_{k'}$, where $1\le k'\le k$. Retain the same nearest-center location $c(x)$ as in the definition of the shift vector; removing repeated copies of that location does not change $c(x)$. If $k'=1$ or $S_\alpha(C)=\emptyset$, there is nothing to prove.

For every \(p\in S_\alpha(C)\cap X_j^{\mathrm{out}}\cap X_{m,\phi,\beta}\), define
\[
    x_p:=\frac{p-a_j}{\norm{p-a_j}_2}.
\]
Since \(p\in S_\alpha(C)\),
\[
    d^2(p,c(a_j))-d^2(p,c(p))\ge\alpha R^2t/16.
\]
Also, since \(c(a_j)\) is the closest center to \(a_j\),
\[
    d^2(a_j,c(p))-d^2(a_j,c(a_j))\ge 0.
\]
Expanding \(p=a_j+(p-a_j)\), we get
\[
    2\ip{p-a_j}{c(p)-c(a_j)}
    =d^2(p,c(a_j))-d^2(p,c(p))+d^2(a_j,c(p))-d^2(a_j,c(a_j))
    \ge\alpha R^2t/16.
\]
Using \(\norm{p-a_j}_2=\Theta(R)\) on the fixed block, this implies
\[
    \ip{x_p}{\frac{c(p)-c(a_j)}{Rt}}\gtrsim \alpha.
\]
Moreover,
\[
    \norm{c(p)-c(a_j)}_2\le d(p,c(p))+d(p,c(a_j))\le 2d(p,c(a_j))\le O(tR).
\]

Define a directed graph $G_C=([k'],E_C)$ as follows. For $e=(i,i')\in[k']\times[k']$, put $e\in E_C$ if there exists $p\in S_\alpha(C)$ such that $c(a(p))=c_i$ and $c(p)=c_{i'}$. For such an edge, set
\[
    S_e:=\{p\in S_\alpha(C):c(a(p))=c_i,\ c(p)=c_{i'}\},
    \qquad
    z_e:=\frac{c_{i'}-c_i}{Rt}.
\]
The sets \(S_e\) are disjoint over \(e\in E_C\), and the preceding paragraph gives \(\norm{z_e}_2\lesssim 1\). Hence
\[
    \alpha\abs{S_\alpha(C)}\lesssim \sum_{e\in E_C}\sum_{p\in S_e}\ip{z_e}{x_p}
    =\sum_{e\in E_C}\ip{z_e}{\sum_{p\in S_e}x_p}.
\]

Applying the directional bound separately to the edges of $G_C$ would require counting up to $k(k-1)$ ordered center pairs. Instead, we route their center differences along a sparse subgraph of the underlying undirected graph. Apply Lemma~\ref{lem:sparse-log-spanner} separately to each nontrivial connected component, keep isolated vertices, and take their union to obtain a spanner \(G_C^{\mathrm{sp}}\) with stretch $O(\log k)$ and \(\abs{E(G_C^{\mathrm{sp}})}=O(k)\). Orient every edge of \(G_C^{\mathrm{sp}}\) arbitrarily. For an oriented spanner edge \(\eta=(r,s)\), set
\[
    z_\eta:=\frac{c_s-c_r}{Rt}.
\]
Since \(G_C^{\mathrm{sp}}\) is a subgraph of the underlying shift graph, we have
\[
    \norm{z_\eta}_2\lesssim 1.
\]

For every directed edge \(e=(i,i')\in E_C\), let \(P(e)\) be a simple spanner path from \(i\) to \(i'\) within its connected component, of length $O(\log k)$. Along this path, the center difference telescopes:
\[
    z_e=\sum_{\eta\in P(e)}\tau_{e,\eta}z_\eta,
    \qquad
    \tau_{e,\eta}\in\{-1,+1\}.
\]
Substituting this into the previous bound and then taking absolute values gives
\[
\begin{aligned}
    \alpha\abs{S_\alpha(C)}
    &\lesssim \sum_{\eta\in E(G_C^{\mathrm{sp}})}
    \abs{\ip{z_\eta}{\sum_{e:\eta\in P(e)}\tau_{e,\eta}\sum_{p\in S_e}x_p}} \\
    &\lesssim \sum_{\eta\in E(G_C^{\mathrm{sp}})}
    \norm{\sum_{e:\eta\in P(e)}\tau_{e,\eta}\sum_{p\in S_e}x_p}_2 .
\end{aligned}
\]

For each spanner edge \(\eta\), define
\[
    T_\eta^+:=\bigsqcup_{\substack{e:\eta\in P(e)\\ \tau_{e,\eta}=+1}}S_e,
    \qquad
    T_\eta^-:=\bigsqcup_{\substack{e:\eta\in P(e)\\ \tau_{e,\eta}=-1}}S_e.
\]
Each point belongs to a unique $S_e$, and the simple path $P(e)$ traverses $\eta$ at most once. Thus $T_\eta^+$ and $T_\eta^-$ are disjoint subsets of the original sample, as required by $\mathcal E_s$. Therefore,
\[
    \sum_{e:\eta\in P(e)}\tau_{e,\eta}\sum_{p\in S_e}x_p
    =\sum_{p\in T_\eta^+}x_p-\sum_{p\in T_\eta^-}x_p.
\]
Let
\[
    M_\eta:=\abs{T_\eta^+}+\abs{T_\eta^-}.
\]
By the signed subset-sum event,
\[
\begin{aligned}
    \alpha\abs{S_\alpha(C)}
    &\lesssim \sum_{\eta\in E(G_C^{\mathrm{sp}}):M_\eta>0}
    \norm{\sum_{p\in T_\eta^+}x_p-\sum_{p\in T_\eta^-}x_p}_2 \\
    &\lesssim \sum_{\eta\in E(G_C^{\mathrm{sp}}):M_\eta>0}
    \left(\sqrt{M_\eta}+M_\eta\sqrt{\frac{\log(eN/M_\eta)}{d}}\right).
\end{aligned}
\]

Each shifted point is charged to $O(\log k)$ spanner edges, so
\[
    \sum_{\eta\in E(G_C^{\mathrm{sp}})}M_\eta
    \lesssim \log k\abs{S_\alpha(C)}.
\]
Thus, by Cauchy--Schwarz and \(\abs{E(G_C^{\mathrm{sp}})}=O(k)\),
\[
    \sum_{\eta\in E(G_C^{\mathrm{sp}})}\sqrt{M_\eta}\lesssim \sqrt{k\log k\abs{S_\alpha(C)}}.
\]
Also,
\[
    \sum_{\eta\in E(G_C^{\mathrm{sp}}):M_\eta>0}M_\eta\sqrt{\frac{\log(eN/M_\eta)}{d}}
    \lesssim \log k\abs{S_\alpha(C)}\sqrt{\frac{\log(eN)}{d}}.
\]
Combining the last three displays gives
\[
    \alpha\abs{S_\alpha(C)}\lesssim \sqrt{k\log k\abs{S_\alpha(C)}}+\log k\abs{S_\alpha(C)}\sqrt{\frac{\log(eN)}{d}}.
\]
By the choice of the universal multiplier in $\alpha_\star$, the second term, including its implicit constant, is at most $\alpha|S_\alpha(C)|/2$. Absorbing it, dividing by $\sqrt{|S_\alpha(C)|}$, and squaring gives
\[
    \abs{S_\alpha(C)}\lesssim k\alpha^{-2}\log k.
\]
This proves the lemma.
\end{proof}

\section{Proof of Theorem~\ref{thm:vanilla_bound}: coreset upper bounds for general mixtures}\label{sec:general-mixtures}

Section \ref{sec:general-mixtures} establishes our population coreset upper for general mixture distributions (Theorem \ref{thm:vanilla_bound}) by adapting the framework in Section \ref{sec:upper-bounds}. We consider a mixture $\mu=\sum_{j=1}^k\pi_j\mu_j$ with component means $a_j$ and finite radii $r_j=(\mathbb E_{p\sim\mu_j}d^2(p,a_j))^{1/2}$. Here $X_j$ denotes the full sample from the latent component $\mu_j$, and
\[
    V_\mu=\sum_{j=1}^k\pi_jr_j^2,\qquad
    \zeta_\mu=\frac{\max_jr_j^2}{V_\mu},\qquad
    \pi_{\min}=\min_j\pi_j.
\]
We assume $V_\mu>0$, the comparison condition~\eqref{eq:comparison_condition}, and Assumption~\ref{ass:general-tail}. Components with $r_j=0$ are allowed.

\subsection{Common reductions and chaining}
Fix $\varepsilon\in(0,1/2)$. Compared with the isotropic proof, Assumption~\ref{ass:general-tail} replaces Gaussian radial concentration in the shell bound, and the good event is only $\mathcal E_c$: the directional event $\mathcal E_s$ is no longer used. We bound count failures unconditionally, since the normalized-radius law may depend on the latent component.

The far/close and tiny/huge partitions and the deterministic inner/outer split are unchanged from Section~\ref{sec:proof_main_theorem}, with inner cutoff $\varepsilon^2r_j$. The same component-mean approximations for far components and inner points, together with the tiny-component estimate, leave only $X_j^{\mathrm{out}}$ for $j\in J_{\mathrm{huge}}(C)$ to be bounded by chaining.

Lemma~\ref{lemma:symm} gives
\[
    Y_C=\sum_{p\in X}\frac{g_p d^2(p,C)}{N\cost(\mu,C)},\qquad
    \mathbb E_X\sup_C\frac{|\cost(X,C)-\cost(\mu,C)|}{\cost(\mu,C)}
    \lesssim\mathbb E_X\mathbb E_g\sup_C|Y_C|.
\]
The following lemmas give the shell and count-failure estimates and the outer-point reduction; their proofs are in Appendix~\ref{appendix:general-common}.

\begin{lemma}[Radial shells and count failures]\label{lem:general-shells}
For $\ell\ge2$,
\[
    \Pr[\mathscr E_\ell]\le\min\{1,2N\exp(-\Omega(2^{q(\ell-1)}))\},\qquad
    \Pr[\mathscr E_1]\le1.
\]
The shell factors $2^{2\ell}$ and $2^{2\ell}(1+\ell)^{3/2}$ have polylogarithmic averages in $N$. Moreover,
\[
    \mathbb E_X\!\left[\mathbf 1_{\overline{\mathcal E_c}}\mathbb E_g\sup_C|Y_C|\right]
    \lesssim\left(\zeta_\mu[\log(eN)]^{2/q}+\pi_{\min}^{-1}\right)
    \sqrt{2k}\exp(-N\pi_{\min}/24).
\]
\end{lemma}

\begin{lemma}[Reduction to outer points]\label{lem:general-reductions}
Fix $X\in\mathcal E_c\cap\mathscr E_\ell$ and suppose $N\pi_{\min}\ge\log(2k)$. Then
\begin{align*}
    \mathbb E_g\sup_C|Y_C|
    &\lesssim 2^{2\ell}\varepsilon+\sqrt{\frac{\log(2k)}{N\pi_{\min}}}\\
    &\quad+\mathbb E_g\sup_C\left|\sum_{j\in J_{\mathrm{huge}}(C)}\sum_{p\in X_j^{\mathrm{out}}}
    \frac{g_p d^2(p,C)}{N\cost(\mu,C)}\right|.
\end{align*}
\end{lemma}

\paragraph{Common cost estimate.}
Both variance arguments use the following consequence of the triangle inequality, the count bounds, and~\eqref{eq:comparison_condition}: for $X\in\mathcal E_c\cap\mathscr E_\ell$,
\begin{equation}\label{eq:general-cost-sum}
\begin{aligned}
    \sum_{j\in J_{\mathrm{huge}}(C)}\sum_{p\in X_j^{\mathrm{out}}}
    \bigl(d^2(p,C)+d^2(p,a_j)\bigr)
    &\lesssim N\sum_j\pi_j\bigl(d^2(a_j,C)+2^{2\ell}r_j^2\bigr)\\
    &\lesssim N2^{2\ell}\cost(\mu,C).
\end{aligned}
\end{equation}

\paragraph{Chaining.}
Use the coarse representatives $v^{C,1}$ and the telescoping processes $Y_{C,\mathrm{Init}},\allowbreak Y_{C,h},\allowbreak Y_{C,\mathrm{End}}$ from Section~\ref{sec:chaining} on the outer points of huge close components, with finest level
\[
    h_{\max}=\left\lceil\log_2\!\left(\varepsilon^{-2}
    \sqrt{\zeta_\mu+(k\pi_{\min})^{-1}}\right)\right\rceil.
\]
Both error functions below dominate $d(p,C)d(p,a_j)+d^2(p,a_j)$ up to a constant. Lemma~\ref{lemma:net_size_in_coarse_level} therefore gives a constant-accuracy coarse cover with
\[
    \log|\mathcal V(1)|\lesssim k\bigl(\ell+\log(1/\varepsilon)\bigr).
\]
Choose all representatives independently of the Gaussian signs. For $0\le h\le h_{\max}$ and $\alpha_h=2^{-h}$, independence of the $g_p$ yields
\begin{equation}\label{eq:general-increment-variance}
    \operatorname{Var}_g(Y_{C,h})\lesssim\alpha_h^2
    \sum_{j\in J_{\mathrm{huge}}(C)}\sum_{p\in X_j^{\mathrm{out}}}
    \frac{\err(p,C)^2}{N^2\cost(\mu,C)^2}.
\end{equation}
The realized-pair argument from Lemma~\ref{lemma:main_term_in_chaining}, which handles the query-dependent denominator, gives
\begin{equation}\label{eq:general-finite-maximum}
    \mathbb E_g\sup_C|Y_{C,h}|\lesssim
    \sqrt{\sup_C\operatorname{Var}_g(Y_{C,h})
    \log\!\left(2|\mathcal V(2^{-h})|\,|\mathcal V(2^{-(h+1)})|\right)}.
\end{equation}

\begin{lemma}[Initial and end terms]\label{lem:general-endpoints}
For $X\in\mathcal E_c\cap\mathscr E_\ell$, either covering construction below satisfies
\[
    \mathbb E_g\sup_C|Y_{C,\mathrm{Init}}|\lesssim\sqrt{\frac{\log(2k)}{N\pi_{\min}}},\qquad
    \mathbb E_g\sup_C|Y_{C,\mathrm{End}}|\lesssim2^{2\ell}\varepsilon.
\]
\end{lemma}
The proof is in Appendix~\ref{appendix:general-endpoints}.

The two covering constructions give the following increment bounds, proved in Sections~\ref{sec:general-relative} and~\ref{sec:general-distance}, respectively.

\begin{lemma}[Chaining increment for relative-error coverings]\label{lem:general-relative-main}
For every $X\in\mathcal E_c\cap\mathscr E_\ell$ and $0\le h\le h_{\max}$,
\[
    \mathbb E_g\sup_C|Y_{C,h}|=\widetilde O\!\left(
    2^{2\ell}(1+\ell)^{1/2}\varepsilon^{-1}\sqrt{\frac{k\zeta_\mu}{N}}\right).
\]
\end{lemma}

\begin{lemma}[Chaining increment for distance-sensitive coverings]\label{lem:general-distance-main}
For every $X\in\mathcal E_c\cap\mathscr E_\ell$ and $0\le h\le h_{\max}$,
\[
    \mathbb E_g\sup_C|Y_{C,h}|=\widetilde O\!\left(
    2^{2\ell}(1+\ell)^{3/2}
    \sqrt{\frac{(k\zeta_\mu+\pi_{\min}^{-1})\sqrt{k}}{N}}\right).
\]
\end{lemma}

\begin{proof}[Proof of Theorem~\ref{thm:vanilla_bound}]
Combining Lemmas~\ref{lem:general-reductions} and~\ref{lem:general-endpoints} with the telescoping sum, and then applying Lemma~\ref{lem:general-shells}, gives
\begin{equation}\label{eq:general-shell-average}
\begin{aligned}
    \mathbb E_X\!\left[\mathbf 1_{\mathcal E_c}\mathbb E_g\sup_C|Y_C|\right]
    &\lesssim\varepsilon[\log(eN)]^{2/q}+\sqrt{\frac{\log(2k)}{N\pi_{\min}}}\\
    &\quad+\sum_{\ell\ge1}\sum_{h=0}^{h_{\max}}
    \mathbb E_X\!\left[\mathbf 1_{\mathcal E_c\cap\mathscr E_\ell}
    \mathbb E_g\sup_C|Y_{C,h}|\right].
\end{aligned}
\end{equation}
Replace $\varepsilon$ throughout the construction by $c\varepsilon/[\log(eN)]^{2/q}$ and absorb the shell averages and chain length into the hidden logarithmic factors. Since $k,\zeta_\mu\le\pi_{\min}^{-1}$, the logarithmic accounting in the proof of Theorem~\ref{thm:main_k_gaussian} applies: these losses are fixed powers of $\log(e/(\varepsilon\pi_{\min}))$, independent of $d$.

For the relative-error construction, Lemmas~\ref{lem:general-relative-main} and~\ref{lem:general-shells} bound the increment sum in~\eqref{eq:general-shell-average} by
\[
    \widetilde O\!\left(\varepsilon^{-1}\sqrt{\frac{k\zeta_\mu}{N}}\right).
\]
The term $\widetilde O(k\zeta_\mu\varepsilon^{-4})$ controls these increments, while $\widetilde O(\pi_{\min}^{-1}\varepsilon^{-2})$ controls the component-mean term and count failures. Thus it suffices to take
\[
    N=\widetilde O\!\left(k\zeta_\mu\varepsilon^{-4}+\pi_{\min}^{-1}\varepsilon^{-2}\right).
\]
For the distance-sensitive construction, Lemmas~\ref{lem:general-distance-main} and~\ref{lem:general-shells} bound the increment sum by
\[
    \widetilde O\!\left(\sqrt{\frac{(k\zeta_\mu+\pi_{\min}^{-1})\sqrt{k}}{N}}\right).
\]
This and the component-mean and count-failure terms are sufficiently small multiples of $\varepsilon$ when
\[
    N=\widetilde O\!\left((k\zeta_\mu+\pi_{\min}^{-1})\sqrt{k}\,\varepsilon^{-2}\right).
\]
In either case, choosing the implicit constants sufficiently large, symmetrization and Markov's inequality show that an i.i.d.\ sample of the indicated size is an $\varepsilon$-coreset with probability at least $0.9$. Use the distance-sensitive bound when $\sqrt{k}\le\varepsilon^{-2}$ and the relative-error bound otherwise to obtain the stated sample bound. Larger samples follow by the subsampling and Jensen argument in Appendix~\ref{appendix:finite-generated-dataset}.
\end{proof}

\subsection{Relative-error coverings}\label{sec:general-relative}

We prove Lemma~\ref{lem:general-relative-main}. In this branch, set
\[
    \err(p,C)=d^2(p,C)+d^2(p,a(p)),
\]
where $a(p)$ is the generating mean. The following variance and covering bounds are proved in Appendix~\ref{appendix:general-relative}.

\begin{lemma}[Relative-error variance]\label{lem:general-relative-variance}
For every $X\in\mathcal E_c\cap\mathscr E_\ell$ and every query $C$,
\[
    \sum_{j\in J_{\mathrm{huge}}(C)}\sum_{p\in X_j^{\mathrm{out}}}
    \frac{\err(p,C)^2}{N^2\cost(\mu,C)^2}
    \lesssim\frac{2^{4\ell}\varepsilon^{-2}\zeta_\mu}{N}.
\]
Consequently,
\[
    \sup_C\operatorname{Var}_g(Y_{C,h})
    \lesssim\frac{2^{4\ell}\varepsilon^{-2}\zeta_\mu\alpha_h^2}{N}.
\]
\end{lemma}

\begin{lemma}[Relative-error covering]\label{lem:general-relative-cover}
For every $X\in\mathcal E_c\cap\mathscr E_\ell$ and $\alpha\in(0,1)$, there is a finite family $\mathcal V(\alpha)\subset\mathbb R^N$ such that every query $C$ has a representative $v^{C,\alpha}\in\mathcal V(\alpha)$ satisfying
\[
    |d^2(p,C)-v_p^{C,\alpha}|\le\alpha\err(p,C)
    \qquad (j\in J_{\mathrm{huge}}(C),\ p\in X_j^{\mathrm{out}}),
\]
zero on all other coordinates, and
\[
    \log|\mathcal V(\alpha)|=\widetilde O\!\left(k\alpha^{-2}(1+\ell)\right).
\]
\end{lemma}
A single multiplicative terminal embedding~\cite{narayanan2019optimal}, followed by Lemma~\ref{lemma:general_alpha_net_size}, suffices for this cover; radius groups and distance blocks are unnecessary.

\begin{proof}[Proof of Lemma~\ref{lem:general-relative-main}]
Apply~\eqref{eq:general-finite-maximum} with Lemmas~\ref{lem:general-relative-variance} and~\ref{lem:general-relative-cover}. The factors $\alpha_h^2$ and $\alpha_h^{-2}$ cancel. The coarse-cover bound gives the same estimate at $h=0$.
\end{proof}

\subsection{Distance-sensitive coverings}\label{sec:general-distance}

\paragraph{Radius groups and distance blocks.}
Reuse the radius groups $G(m)$ and distance blocks from Definitions~\ref{def:group-structure} and~\ref{def:certificates-and-blocks}, with the same certificates $\phi^{(C,m)}$ and fixed blocks $X_{m,\phi,\beta}$. Write
\[
    R=2^{-m}2^\ell\max_jr_j,\qquad t=2^\beta.
\]
Then $R/2<d(p,a_j)\le R$ on $G(m)\cap X_j^{\mathrm{out}}$, and fixing $m,\phi$ fixes the blocks as $C$ varies over $\mathcal C_{m,\phi}$.

\paragraph{Distance-sensitive error.}
For $p\in G(m)\cap X_j^{\mathrm{out}}$, set
\begin{equation}\label{eq:general-distance-error}
    \err(p,C)=\bigl(d(p,C)d(p,a_j)+d^2(p,a_j)\bigr)
    \sqrt{1+\frac{\cost(\mu,C)}{k\pi_{\min}R^2}}.
\end{equation}
The cover approximates the original costs: it covers $d^2(p,C)-d^2(a_j,C)$ and adds rounded component-mean costs. The enlargement in~\eqref{eq:general-distance-error} permits coarser componentwise additive embeddings and contributes only an additive $(k\pi_{\min})^{-1}$ to the variance coefficient.

\begin{lemma}[Distance-sensitive variance]\label{lem:general-distance-variance}
For every $X\in\mathcal E_c\cap\mathscr E_\ell$ and every query $C$,
\[
    \sum_{j\in J_{\mathrm{huge}}(C)}\sum_{p\in X_j^{\mathrm{out}}}
    \frac{\err(p,C)^2}{N^2\cost(\mu,C)^2}
    \lesssim\frac{2^{4\ell}}{N}\left(\zeta_\mu+\frac1{k\pi_{\min}}\right).
\]
Consequently,
\[
    \sup_C\operatorname{Var}_g(Y_{C,h})
    \lesssim\frac{2^{4\ell}\alpha_h^2}{N}\left(\zeta_\mu+\frac1{k\pi_{\min}}\right).
\]
\end{lemma}
The proof is in Appendix~\ref{appendix:general-distance}.

\begin{lemma}[Distance-sensitive covering]\label{lem:general-distance-cover}
For every $X\in\mathcal E_c\cap\mathscr E_\ell$ and $\alpha\in(0,1)$, there is a finite family $\mathcal V(\alpha)\subset\mathbb R^N$ such that every query $C$ has a representative $v^{C,\alpha}\in\mathcal V(\alpha)$ satisfying
\[
    |d^2(p,C)-v_p^{C,\alpha}|\le\alpha\err(p,C)
    \qquad (j\in J_{\mathrm{huge}}(C),\ p\in X_j^{\mathrm{out}}),
\]
with the error in~\eqref{eq:general-distance-error}, and $v_p^{C,\alpha}=0$ on all other coordinates. Moreover,
\[
    \log|\mathcal V(\alpha)|=\widetilde O\!\left(k^{3/2}\alpha^{-2}(1+\ell)^3\right).
\]
\end{lemma}
The proof is in Appendix~\ref{appendix:general-distance}. 

\begin{proof}[Proof of Lemma~\ref{lem:general-distance-main}]
Apply~\eqref{eq:general-finite-maximum} with Lemmas~\ref{lem:general-distance-variance} and~\ref{lem:general-distance-cover}. The factors $\alpha_h^2$ and $\alpha_h^{-2}$ cancel, and $k^{3/2}(\zeta_\mu+(k\pi_{\min})^{-1})=(k\zeta_\mu+\pi_{\min}^{-1})\sqrt{k}$. At $h=0$, use the common coarse-cover bound.
\end{proof}

\section{Proof of Theorem \ref{thm:gaussian-sampling-lower-bound}: lower bounds}\label{sec:lower-bounds}

Our construction adapts the shared-center geometry and opposite-query
argument of Hu et al.~\cite[Sections~2.2--2.3]{HuHuangLiXu2026TightCoresetBounds}.
Here Gaussian orthants replace the discrete groups: center coordinates
control their costs independently, and two opposite choices make the
sampling errors add. We prove the Gaussian geometry and fluctuation
estimates below. The resulting lower bound concerns uniform sampling,
whereas the cited result allows arbitrary coreset reweighting.

Theorem~\ref{thm:gaussian-sampling-lower-bound} is stated in Section~\ref{sec:contributions}. We prove it in this section.

All implicit constants below are absolute; logarithmic factors are
shown explicitly. Recall that $\cost(X,C)=N^{-1}\sum_{p\in X}d^2(p,C)$, and write
$\Delta(C):=\cost(X,C)-\cost(\mu,C)$ for the sampling error.
We allow singular covariance matrices and give every auxiliary coordinate
zero variance.

\subsection{The construction}
\label{sec:construction}

Initially let $k$ be a perfect square. Choose an integer $r=\Theta(\log k)$
with $2^r=\Theta(k)$ and $2^r+1\le k$. In $\mathbb R^{r+\sqrt k+2}$,
use an $r$-dimensional Gaussian block, a common mean coordinate,
$\sqrt k$ component coordinates, and one norm-adjustment coordinate.
Writing $e_j$ for the standard basis of the component block, set
\begin{equation}
 \Sigma=r^{-1}I_r\oplus0_{\sqrt k+2},\qquad
 a_j=\begin{cases}
 (0,\sqrt3\,k^{1/4},\sqrt2\,k^{1/4}e_j,0),&j\le\sqrt k,\\
 0,&j>\sqrt k.
 \end{cases}
 \label{eq:mixture}
\end{equation}
A point from component $j$ is $p=a_j+(g/\sqrt r,0)$, with
$g\sim\mathcal N(0,I_r)$. The first $\sqrt k$ components are active;
the others have mean zero. In particular,
\begin{equation}
 \mathbb E\|p\|^2=1+\frac1k\sum_{j\le\sqrt k}\|a_j\|^2=6.
 \label{eq:cost-upper}
\end{equation}

For each odd sign vector $\sigma\in\{-1,1\}^r$, choose
$y_{j\sigma}\in\{0,-\sqrt2\}$ freely; for even $\sigma$, set all
$y_{j\sigma}=0$. Include the origin and the centers
\begin{equation}
 c_\sigma=\left(\frac{k^{1/4}}{\sqrt r}\sigma,\,
 \sqrt3\,k^{1/4},\,(y_{j\sigma})_{j\le\sqrt k},\,
 \sqrt{2\sqrt k-\sum_{j\le\sqrt k}y_{j\sigma}^2}\right).
 \label{eq:centers}
\end{equation}
Parity refers to the number of negative coordinates. Every choice is
legal, uses at most $k$ centers, and has $\|c_\sigma\|^2=6\sqrt k$.
Since $0\in C$, every constructed query satisfies $\cost(\mu,C)\le6$.

\begin{claim}[Point-to-center distance expansion]
\label{clm:center-distances}
For $p$ from an active component $j$,
\begin{equation}
 \|p-c_\sigma\|^2-\|p\|^2
 =-2\sqrt2\,k^{1/4}y_{j\sigma}
   -\frac{2k^{1/4}}r\langle\sigma,g\rangle.
 \label{eq:distance}
\end{equation}
\end{claim}
\begin{proof}
Expand the squared distance, using
$2\langle a_j,c_\sigma\rangle=6\sqrt k+2\sqrt2\,k^{1/4}y_{j\sigma}$.
\end{proof}

\begin{lemma}[Orthant locality]
\label{lem:orthant-costs}
For an active component $j$ and $\sigma=\operatorname{sign}(g)$,
\begin{equation}
 d^2(p,C)=\|p\|^2-\frac{2k^{1/4}}r\sum_{\ell=1}^r|g_\ell|
 +4k^{1/4}\min\{1,r^{-1}\min_\ell|g_\ell|\}
     \mathbf1[y_{j\sigma}=-\sqrt2].
 \label{eq:orthant-cost}
\end{equation}
Thus changing $y_{j\sigma}$ affects only its own component--orthant
cell, with cost difference
\begin{equation}
 4k^{1/4}\min\{1,r^{-1}\min_\ell|g_\ell|\}.
 \label{eq:cost-gap}
\end{equation}
Even orthants and inactive components have fixed costs.
\end{lemma}
\begin{proof}
By~\eqref{eq:distance}, a mismatched sign coordinate incurs penalty
$4k^{1/4}|g_\ell|/r$. An even matching center is therefore nearest.
For an odd orthant, the best even center flips a coordinate minimizing
$|g_\ell|$; every other odd center differs in at least two coordinates
and is no better. The origin is also no better, since $r\ge2$.
Comparing the matching odd center with this even center gives the formula.
For inactive components, all center norms and Gaussian projections are fixed.
\end{proof}

\begin{lemma}[Gap moments]
\label{clm:gap-moments}
The cost difference~\eqref{eq:cost-gap} is independent of the component
and orthant, and, for $h\in\{1,2,4\}$,
\begin{equation}
 \mathbb E_g\bigl[(4k^{1/4}\min\{1,r^{-1}\min_\ell|g_\ell|\})^h\bigr]
 =\Theta(k^{h/4}r^{-2h}).
 \label{eq:gap-moments}
\end{equation}
\end{lemma}
\begin{proof}
Gaussian signs and absolute values are independent. Moreover,
$\Pr(\min_\ell|g_\ell|>t)\le e^{-crt}$ for $t\ge0$, while
$\Pr(\min_\ell|g_\ell|>c'/r)\ge1/2$ for sufficiently small $c'>0$.
Tail integration gives the claimed moments.
\end{proof}

\subsection{Sampling error}

We use the opposite-query comparison of
\cite[Proposition~2.3]{HuHuangLiXu2026TightCoresetBounds}.
If $C,C'$ differ only by changing $y_{j\sigma}$ from zero to $-\sqrt2$
in an odd active cell, locality gives
\begin{equation}
 \Delta(C')-\Delta(C)=4k^{1/4}\left(
 \frac1N\mathop{\textstyle\sum}\limits_{p\in X_j,\,\operatorname{sign}(g)=\sigma}
       \min\{1,r^{-1}\min_\ell|g_\ell|\}
 -\tfrac{\mathbb E_g\min\{1,r^{-1}\min_\ell|g_\ell|\}}{k2^r}
 \right).
 \label{eq:cell-error}
\end{equation}
The population term uses cell probability $1/(k2^r)$ and
Lemma~\ref{clm:gap-moments}. Choose one query to maximize each cell's
signed error and the other to minimize it. These choices are simultaneous,
and their common costs cancel. Hence, over the constructed queries,
\begin{align}
 &\sup_C|\Delta(C)|\ge\frac12\left(\max_C\Delta(C)-\min_C\Delta(C)\right)\notag\\
 &\quad=2k^{1/4}\mathop{\textstyle\sum}\limits_{j\le\sqrt k,\,\sigma\text{ odd}}
 \left|\frac1N\mathop{\textstyle\sum}\limits_{p\in X_j,\,\operatorname{sign}(g)=\sigma}
       \min\{1,r^{-1}\min_\ell|g_\ell|\}
 -\frac{\mathbb E_g\min\{1,r^{-1}\min_\ell|g_\ell|\}}{k2^r}\right|.
 \label{eq:error-range}
\end{align}
Here the family is fixed before sampling; only the two selected queries
depend on $X$. Active component labels are determined by their mean
coordinates, and population costs hold the selected query fixed.

There are $\Theta(k^{3/2})$ adjustable cells, of total probability
$1/(2\sqrt k)$. For $N\le c k^2$ with sufficiently small $c>0$,
Markov's inequality shows that at least half are empty with probability
at least $0.99$. Their contributions to~\eqref{eq:error-range} yield
\[
 \sup_C|\Delta(C)|
 =\Omega(k^{-1/4}r^{-2}),
\]
proving the quadratic sample bound at accuracy $\Theta(k^{-1/4}/\log^2k)$.

\begin{lemma}[Frequency fluctuations]
\label{lem:frequency}
For every $N\ge1$, with probability at least $0.99$, the right-hand side
of~\eqref{eq:error-range} is at least a constant times
\[
 \min\{k^{-1/4}r^{-2},\,k^{3/4}/(\sqrt N\,r^2)\}.
\]
\end{lemma}
\begin{proof}
The empty-cell argument handles $N\le c k^2$. Otherwise, the second
and fourth moments of $N$ times~\eqref{eq:cell-error} are
\[
 \Theta\left(\frac{N}{k^{3/2}r^4}\right),\qquad
 O\left(\frac{N}{kr^8}+\frac{N^2}{k^3r^8}\right)
 =O\left(\frac{N^2}{k^3r^8}\right).
\]
H\"older's inequality gives expected absolute cell error
$\Omega(k^{-3/4}/(\sqrt N\,r^2))$, so summing gives expectation
$\Omega(k^{3/4}/(\sqrt N\,r^2))$ in~\eqref{eq:error-range}.
Replacing one observation changes the right-hand side of~\eqref{eq:error-range}
by at most the sum of the old and new cell-gap contributions divided by $N$.
A contribution is nonzero only in an adjustable cell, whose total probability
is $1/(2\sqrt k)$. By~\eqref{eq:gap-moments}, the expected squared change is
therefore $O(1/(N^2r^4))$. Efron--Stein bounds the variance by
$O(1/(Nr^4))$. The relative variance is
$O(k^{-3/2})$, and Chebyshev completes the proof.
\end{proof}

Since $\cost(\mu,C)\le6$, we obtain failure for sufficiently small
absolute constants whenever
\[
 \varepsilon\le c\,\frac{k^{-1/4}}{\log^2k},\qquad
 N\le c'\,\frac{k^{3/2}}{\varepsilon^2\log^4k}.
\]

\subsection{Other accuracies and the comparison condition}
\label{sec:accuracies}

\paragraph{Smaller copies.}
For a sufficiently large square $q\le k$, take as many $q$-component
copies as fit within $k$ components, replacing $k$ by $q$ in the construction.
Give each component weight $1/k$ and put any remaining components at the
first copy's inactive mean. All copies share the same coordinate blocks;
translate copy $b$ by $4bq^{1/4}$ along one additional zero-variance axis.
Each copy uses at most $q$ centers, and the total population cost remains
at most six. Exact separation is verified below.

There are $\Theta(k\sqrt q)$ adjustable cells, each of probability
$\Theta(1/(kq))$. Applying the same argument gives, with probability
at least $0.99$,
\begin{equation}
 \sup_C|\Delta(C)|
 \ge c\min\left\{\frac{q^{-1/4}}{\log^2q},
                  \frac{\sqrt k\,q^{1/4}}{\sqrt N\log^2q}\right\}.
 \label{eq:copies-frequency}
\end{equation}
The empty-cell threshold is now $N=\Theta(kq)$; above it, the variance
is $O(1/(N\log^4q))$ and the relative variance is $O((k\sqrt q)^{-1})$.

Choose $q$, up to rounding to a square, as a sufficiently small constant times
\[
 \min\{k,\varepsilon^{-4}\log^{-8}(k/\varepsilon)\}.
\]
Then $\varepsilon$ is small enough for the first term
in~\eqref{eq:copies-frequency}, and
\begin{equation}
 \frac{k\sqrt q}{\varepsilon^2\log^4q}
 =\Omega\left(
 \frac{\min\{k^{3/2}\varepsilon^{-2},\,k\varepsilon^{-4}\}}
      {\log^8(k/\varepsilon)}\right).
 \label{eq:full-accuracy}
\end{equation}
This also handles nonsquare $k$.

If no sufficiently large $q$ is available, the threshold
in~\eqref{eq:sample-range} is $O(k)$; shrink its constant so that
$N\le k/4$. Use instead means
$a_j=(0,0,0,0,j\sqrt{200/(1-\varepsilon)})$ and
$\Sigma=\tfrac14 I_4\oplus0$. Take the observed component means as centers.
At least $3k/4$ components are absent, giving population cost at least
$150/(1-\varepsilon)$, whereas Markov bounds the empirical cost by
$100$ with probability $0.99$. This proves failure in the remaining range.

\paragraph{Separation and cost comparison.}
\label{sec:regularity}
For a point in copy $b$, a center in copy $b'\ne b$ is farther than
the local center with the same orthant index: their Gaussian projections
and local norms agree, so the squared-distance difference is at least
\begin{equation}
 16(b-b')^2\sqrt q-4q^{1/4}>0.
 \label{eq:perturbation}
\end{equation}
Cross-copy origins are also farther. Thus locality holds exactly,
including for the duplicated inactive components.

Finally, a Gaussian with covariance $r^{-1}I_r$ places at most
$(\sqrt e\,\rho)^r$ mass in any radius-$\rho$ ball, by the density and
ball-volume bounds. For a sufficiently small absolute $\rho>0$, an
ambient ball meets at most one copy, of mass at most $2q/k$.
Since $2^r=\Theta(q)$, the union of balls around any $k$ centers has
mass at most $2q(\sqrt e\,\rho)^r\le1/2$. Hence
$\cost(\mu,C)=\Omega(1)$ for every query, while
$V_\mu=\mathsf{tr}\Sigma=1$. The single-copy case uses $q=k$;
the coarse construction uses the same argument with $r=4$ and mass
$1/k$ per separated component. All parameters are chosen before sampling,
completing the proof of Theorem~\ref{thm:gaussian-sampling-lower-bound}.

\newcommand{\etalchar}[1]{$^{#1}$}

\appendix

\section{Standard tools}

\begin{lemma}[Relaxed triangle inequality]
Let $a, b, c$ belong to a metric space $(\mathcal X,d)$. For every $t\in(0,1]$, the following inequality holds:
\[
d^2(a, b) \leq (1 + t) d^2(a, c) + \left(1 + \frac{1}{t}\right) d^2(b, c).
\]
\end{lemma}

\begin{lemma}[Gaussian maximal inequality, Lemma 2.3 in \cite{massart2007concentration}, Lemma A.21 in \cite{DBLP:conf/stoc/Huang0024}]
\label{lemma:gaussian_maximal_ineq}
    Let $g_i \sim N(0, \sigma_i^2)$ for each $i \in [n]$ be Gaussian random variables (not need to be independent) and suppose $\sigma = \max_{i \in [n]} \sigma_i$. Then
\[
\mathbb{E}\left[\max_{i \in [n]} |g_i|\right] \leq 2\sigma \cdot \sqrt{2 \log(2n)}.
\]
\end{lemma}

\begin{lemma}[Variance of the combination of Gaussians]
For independent centered Gaussian random variables $g_1,\ldots,g_n$ and every deterministic vector $v\in\mathbb R^n$,
\[
    \operatorname{Var}\left(\sum_{i=1}^n v_i g_i\right)
    =\sum_{i=1}^n v_i^2\operatorname{Var}(g_i).
\]
\end{lemma}

\begin{lemma}[Symmetrization, Lemma 7.4 in \cite{van2014probability}]\label{lemma:symm}
     Let $X_1, \ldots, X_n$ be i.i.d. random variables in $\mathbb{X}$ with distribution $\mu$, and let $\mathcal{F}$ be a class of functions on $\mathbb{X}$. Then

$$
\begin{aligned}
\mathbb{E}\left[\sup _{f \in \mathcal{F}} \sum_{k=1}^n(f\left(X_k\right)-\mu f)\right] & \leq 2 \mathbb{E}\left[\sup _{f \in \mathcal{F}} \sum_{k=1}^n \varepsilon_k f\left(X_k\right)\right] \\
& \leq \sqrt{2 \pi} \mathbb{E}\left[\sup _{f \in \mathcal{F}} \sum_{k=1}^n g_k f\left(X_k\right)\right]
\end{aligned}
$$

where $\varepsilon_1, \ldots, \varepsilon_n$ are i.i.d. symmetric Bernoulli random variables and $g_1, \ldots, g_n$ are i.i.d. $N(0,1)$ random variables independent of $X$.
\end{lemma}

\begin{lemma}[Log-Sobolev concentration, Theorem 5.3 in \cite{ledoux2001concentration}]
Let $\mu$ be a probability measure on the Borel sets of a metric space $(X,d)$. Assume that for some constant $C>0$ and every locally Lipschitz function $f$ on $X$,
\[
    \operatorname{Ent}_{\mu}(f^2) \le 2C \int |\nabla f|^2\,d\mu .
\]
Then every $1$-Lipschitz function $F:X\to\mathbb R$ is integrable and, for every $r\ge 0$,
\[
    \mu\!\left(\left\{x\in X:F(x)\ge \int F\,d\mu+r\right\}\right) \le \exp\!\left(-\frac{r^2}{2C}\right).
\]
In particular, the concentration function satisfies
\[
    \alpha_{(X,d,\mu)}(r) \le \exp\!\left(-\frac{r^2}{8C}\right),\qquad r>0 .
\]
\end{lemma}
\begin{corollary}[Lipschitz concentration form]\label{lem:Lipschitz_concentration}
Under the assumptions of the preceding lemma, if $F:X\to\mathbb R$ is $L$-Lipschitz with $L>0$, then for every $r\ge 0$,
\[
    \mu\!\left(\left\{x\in X:F(x)\ge \mathbb E_{\mu}F+r\right\}\right) \le \exp\!\left(-\frac{r^2}{2CL^2}\right).
\]
Equivalently, if $X_0\sim\mu$, then
\[
    \Pr\!\left[F(X_0)\ge \mathbb E F(X_0)+r\right] \le \exp\!\left(-\frac{r^2}{2CL^2}\right).
\]
\end{corollary}
\begin{proof}
Apply the preceding lemma to the $1$-Lipschitz function $F/L$:
\[
    \mu\!\left(\frac{F}{L}\ge\frac{\mathbb E_\mu F}{L}+\frac rL\right)
    \le\exp\!\left(-\frac{r^2}{2CL^2}\right).
\]
Multiplying the inequality defining the event by $L$ gives the result.
\end{proof}

\begin{definition}[$\rho$-spanner]\label{def:rho-spanner}
Let $G=(V,E)$ be an unweighted graph, and let $d_G(\cdot,\cdot)$ denote the shortest-path distance in $G$. A subgraph $G'=(V,E')$ with $E'\subseteq E$ is a $\rho$-spanner of $G$ if for every $u,v\in V$ connected in $G$,
\[
    d_{G'}(u,v)\le \rho\cdot d_G(u,v).
\]
\end{definition}

\begin{lemma}[Sparse $O(\log|V|)$-spanner, Theorem 1 in \cite{althofer1993sparse}]\label{lem:sparse-log-spanner}
    For any graph $G=(V,E)$ with $|V|\ge1$ and every $R>0$, there is a $(2R+1)$-spanner $G'$ such that the number of edges in $G'$ is less than $\abs{V}\cdot\ceil{\abs{V}^{1/R}}$.
\end{lemma}
A singleton graph has an empty spanner. For $|V|\ge2$, taking $R=\log\abs{V}$ gives a $\rho$-spanner with $\rho=O(\log\abs{V})$ and
\[
    |E(G')|\le \abs{V}\cdot\ceil{\abs{V}^{1/R}}=O(\abs{V}).
\]

\section{Gaussian-mixture structural lemmas}\label{appendix:technical_lemmas}

\begin{lemma}[Component-count event]\label{lemma:sample_count}
If $N\ge2$ and $N\pi_{\min}\gtrsim\log(kN)$ with a large enough implicit constant, then
\[
    \Pr\!\left[\forall j\in[k],\ \tfrac12N\pi_j\le|X_j|\le\tfrac32N\pi_j\right]
    =\Pr[\mathcal E_c]\ge1-\tfrac12N^{-2}.
\]
Here $X_j$ is the full original component sample.
\end{lemma}

\begin{proof}
For each $j\in[k]$, $|X_j|$ has distribution $\operatorname{Bin}(N,\pi_j)$. The multiplicative Chernoff bound with relative deviation $1/2$, followed by a union bound, gives
\[
    \Pr[\overline{\mathcal E_c}]
    \le\sum_{j=1}^k2\exp(-N\pi_j/12)
    \le2k\exp(-N\pi_{\min}/12)
    \le\tfrac12N^{-2}.
\]
The final inequality uses the stated sample-size condition.
\end{proof}

\begin{lemma}[Tensorized Lipschitz concentration on product of spheres]
\label{lem:product-sphere-lipschitz-concentration}
Let $x_1,\ldots,x_N$ be independent and uniformly distributed on $S^{d-1}$,
where $d\ge 2$. Equip $(S^{d-1})^N$ with the product Euclidean metric
\[
    d_{\mathrm{prod}}(X,Y)
    :=
    \left(\sum_{\ell=1}^N \|x_\ell-y_\ell\|_2^2\right)^{1/2}.
\]
There exists a universal constant $c_{\mathrm{sp}}>0$ such that, for every
$L$-Lipschitz function
$f:((S^{d-1})^N,d_{\mathrm{prod}})\to \mathbb R$ with $L>0$ and every $t\ge 0$,
\[
    \Pr\left[
        f(x_1,\ldots,x_N)\ge \mathbb E f(x_1,\ldots,x_N)+t
    \right]
    \le
    \exp\left(-c_{\mathrm{sp}}\frac{d t^2}{L^2}\right).
\]
\end{lemma}
\begin{proof}
The logarithmic Sobolev inequality on the unit sphere $S^{d-1}$
has coefficient $2/(d-1)$, including the unit circle when $d=2$;
see~\cite[Section~7.3, p.~206]{ledoux1999concentrationlogSobolev}.
Entropy tensorization~\cite[Proposition~2.2]{ledoux1999concentrationlogSobolev}
therefore gives
\[
    \operatorname{Ent}(h^2)
    \le \frac{2}{d-1}\,
    \mathbb E\sum_{i=1}^N\|\nabla_i h\|_2^2
\]
for locally Lipschitz $h$ on $(S^{d-1})^N$, where $\nabla_i$
is the tangential gradient in the $i$-th coordinate and expectation
is under the product of normalized surface measures.
The product Euclidean distance is at most the product geodesic
distance. Hence $f$ is also $L$-Lipschitz for the product
Riemannian metric, and
$\sum_i\|\nabla_i f\|_2^2\le L^2$ almost everywhere.
Corollary~\ref{lem:Lipschitz_concentration}, with $C=1/(d-1)$,
now yields
\[
    \Pr[f\ge\mathbb Ef+t]
    \le \exp\!\left(-\frac{(d-1)t^2}{2L^2}\right)
    \le \exp\!\left(-\frac{dt^2}{4L^2}\right).
\]
Thus $c_{\mathrm{sp}}=1/4$ suffices.
\end{proof}

\begin{lemma}[Signed subset-sum event $\mathcal E_s$]\label{lem:signed-subset-sum-good-event}
Let $N\ge2$, $d\ge1$, and let $x_1,\ldots,x_N$ be independent random vectors uniformly distributed on $S^{d-1}$. Use the event $\mathcal E_s$ of Definition~\ref{def:signed-subset-sum-good-event}, with normalized directions $x_i$ and coefficient $C_2$. There exists a universal constant $c_0>0$ such that, for every $C_2\ge 1$ satisfying $c_0C_2^2>3$,
\[
    \Pr[\mathcal E_{\mathrm{s}}^c]\le N^{-(c_0C_2^2-3)}.
\]
Consequently, for every fixed $q>0$, choosing $C_2\ge \sqrt{(q+3)/c_0}$ gives $\Pr[\mathcal E_{\mathrm{s}}]\ge 1-N^{-q}$.
\end{lemma}

\begin{proof}
If $d=1$, every signed sum has absolute value at most its number of terms. Since $\log(eN/m)\ge1$ for $1\le m\le N$ and $C_2\ge1$, the event $\mathcal E_s$ then holds deterministically. Assume henceforth that $d\ge2$. Empty subsets satisfy the required inequality automatically.
Fix two disjoint subsets $T_+,T_-\subseteq[N]$, write $T:=T_+\cup T_-$, and define $\sigma_\ell=1$ for $\ell\in T_+$ and $\sigma_\ell=-1$ for $\ell\in T_-$. Consider $F_T(x_1,\ldots,x_N):=\left\|\sum_{\ell\in T}\sigma_\ell x_\ell\right\|_2$. With respect to the product Euclidean metric on $(S^{d-1})^N$, $F_T$ is $\sqrt{|T|}$-Lipschitz, since
\[
    |F_T(X)-F_T(X')|\le \left\|\sum_{\ell\in T}\sigma_\ell(x_\ell-x'_\ell)\right\|_2\le \sum_{\ell\in T}\|x_\ell-x'_\ell\|_2\le \sqrt{|T|}\,d_{\mathrm{prod}}(X,X').
\]
Moreover, since the $x_\ell$'s are independent and uniformly distributed on $S^{d-1}$, we have $\mathbb E x_\ell=0$ and $\mathbb E\langle x_\ell,x_m\rangle=0$ for $\ell\ne m$. Hence
\[
    \mathbb E F_T\le (\mathbb E F_T^2)^{1/2}=\left(\sum_{\ell\in T}\mathbb E\|x_\ell\|_2^2+\sum_{\ell\ne m\in T}\sigma_\ell\sigma_m\mathbb E\langle x_\ell,x_m\rangle\right)^{1/2}=\sqrt{|T|}.
\]
By concentration on the product of spheres, there is a universal constant $c>0$ such that, for every $t\ge 0$,
\[
    \Pr[F_T\ge \sqrt{|T|}+t]\le \Pr[F_T\ge \mathbb E F_T+t]
    \le \exp\!\left(-c\,\frac{dt^2}{|T|}\right).
\]
Taking $t=C_2|T|\sqrt{\log(eN/|T|)/d}$, we get
\[
    \Pr\!\left[F_T\ge \sqrt{|T|}+C_2|T|\sqrt{\frac{\log(eN/|T|)}{d}}\right]\le \exp\!\left(-cC_2^2|T|\log(eN/|T|)\right).
\]
Since $C_2\ge1$, the threshold just bounded is at most the defining threshold $C_2\bigl(\sqrt{|T|}+|T|\sqrt{\log(eN/|T|)/d}\bigr)$ for $\mathcal E_s$. For $m=|T|$, the number of signed choices $(T_+,T_-)$ with union size $m$ is at most $\binom Nm2^m\le (2eN/m)^m\le \exp(3m\log(eN/m))$. Therefore,
\[
    \Pr[\overline{\mathcal E_{\mathrm{s}}}]\le \sum_{m=1}^N\exp(-(cC_2^2-3)m\log(eN/m)).
\]
Let $a:=cC_2^2-3$. Since $m\log(eN/m)\ge \log(eN)$ for every $1\le m\le N$, if $a>1$ then
\[
    \sum_{m=1}^N\exp(-a m\log(eN/m))\le N\exp(-a\log(eN))\le N^{-(a-1)}=N^{-(cC_2^2-4)}.
\]
Taking $c_0=c/2$, the hypothesis $c_0C_2^2>3$ ensures $a>1$ and $cC_2^2-4\ge c_0C_2^2-3$, completing the proof.
\end{proof}

\begin{lemma}[Empirical mean of an isotropic Gaussian]\label{lemma:sample-mean-norm}
Let $x_1,\ldots,x_m \stackrel{i.i.d.}{\sim} \mathcal N(a,\sigma I_d)$, and let
\[
\bar x:=\frac1m\sum_{j=1}^m x_j.
\]
Then with probability at least $1-2e^{-d/32}$,
\[
\|\bar x-a\|^2=\Theta\!\left(\frac{d\sigma}{m}\right)
\]
\end{lemma}

\begin{proof}
Since sums of independent Gaussians are Gaussian,
\[
\bar x-a=\frac1m\sum_{j=1}^m (x_j-a)
\sim \mathcal N\!\left(0,\frac{\sigma}{m}I_d\right).
\]
Thus
\[
\frac{m}{\sigma}\|\bar x-a\|^2 \sim \chi_d^2.
\]
Now apply the standard $\chi^2$ concentration inequality: for $0<\delta<1$,
\[
\Pr\!\left(\left|\chi_d^2-d\right|\ge \delta d\right)\le 2\exp\!\left(-\frac{\delta^2 d}{8}\right).
\]
Taking $\delta=\frac12$, we obtain
\[
\Pr\!\left(\frac d2\le \chi_d^2\le \frac{3d}{2}\right)\ge 1-2e^{-d/32}.
\]
Multiplying by $\sigma/m$ gives
\[
\Pr\!\left(
\frac{d\sigma}{2m}\le \|\bar x-a\|^2\le \frac{3d\sigma}{2m}
\right)\ge 1-2e^{-d/32},
\]
as claimed.
\end{proof}

\subsection{Expected-cost lower bounds for isotropic Gaussians}
\label{subsec:quantization-lower-bound}

In this subsection we record the denominator comparisons used later. Recall that
\[
    V_\mu=\sum_{j\in[k]}\pi_j r_j^2=\sum_{j\in[k]}\pi_j\sigma_j d
\]
is the assigned cost to the true component means.

The exact factor $k^{-2/d}$ below implies~\eqref{eq:comparison_condition} whenever $d\ge\log k$.

\begin{lemma}[One-component quantization lower bound]
\label{lemma:one_component_quantization_lower_bound}
Let \(\nu=\mathcal N(a,\sigma I_d)\). For center set \(C\in(\mathbb R^d)^k\),
\[
    \mathbb E_{p\sim \nu} d^2(p,C)\ge \sigma d\cdot k^{-2/d}.
\]
\end{lemma}
 
\begin{proof}[Proof of Lemma~\ref{lemma:one_component_quantization_lower_bound}]
Let \(p\sim\nu\), and let \(q\) be a nearest center to \(p\) in \(C\), with an arbitrary fixed tie-breaking rule. Then \(q\) takes at most \(k\) values. Write
\[
    D:=\mathbb E\|p-q\|_2^2=\mathbb E_{p\sim\nu}d^2(p,C).
\]
Let \(h(\cdot)\) denote differential entropy. Since \(p\) is determined by the pair \((q,p-q)\),
\[
    h(p)\le H(q)+h(p-q\mid q).
\]
The discrete entropy satisfies \(H(q)\le \log k\). Also, by the maximum-entropy property of the Gaussian and Jensen's inequality,
\[
    h(p-q\mid q)\le \frac d2\log\left(\frac{2\pi eD}{d}\right).
\]
Therefore
\[
    h(p)\le \log k+\frac d2\log\left(\frac{2\pi eD}{d}\right).
\]
Rearranging gives
\[
    D\ge \frac{d}{2\pi e}\exp\left(\frac{2h(p)}{d}\right)k^{-2/d}.
\]
For \(p\sim\mathcal N(a,\sigma I_d)\),
\[
    h(p)=\frac d2\log(2\pi e\sigma).
\]
Substituting this into the previous display yields
\[
    D\ge \sigma d\cdot k^{-2/d}.
\]
This proves the lemma.
\end{proof}

\begin{corollary}[Mixture quantization lower bound]
\label{cor:mixture_quantization_lower_bound}
Let \(\mu=\sum_{j\in[k]}\pi_j\mathcal N(a_j,\sigma_j I_d)\). For every center set $C\in (\mathbb{R}^d)^k$,
\[
    \cost(\mu,C)\ge k^{-2/d}V_\mu=k^{-2/d}\sum_{j\in[k]}\pi_j r_j^2.
\]
In particular, if \(d\ge \log k\), then
\[
    \cost(\mu,C)\gtrsim V_\mu.
\]
\end{corollary}

\begin{proof}[Proof of Corollary~\ref{cor:mixture_quantization_lower_bound}]
Fix a center set \(C\in(\mathbb R^d)^k\). Applying Lemma~\ref{lemma:one_component_quantization_lower_bound} to each component \(\mu_j=\mathcal N(a_j,\sigma_j I_d)\), gives
\[
    \mathbb E_{p\sim\mu_j}d^2(p,C)\ge \sigma_jd\cdot k^{-2/d}=r_j^2 k^{-2/d}.
\]
Multiplying by \(\pi_j\) and summing over \(j\in[k]\), we obtain
\[
    \cost(\mu,C)=\sum_{j\in[k]}\pi_j\mathbb E_{p\sim\mu_j}d^2(p,C)\ge k^{-2/d}\sum_{j\in[k]}\pi_jr_j^2=k^{-2/d}V_\mu.
\]
For $d\ge\log k$ we have $k^{-2/d}=\exp(-2\log k/d)\ge e^{-2}$, giving the second assertion.
\end{proof}
 
\begin{corollary}[Expected-cost lower bound]\label{cor:mean_radius_lower_bound}
Let $\mu=\sum_{j\in[k]}\pi_j\mathcal N(a_j,\sigma_j I_d)$. For every integer $\ell\ge0$ and every $C\in(\mathbb R^d)^k$,
\[
    \cost(\mu,C)\gtrsim 2^{-2\ell}k^{-2/d}\sum_{j\in[k]}\pi_j\left(d^2(a_j,C)+2^{2\ell}r_j^2\right).
\]
If $d\ge\log k$, the factor $k^{-2/d}$ can be omitted. In particular, taking $\ell=0$ gives
\[
    \sum_{j\in[k]}\pi_jd^2(a_j,C)\lesssim\cost(\mu,C),\qquad V_\mu\lesssim\cost(\mu,C).
\]
\end{corollary}

\begin{proof}
Integrating $d^2(a_j,C)\le2d^2(p,C)+2d^2(p,a_j)$ over each component and summing gives
\[
    \sum_{j\in[k]}\pi_jd^2(a_j,C)\le2\cost(\mu,C)+2V_\mu.
\]
By Corollary~\ref{cor:mixture_quantization_lower_bound}, $k^{-2/d}V_\mu\le\cost(\mu,C)$. Multiplying the preceding display by $2^{-2\ell}k^{-2/d}\le1$ bounds its left-hand side by $4\cost(\mu,C)$. Adding $k^{-2/d}\sum_j\pi_jr_j^2\le\cost(\mu,C)$ proves the first assertion. For $d\ge\log k$ use $k^{-2/d}\ge e^{-2}$.
\end{proof}

\section{Missing proofs in the main part}
We first prove Corollary~\ref{cor:finite-generated-dataset}, then supply the missing proofs from Sections~\ref{sec:proof_main_theorem} and~\ref{sec:chaining}. We use their definitions, the accuracy parameter $\varepsilon\in(0,1)$, and the same convention for $\lesssim$.

\subsection{From population approximation to finite-data coresets}\label{appendix:finite-generated-dataset}

\begin{proof}[Proof of Corollary~\ref{cor:finite-generated-dataset}]
The proof of Theorem~\ref{thm:main_k_gaussian}, applied at accuracy $\varepsilon/6$, gives a sample size $m$ of the stated order for which
\[
    \mathbb E_{S\sim\mu^m}\sup_C
    \frac{|\cost(S,C)-\cost(\mu,C)|}{\cost(\mu,C)}
    \le\varepsilon/60.
\]
Fix any $N\ge m$, draw $X\sim\mu^N$, and select $S$ as in the corollary. With weight $1/m$ per selected observation,
$\cost(S,C)=m^{-1}\sum_{p\in S}\cost(p,C)$.
The subsample has marginal law $\mu^m$, and, conditional on $X$,
\[
    \mathbb E_{S\mid X}\cost(S,C)
    =\frac1N\sum_{p\in X}\cost(p,C)
    =\cost(X,C).
\]
The triangle inequality and moving the supremum outside this conditional expectation therefore give
\[
    \mathbb E_X\sup_C
    \frac{|\cost(X,C)-\cost(\mu,C)|}{\cost(\mu,C)}
    \le
    \mathbb E_{X,S}\sup_C
    \frac{|\cost(S,C)-\cost(\mu,C)|}{\cost(\mu,C)}
    \le\varepsilon/60.
\]
For each of the two relative errors, Markov's inequality bounds the probability of exceeding $\varepsilon/3$ by $1/20$. A union bound therefore shows that both errors are at most $\varepsilon/3$ with probability at least $0.9$; independence of the two events is not required. On their intersection, for every $C$,
\[
    \frac{1-\varepsilon/3}{1+\varepsilon/3}
    \le
    \frac{\cost(S,C)}{\cost(X,C)}
    \le
    \frac{1+\varepsilon/3}{1-\varepsilon/3}.
\]
For $0<\varepsilon<1$, these bounds lie in $[1-\varepsilon,1+\varepsilon]$. If $N<m$, keeping the whole dataset with weight $1/N$ gives equality of costs.
\end{proof}

\subsection{Omitted proofs from Section~\ref{sec:proof_main_theorem}}
In this subsection, $X_j$ denotes the full original component sample; inner subsets are written explicitly as $X_j^{\mathrm{in}}$.
\subsubsection{Proof of Lemma~\ref{lem:radius-shell-prob}: radius shell probability}\label{appendix:radius-shell-prob}

\paragraph{Lemma~\ref{lem:radius-shell-prob} (restated).}
For every $\ell\ge2$,
\[
    \Pr[\mathscr E_\ell]\le\min\{1,N\exp(-d4^{\ell-1}/8)\}.
\]
For the first shell, $\Pr[\mathscr E_1]\le1$.

\begin{proof}
For $p\sim\mu_j$ and $t\ge2$, the moment generating function of the squared normalized Gaussian radius gives
\[
    \Pr[d(p,a_j)>tr_j]
    \le e^{-dt^2/4}2^{d/2}\le e^{-dt^2/8}.
\]
Since $\mathscr E_\ell\subseteq\overline{\mathcal E_{\ell-1}}$, the union bound over the $N$ original sample coordinates proves the claim.
\end{proof}

\subsubsection{Proof of Lemma~\ref{lemma:far_clusters}: contribution of far clusters}\label{appendix:far_clusters}

\paragraph{Lemma~\ref{lemma:far_clusters} (restated).}
If $N\pi_{\min}\gtrsim\varepsilon^{-2}\log(2k)$, then, for every $\ell\ge1$,
\[
    \mathbb E_{X\mid\mathscr E_\ell}\mathbb E_g\sup_C
    \left|\sum_{p\in X_{J_{\mathrm{far}}(C)}}
    \frac{g_p d^2(p,C)}{N\cost(\mu,C)}\right|\lesssim\varepsilon.
\]

\begin{proof}
For $j\in J_{\mathrm{far}}(C)$, the $L_2$ triangle inequality gives
\[
    (\mathbb E_{p\sim\mu_j}d^2(p,C))^{1/2}
    \ge d(a_j,C)-r_j\ge(1-\varepsilon/2)d(a_j,C).
\]
Thus $\sum_{j\in J_{\mathrm{far}}(C)}\pi_jd^2(a_j,C)\lesssim\cost(\mu,C)$. On the shell, $d(p,a_j)\le\varepsilon d(a_j,C)$ and
\[
    |d^2(p,C)-d^2(a_j,C)|\le3\varepsilon d^2(a_j,C).
\]
Group the component-mean terms by component and bound the residual using absolute signs. This yields
\[
    \sup_C\left|\sum_{p\in X_{J_{\mathrm{far}}(C)}}
    \frac{g_p d^2(p,C)}{N\cost(\mu,C)}\right|
    \lesssim\max_j\frac{|\sum_{p\in X_j}g_p|}{N\pi_j}
    +\varepsilon\max_j\frac{\sum_{p\in X_j}|g_p|}{N\pi_j}.
\]
The shell depends only on normalized radii and is independent of the latent labels and signs. Bernstein's inequality and integration of the union bound give
\begin{align*}
    \mathbb E\max_j\frac{|\sum_{p\in X_j}g_p|}{N\pi_j}
    &\lesssim\sqrt{\frac{\log(2k)}{N\pi_{\min}}}
    +\frac{\log(2k)}{N\pi_{\min}},\\
    \mathbb E\max_j\frac{\sum_{p\in X_j}|g_p|}{N\pi_j}
    &\lesssim1+\sqrt{\frac{\log(2k)}{N\pi_{\min}}}
    +\frac{\log(2k)}{N\pi_{\min}}.
\end{align*}
For completeness, each summand is a Bernoulli variable of mean $\pi_j$ times $g_p$ or $|g_p|$. After centering, its moment generating function is at most $\exp(O(\pi_j t^2))$ for $|t|$ below a universal constant. After normalization by $N\pi_j$, its tail at $t>0$ is at most $2\exp(-\Omega(N\pi_j\min\{t^2,t\}))$. Integration gives the displayed maxima. Under the sample-size assumption, their nonconstant terms are $\lesssim\varepsilon$.
\end{proof}

\subsubsection{Proof of Lemma~\ref{lemma:tiny_clusters}: contribution of tiny close clusters}\label{appendix:tiny_clusters}

\paragraph{Lemma~\ref{lemma:tiny_clusters} (restated).}
For every deterministic sample in $\mathscr E_\ell$, under the comparison condition,
\[
    \mathbb E_g\sup_C\left|\sum_{p\in X_{J_{\mathrm{tiny}}(C)}}
    \frac{g_p d^2(p,C)}{N\cost(\mu,C)}\right|
    \lesssim2^{2\ell}\varepsilon/k^2.
\]

\begin{proof}
For $j\in J_{\mathrm{tiny}}(C)$ and $p\in X_j$, recall that $\zeta_{\mu}= \max_{i\in[k]}r_i^{2}/V_\mu$. Closeness and the tiny threshold give
\[
    d^2(p,C)\le4\varepsilon^{-2}2^{2\ell}r_j^2
    \le 4\varepsilon^{-2}2^{2\ell}(k^{-1}\varepsilon^{3/2}\zeta_\mu^{-1/2}\max_{i\in[k]}r_i)^{2}
    \le4k^{-2}\varepsilon2^{2\ell}V_\mu.
\]
Divide by $N\cost(\mu,C)$, use $V_\mu/\cost(\mu,C)\lesssim1$, and take $\mathbb E_g\sum_p|g_p|\le N$.
\end{proof}

\subsubsection{Proof of Lemma~\ref{lemma:prob-good-event}: probability of the good event}\label{appendix:prob-good-event}

\paragraph{Lemma~\ref{lemma:prob-good-event} (restated).}
If $N\ge2$ and $N\pi_{\min}\gtrsim\log(kN)$ with a large enough implicit constant, then $C_2$ can be chosen universally so that
\[
    \Pr[\overline{\mathcal E}]\le N^{-2}.
\]
These hypotheses hold for the prescribed sample size.

\begin{proof}
Chernoff's inequality gives $\Pr[\overline{\mathcal E_c}]\le2k\exp(-N\pi_{\min}/12)\le(2N^2)^{-1}$. For $d=1$, the norm of a signed sum of unit directions is at most its number of terms, so $\mathcal E_s$ holds deterministically for $C_2\ge1$. For $d\ge2$, the product-sphere concentration used in Lemma~\ref{lem:signed-subset-sum-good-event} gives, for each fixed signed subset of size $m$, failure probability at most $\exp(-\Omega(C_2^2m\log(eN/m)))$. There are at most $(2eN/m)^m$ such subsets. Taking $C_2$ sufficiently large therefore gives
\[
    \Pr[\overline{\mathcal E_s}]
    \le\sum_{m=1}^N\exp(-4m\log(eN/m))
    \le N^{-3}\le(2N^2)^{-1}.
\]
The union bound completes the proof.
\end{proof}

\subsubsection{Proof of Lemma~\ref{lemma:worst_case_bound}: worst-case bound on a shell}\label{appendix:worst_case_bound}

\paragraph{Lemma~\ref{lemma:worst_case_bound} (restated).}
Under the comparison condition, every deterministic sample in $\mathscr E_\ell$ satisfies
\[
    \mathbb E_g\sup_C\left|\sum_{p\in X_{J_{\mathrm{huge}}(C)}}
    \frac{g_p d^2(p,C)}{N\cost(\mu,C)}\right|
    \lesssim\zeta_\mu2^{2\ell}\varepsilon^{-2}.
\]

\begin{proof}
On huge close components, $d^2(p,C)\le4\varepsilon^{-2}2^{2\ell}\max_jr_j^2$. Use
\[
    \frac{\max_jr_j^2}{\cost(\mu,C)}
    =\zeta_\mu\frac{V_\mu}{\cost(\mu,C)}
    \lesssim\zeta_\mu
\]
and $\mathbb E_g\sum_p|g_p|\le N$.
\end{proof}

\subsubsection{Proof of Lemma~\ref{lemma:inner_truncation}: contribution of inner points}\label{appendix:inner-truncation}

Use the definition of inner and outer points in Section~\ref{sec:proof_main_theorem}; here $X_j$ denotes the full original component sample, and $X_j^{\mathrm{in}}$ its inner subset.

\paragraph{Lemma~\ref{lemma:inner_truncation} (restated).}
Under the standing assumptions, for every $\ell\ge1$,
\[
    \mathbb E_{X\mid\mathscr E_\ell}\mathbb E_g\sup_C\left|
    \sum_{j\in J_{\mathrm{huge}}(C)}
    \sum_{p\in X_j^{\mathrm{in}}}
    \frac{g_p d^2(p,C)}{N\cost(\mu,C)}\right|
    \lesssim\varepsilon.
\]

\begin{proof}
For each inner point, write $d^2(p,C)=d^2(a_j,C)+(d^2(p,C)-d^2(a_j,C))$. We bound the component-mean contribution and the residual separately. The fixed cutoff gives
\[
    |d^2(p,C)-d^2(a_j,C)|
    \le2d(a_j,C)d(p,a_j)+d^2(p,a_j)
    \le\varepsilon^2\bigl(d^2(a_j,C)+2r_j^2\bigr).
\]
Taking absolute values, enlarging the nonnegative sums from inner points to all original component points, and using the weighted triangle inequality and~\eqref{eq:comparison_condition} therefore gives
\begin{align*}
    &\sup_C\left|
    \sum_{j\in J_{\mathrm{huge}}(C)}\sum_{p\in X_j^{\mathrm{in}}}
    \frac{g_p\bigl(d^2(p,C)-d^2(a_j,C)\bigr)}{N\cost(\mu,C)}\right|\\
    &\qquad\lesssim\varepsilon^2\max_j\frac{\sum_{p\in X_j}|g_p|}{N\pi_j}.
\end{align*}
For the component-mean contribution, take absolute values componentwise. The weighted triangle inequality and \eqref{eq:comparison_condition} give
\[
    \sum_{j\in[k]}\pi_jd^2(a_j,C)\le2\cost(\mu,C)+2V_\mu\lesssim\cost(\mu,C).
\]
Therefore,
\[
    \sup_C\left|\sum_{j\in J_{\mathrm{huge}}(C)}
    \sum_{p\in X_j^{\mathrm{in}}}
    \frac{g_p d^2(a_j,C)}{N\cost(\mu,C)}\right|
    \lesssim\max_j\frac{
    \left|\sum_{p\in X_j^{\mathrm{in}}}g_p\right|}{N\pi_j}.
\]
For each fixed sample, the inner and outer Gaussian sign sums are independent and the outer sums are centered. The maximum of the normalized absolute component sums is convex, so conditional Jensen's inequality over the outer signs gives
\[
    \mathbb E_g\max_j\frac{\left|\sum_{p\in X_j^{\mathrm{in}}}g_p\right|}{N\pi_j}
    \le\mathbb E_g\max_j\frac{\left|\sum_{p\in X_j}g_p\right|}{N\pi_j}.
\]
The shell depends only on normalized radii and is independent of the latent labels and Gaussian signs. Thus the full-component maxima have the same expectations after conditioning on the shell. The bounds proved in Appendix~\ref{appendix:far_clusters} give
\begin{align*}
    \mathbb E_{X\mid\mathscr E_\ell}\mathbb E_g\max_j\frac{\left|\sum_{p\in X_j}g_p\right|}{N\pi_j}
    &\lesssim\sqrt{\frac{\log(2k)}{N\pi_{\min}}}+\frac{\log(2k)}{N\pi_{\min}},\\
    \mathbb E_{X\mid\mathscr E_\ell}\mathbb E_g\max_j\frac{\sum_{p\in X_j}|g_p|}{N\pi_j}
    &\lesssim1+\sqrt{\frac{\log(2k)}{N\pi_{\min}}}+\frac{\log(2k)}{N\pi_{\min}}
    \lesssim1.
\end{align*}
Here the last inequality uses the prescribed sample size, which ensures $N\pi_{\min}\gtrsim\varepsilon^{-2}\log(2k)$.

Combining the component-mean and residual bounds with the triangle inequality gives
\[
    \mathbb E_{X\mid\mathscr E_\ell}\mathbb E_g\sup_C\left|
    \sum_{j\in J_{\mathrm{huge}}(C)}
    \sum_{p\in X_j^{\mathrm{in}}}
    \frac{g_p d^2(p,C)}{N\cost(\mu,C)}\right|
    \lesssim\sqrt{\frac{\log(2k)}{N\pi_{\min}}}+\frac{\log(2k)}{N\pi_{\min}}+\varepsilon^2
    \lesssim\varepsilon.
\]
\end{proof}

\subsection{Omitted proofs from Section~\ref{sec:chaining}}\label{sec:variance-bound}

As in Section~\ref{sec:chaining}, fix $\ell\ge1$ and the original sample $X\in\mathcal E\cap\mathscr E_\ell$, under the standing comparison condition. Every $p\in X_j^{\mathrm{out}}$ satisfies $\varepsilon^2r_j<d(p,a_j)\le2^\ell r_j$; the lower bound comes from the deterministic split, not from the shell event. The mask $\mathsf M(C)=\bigcup_{j\in J_{\mathrm{huge}}(C)}X_j^{\mathrm{out}}$ consists only of outer points in huge close components. All representatives are zero elsewhere on the original $N$ coordinates. The outer subsets satisfy $|X_j^{\mathrm{out}}|\le|X_j|\le3N\pi_j/2$; no lower bound on their sizes is used.

\subsubsection{Proof of Lemma~\ref{lemma:error-sum-high-dimensional}: error-sum bound}

\begin{proof}
On $\mathscr E_\ell$, every retained point satisfies $d(p,a_j)\le2^\ell r_j$. The triangle inequality and the inherited count upper bounds give
\begin{align*}
    \sum_{j\in J_{\mathrm{huge}}(C)}\sum_{p\in X_j^{\mathrm{out}}}\err(p,C)^2
    &\lesssim\sum_{j\in J_{\mathrm{huge}}(C)}\sum_{p\in X_j^{\mathrm{out}}}d^2(p,a_j)\bigl(d^2(p,C)+d^2(p,a_j)\bigr)\\
    &\lesssim\sum_{j\in J_{\mathrm{huge}}(C)}|X_j^{\mathrm{out}}|2^{2\ell}r_j^2\bigl(d^2(a_j,C)+2^{2\ell}r_j^2\bigr)\\
    &\lesssim N2^{2\ell}\max_{i\in[k]}r_i^2\sum_{j\in[k]}\pi_j\bigl(d^2(a_j,C)+2^{2\ell}r_j^2\bigr).
\end{align*}
The weighted triangle inequality gives
\[
    \sum_{j\in[k]}\pi_jd^2(a_j,C)\le2\cost(\mu,C)+2V_\mu.
\]
Using $\max_{i\in[k]}r_i^2=\zeta_\mu V_\mu$ and $\sum_j\pi_jr_j^2=V_\mu$, we obtain
\begin{align*}
    \sum_{j\in J_{\mathrm{huge}}(C)}\sum_{p\in X_j^{\mathrm{out}}}\frac{\err(p,C)^2}{N^2\cost(\mu,C)^2}
    &\lesssim\frac{\zeta_\mu}{N}\left(2^{2\ell}\frac{V_\mu}{\cost(\mu,C)}+2^{4\ell}\frac{V_\mu^2}{\cost(\mu,C)^2}\right)\\
    &\lesssim\frac{2^{4\ell}\zeta_\mu}{N}.
\end{align*}
The last inequality is exactly where \eqref{eq:comparison_condition} is used. All sums on the left concern the retained outer coordinates, and every denominator uses the original $N$.
\end{proof}

\subsubsection{Proof of Lemma~\ref{lemma:variance_bound}: variance bound}

\begin{proof}[Proof of Lemma~\ref{lemma:variance_bound}]
For $0\le h\le h_{\max}$, let $\alpha_h=2^{-h}$. On outer points of huge close components,
\[
    v_p^{C,2^{-(h+1)}}-v_p^{C,2^{-h}}=(v_p^{C,2^{-(h+1)}}-v_p^C)+(v_p^C-v_p^{C,2^{-h}}),
\]
so the common local-error convention gives
\[
    |v_p^{C,2^{-(h+1)}}-v_p^{C,2^{-h}}|\lesssim \alpha_h\err(p,C).
\]
For $h=0$ use the constant-accuracy coarse cover; for $h\ge1$ use the fine covers at both scales and $\alpha_{h+1}=\alpha_h/2$. Both representatives vanish outside the active outer coordinates, including when the construction changes between scales.
The representatives are chosen independently of the Gaussian signs. Independence of the $g_p$ therefore gives
\begin{align*}
    \operatorname{Var}_g(Y_{C,h})
    &=\sum_{p\in X}\frac{(v_p^{C,2^{-(h+1)}}-v_p^{C,2^{-h}})^2}{N^2\cost(\mu,C)^2}\\
    &\lesssim\alpha_h^2\sum_{j\in J_{\mathrm{huge}}(C)}\sum_{p\in X_j^{\mathrm{out}}}\frac{\err(p,C)^2}{N^2\cost(\mu,C)^2}\\
    &\lesssim\frac{2^{4\ell}\zeta_\mu\alpha_h^2}{N}.
\end{align*}
The last inequality is~\eqref{eq:error_sum}. The estimate holds uniformly over $C$, including the first increment, and proves the variance bound.
\end{proof}

\subsection{Omitted proofs from Section~\ref{sec:general-mixtures}}\label{appendix:general-mixtures}

\subsubsection{Common reductions}\label{appendix:general-common}

\begin{proof}[Proof of Lemma~\ref{lem:general-shells}]
For $\ell\ge2$, the event $\mathscr E_\ell$ implies that some sample point violates the radius bound at level $\ell-1$. Applying Assumption~\ref{ass:general-tail} conditional on each point's latent label and taking a union bound gives
\[
    \Pr[\mathscr E_\ell]\le\min\{1,2N\exp(-\Omega(2^{q(\ell-1)}))\}.
\]
Components with zero radius never violate the bound. Splitting the shell sums at the first index for which $2^{q(\ell-1)}$ exceeds a large enough constant multiple of $\log(eN)$ gives
\begin{align*}
    \sum_{\ell\ge1}\Pr[\mathscr E_\ell]2^{4\ell}
    &\lesssim[\log(eN)]^{4/q},\\
    \sum_{\ell\ge1}\Pr[\mathscr E_\ell]2^{2\ell}(1+\ell)^{3/2}
    &\lesssim[\log(eN)]^{2/q}[1+\log\log(eN)]^{3/2}.
\end{align*}
The initial parts are bounded by the largest shell factors in their ranges, since the shell probabilities sum to at most one. After shifting the index, each remaining tail is bounded by the respective right-hand side times the convergent series
\[
    \sum_{i\ge0}2^{4i}(1+i)^{3/2}\exp(-\Omega(2^{qi})).
\]
Since the shells partition the sample space, Cauchy--Schwarz also gives
\[
    \sum_{\ell\ge1}\Pr[\mathscr E_\ell]2^{2\ell}
    \le\left(\sum_{\ell\ge1}\Pr[\mathscr E_\ell]2^{4\ell}\right)^{1/2}
    \lesssim[\log(eN)]^{2/q}.
\]

Chernoff's inequality for the latent component counts gives $\Pr[\overline{\mathcal E_c}]\le2k\exp(-N\pi_{\min}/12)$. The triangle inequality and the comparison condition imply
\[
    \sum_j\pi_jd^2(a_j,C)\le2\cost(\mu,C)+2V_\mu\lesssim\cost(\mu,C).
\]
Consequently, on $\mathscr E_\ell$, every sample coordinate satisfies
\[
    \frac{d^2(p,C)}{\cost(\mu,C)}
    \lesssim2^{2\ell}\zeta_\mu+\pi_{\min}^{-1}.
\]
Taking absolute values in $Y_C$ gives the same upper bound for $\mathbb E_g\sup_C|Y_C|$. Cauchy--Schwarz and the fourth-moment bound above now give
\begin{align*}
    \mathbb E_X\!\left[\mathbf 1_{\overline{\mathcal E_c}}\mathbb E_g\sup_C|Y_C|\right]
    &\lesssim\left(\zeta_\mu
    \Bigl(\sum_{\ell\ge1}\Pr[\mathscr E_\ell]2^{4\ell}\Bigr)^{1/2}
    +\pi_{\min}^{-1}\right)\Pr[\overline{\mathcal E_c}]^{1/2}\\
    &\lesssim\left(\zeta_\mu[\log(eN)]^{2/q}+\pi_{\min}^{-1}\right)
    \sqrt{2k}\exp(-N\pi_{\min}/24).
\end{align*}
\end{proof}

\begin{proof}[Proof of Lemma~\ref{lem:general-reductions}]
Fix $X\in\mathcal E_c\cap\mathscr E_\ell$. The count upper bounds and the Gaussian maximal inequality give
\[
    \mathbb E_g\max_j\frac{|\sum_{p\in X_j}g_p|}{N\pi_j}
    \lesssim\sqrt{\frac{\log(2k)}{N\pi_{\min}}}.
\]
Each function $\sum_{p\in X_j}|g_p|/(N\pi_j)$ has bounded mean and Gaussian Lipschitz constant at most $\sqrt{3/(2N\pi_j)}$. Integrating the union bound for their Gaussian concentration tails yields
\[
    \mathbb E_g\max_j\frac{\sum_{p\in X_j}|g_p|}{N\pi_j}
    \lesssim1+\sqrt{\frac{\log(2k)}{N\pi_{\min}}}\lesssim1.
\]
For a far component, the shell bound gives $d(p,a_j)\le\varepsilon d(a_j,C)$, and hence
\[
    |d^2(p,C)-d^2(a_j,C)|\le3\varepsilon d^2(a_j,C).
\]
Using $\sum_j\pi_jd^2(a_j,C)\lesssim\cost(\mu,C)$, separate the component-mean cost and its residual to obtain
\begin{align*}
    \mathbb E_g\sup_C\left|\sum_{p\in X_{J_{\mathrm{far}}(C)}}
    \frac{g_p d^2(p,C)}{N\cost(\mu,C)}\right|
    &\lesssim\mathbb E_g\max_j\frac{|\sum_{p\in X_j}g_p|}{N\pi_j}
    +\varepsilon\mathbb E_g\max_j\frac{\sum_{p\in X_j}|g_p|}{N\pi_j}\\
    &\lesssim\sqrt{\frac{\log(2k)}{N\pi_{\min}}}+\varepsilon.
\end{align*}
The deterministic estimate of Lemma~\ref{lemma:tiny_clusters} bounds the tiny close components by $O(2^{2\ell}\varepsilon/k^2)$.

For the inner points, apply the argument of Appendix~\ref{appendix:inner-truncation} using the fixed cutoff, the component-count upper bounds, and the comparison condition. It gives
\[
    \mathbb E_g\sup_C\left|
    \sum_{j\in J_{\mathrm{huge}}(C)}\sum_{p\in X_j^{\mathrm{in}}}
    \frac{g_p d^2(p,C)}{N\cost(\mu,C)}\right|
    \lesssim\sqrt{\frac{\log(2k)}{N\pi_{\min}}}+\varepsilon^2.
\]
The remaining coordinates are exactly the outer points of huge close components. Combining these bounds proves the lemma.
\end{proof}

\begin{proof}[Proof of Lemma~\ref{lem:general-endpoints}]\label{appendix:general-endpoints}
The coarse representative in~\eqref{eq:initial-representative} is constant on each $X_j^{\mathrm{out}}$, with value between zero and $d^2(a_j,C)$. Thus the mean-cost comparison and the Gaussian maximal inequality give
\[
    \mathbb E_g\sup_C|Y_{C,\mathrm{Init}}|
    \lesssim\mathbb E_g\max_j\frac{|\sum_{p\in X_j^{\mathrm{out}}}g_p|}{N\pi_j}
    \lesssim\sqrt{\frac{\log(2k)}{N\pi_{\min}}}.
\]
For the end term, Cauchy--Schwarz on the original $N$ coordinates gives
\begin{equation}\label{eq:general-terminal-bound}
    \mathbb E_g\sup_C|Y_{C,\mathrm{End}}|
    \lesssim\sqrt N\,2^{-(h_{\max}+1)}
    \left(\sup_C\sum_{j\in J_{\mathrm{huge}}(C)}\sum_{p\in X_j^{\mathrm{out}}}
    \frac{\err(p,C)^2}{N^2\cost(\mu,C)^2}\right)^{1/2}.
\end{equation}
Lemmas~\ref{lem:general-relative-variance} and~\ref{lem:general-distance-variance} bound this, for either branch, by
\[
    O\!\left(2^{2\ell}2^{-(h_{\max}+1)}\varepsilon^{-1}
    \sqrt{\zeta_\mu+(k\pi_{\min})^{-1}}\right)
    \lesssim2^{2\ell}\varepsilon.
\]
\end{proof}

\subsubsection{Relative-error estimates}\label{appendix:general-relative}

\begin{proof}[Proof of Lemma~\ref{lem:general-relative-variance}]
Closeness and the shell bound imply, on every active coordinate,
\[
    d^2(p,C)+d^2(p,a_j)
    \lesssim\varepsilon^{-2}2^{2\ell}\max_i r_i^2.
\]
Multiply~\eqref{eq:general-cost-sum} by this upper bound and divide by $N^2\cost(\mu,C)^2$. Since $\max_i r_i^2=\zeta_\mu V_\mu$, the comparison condition gives
\[
    \sum_{j\in J_{\mathrm{huge}}(C)}\sum_{p\in X_j^{\mathrm{out}}}
    \frac{\err(p,C)^2}{N^2\cost(\mu,C)^2}
    \lesssim\frac{2^{4\ell}\varepsilon^{-2}\zeta_\mu}{N}.
\]
Equation~\eqref{eq:general-increment-variance} proves the variance claim.
\end{proof}

\begin{proof}[Proof of Lemma~\ref{lem:general-relative-cover}]
Apply a multiplicative terminal embedding~\cite{narayanan2019optimal} of distortion $1+c\alpha$ to $X\cup A$, into $O(\alpha^{-2}\log(N+k))$ dimensions. Write $f$ for this fixed map. For every query $C$,
\[
    |d^2(f(p),f(C))-d^2(p,C)|\lesssim c\alpha d^2(p,C).
\]
For each candidate active subset $J\subseteq\{j:r_j>0\}$, apply Lemma~\ref{lemma:general_alpha_net_size} to the embedded outer points, using component balls of radius $2^{\ell+1}r_j$, center-distance factor $\varepsilon^{-1}$, and additive cost tolerance $c\alpha\varepsilon^4r_j^2$. These choices are valid because
\[
    d(f(p),f(a_j))\le2^{\ell+1}r_j,\qquad
    d(f(a_j),f(C))\le\varepsilon^{-1}2^{\ell+1}r_j
\]
for active components. The componentwise ratios in the first entropy bound of that lemma are all $O(2^{2\ell}\alpha^{-1}\varepsilon^{-6})$. Taking the union of the zero-padded covers over the at most $2^k$ subsets therefore gives
\[
    \log|\mathcal V(\alpha)|
    \lesssim k\log k+k\alpha^{-2}\log(N+k)
    \log\!\left(1+2^{2\ell}\alpha^{-1}\varepsilon^{-6}\right).
\]
Empty active subsets use the zero vector. For every active outer coordinate, the cutoff gives $\varepsilon^4r_j^2<d^2(p,a_j)$. Combining the embedding and discretization errors, and choosing the absolute constant $c$ sufficiently small, yields
\[
    |d^2(p,C)-v_p^{C,\alpha}|
    \le\alpha\bigl(d^2(p,C)+d^2(p,a_j)\bigr).
\]
This proves the covering property and the stated entropy bound.
\end{proof}

\subsubsection{Distance-sensitive estimates}\label{appendix:general-distance}

\begin{proof}[Proof of Lemma~\ref{lem:general-distance-variance}]
For the part of~\eqref{eq:general-distance-error} without the enlargement, use
\[
    \bigl(d(p,C)d(p,a_j)+d^2(p,a_j)\bigr)^2
    \le2d^2(p,a_j)\bigl(d^2(p,C)+d^2(p,a_j)\bigr).
\]
The shell bound and~\eqref{eq:general-cost-sum} show that its normalized square sum is at most
\[
    O\!\left(\frac{2^{4\ell}\max_jr_j^2}{N\cost(\mu,C)}\right)
    \lesssim\frac{2^{4\ell}\zeta_\mu}{N}.
\]
For the additional part, $d(p,a_j)\le R$ on $G(m)$ gives
\[
    \frac{\bigl(d(p,C)d(p,a_j)+d^2(p,a_j)\bigr)^2}{R^2}
    \le2\bigl(d^2(p,C)+d^2(p,a_j)\bigr).
\]
After multiplying by $\cost(\mu,C)/(k\pi_{\min})$ and normalizing, sum over the disjoint radius groups. By~\eqref{eq:general-cost-sum}, the contribution is at most
\[
    \frac{2}{N^2k\pi_{\min}\cost(\mu,C)}
    \sum_{j\in J_{\mathrm{huge}}(C)}\sum_{p\in X_j^{\mathrm{out}}}
    \bigl(d^2(p,C)+d^2(p,a_j)\bigr)
    \lesssim\frac{2^{2\ell}}{Nk\pi_{\min}}.
\]
Adding the two contributions proves the square-sum bound;~\eqref{eq:general-increment-variance} gives the variance bound.
\end{proof}

The following proof of Lemma~\ref{lem:general-distance-cover} adapts the two constructions of~\cite[Lemma~B.11]{DBLP:conf/stoc/Huang0024}. On each fixed distance block, both use the error~\eqref{eq:general-distance-error}, with logarithmic sizes proportional to $k\alpha^{-2}t^2$ and $k\alpha^{-2}(1+k/t^2)$. Taking the smaller gives $\min\{t^2,1+k/t^2\}\le2\sqrt{k}$ for every block.

\begin{proof}[Proof of Lemma~\ref{lem:general-distance-cover}]
Fix a nonempty block $X_{m,\phi,\beta}$, and let $J=\{j:\phi_j=\beta\}$, $R=2^{-m}2^\ell\max_i r_i$, and $t=2^\beta$. All constructions below are fixed for this block and cover every $C\in\mathcal C_{m,\phi}$. Lemma~\ref{prop:scale_properties} gives, for $p\in X_j^{\mathrm{out}}\cap X_{m,\phi,\beta}$,
\[
    \frac{tR^2}{4}\le d(p,C)d(p,a_j)+d^2(p,a_j)\le3tR^2,
\]
and the triangle inequality gives
\[
    |d^2(p,C)-d^2(a_j,C)|\le3tR^2.
\]
We construct two covers of these centered differences. In each construction, restore the original costs by adding
\[
    c\alpha R^2\left\lfloor\frac{d^2(a_j,C)}{c\alpha R^2}\right\rfloor
\]
on component $j$. Since $d^2(a_j,C)\le t^2R^2$, these mean-cost choices add $O(k\log(et/\alpha))$ to the logarithmic size and at most $c\alpha R^2$ to the error.

For the first cover, apply a multiplicative terminal embedding~\cite{narayanan2019optimal} to the block points and the means $(a_j)_{j\in J}$, with distortion $1+c\alpha/t$. Its dimension is
\[
    O\!\left(t^2\alpha^{-2}\log(N+k)\right).
\]
Every point-to-query or mean-to-query distance under consideration is at most $(t+1)R$. Thus the squared-distance error at each terminal is $O(c\alpha tR^2)$. In the embedded space, the component balls have radius at most $2R$ and the nearest-query distances from their means are at most $2tR$. Apply Lemma~\ref{lemma:general_alpha_net_size} with these radii, center-distance factor $t$, and additive cost tolerance $c\alpha R^2$, including each mean as a terminal in its component. Subtract each mean coordinate from its component's point coordinates. The resulting centered-difference error, including the embedding error, is $O(c\alpha tR^2)$. After restoring the rounded means, this is at most $\alpha\err(p,C)$ for sufficiently small $c$. The logarithmic size is at most
\[
    O\!\left(kt^2\alpha^{-2}\log(N+k)\log(et/\alpha)\right).
\]
This construction also handles $t=1$.

For the second cover, suppose $t>1$. The population mean-cost comparison and the distance class give
\[
    \cost(\mu,C)\gtrsim\sum_{j\in J}\pi_jd^2(a_j,C)
    \gtrsim |J|\pi_{\min}t^2R^2.
\]
Hence the same error function satisfies
\[
    \err(p,C)\gtrsim tR^2\sqrt{1+|J|t^2/k}.
\]
If $\alpha\sqrt{1+|J|t^2/k}$ exceeds a sufficiently large constant, the zero vector approximates the centered differences within half the allowed error; only the rounded means need to be recorded.

Otherwise, for every $j\in J$, apply Lemma~\ref{lem:additive_terminal_embedding} to
\[
    \{0\}\cup\{p-a_j:p\in X_j^{\mathrm{out}}\cap X_{m,\phi,\beta}\}\subset B(0,R)
\]
with accuracy $c\alpha\sqrt{1+|J|t^2/k}$. Choose $c$ so this accuracy is below $1/4$, and write $f_j(x)$ for the image of $x-a_j$. The dimension is
\[
    O\!\left(\frac{\alpha^{-2}}{1+|J|t^2/k}\log(N+k)\right).
\]
For every point terminal and the mean terminal,
\[
    |d(p,C)-d(f_j(p),f_j(C))|
    \le c\alpha R\sqrt{1+|J|t^2/k}.
\]
Squaring and subtracting the mean cost therefore introduces error
\[
    O\!\left(c\alpha tR^2\sqrt{1+|J|t^2/k}\right).
\]
The embedded terminals lie in $B(f_j(a_j),2R)$ and $d(f_j(a_j),f_j(C))\le2tR$. Apply the single-ball bound of Lemma~\ref{lemma:general_alpha_net_size} with center-distance factor $t$ and additive cost tolerance $c\alpha R^2$, again including the mean terminal. Subtract the discretized mean coordinate and restore the rounded original mean cost. The total error is at most $\alpha\err(p,C)$ for sufficiently small $c$.

Taking the product of these $|J|$ single-ball covers and including the mean-cost grids gives logarithmic size at most
\[
    O\!\left(
    \frac{k|J|\alpha^{-2}}{1+|J|t^2/k}\log(N+k)\log(et/\alpha)
    +k\log(et/\alpha)\right).
\]
Since $|J|/(1+|J|t^2/k)\le k/t^2$, this is bounded by
\[
    O\!\left(k\alpha^{-2}(1+k/t^2)\log(N+k)\log(et/\alpha)\right).
\]
The same bound includes the zero-centered-difference construction. Taking the smaller of the two covers therefore gives
\[
    \log|\mathcal V_{m,\phi,\beta}(\alpha)|
    \lesssim k\alpha^{-2}\min\{t^2,1+k/t^2\}
    \log(N+k)\log(et/\alpha).
\]
For an empty block, use the singleton zero vector.

For each group $G(m)$ and certificate $\phi\in\Phi_m$, take the product of its block covers and set inactive coordinates to zero. Take the union over certificates and then the product over the disjoint groups. This gives a cover of the original active costs, with zero entries elsewhere, and
\[
    \log|\mathcal V(\alpha)|
    \le\sum_{m=0}^{m_{\max}}\left(
    k\log(\beta_{\max}+2)
    +\max_{\phi\in\Phi_m}\sum_{\beta=0}^{\beta_{\max}}
    \log|\mathcal V_{m,\phi,\beta}(\alpha)|\right).
\]
Here $\beta_{\max}=m+\lceil\log_2(\varepsilon^{-1})\rceil$. For every $t\ge1$,
\[
    \min\{t^2,1+k/t^2\}\le2\sqrt{k}.
\]
The group count, distance-class count, and factor $\log(et/\alpha)$ each contribute at most one factor linear in $1+\ell$, up to logarithms in the remaining parameters. Consequently,
\[
    \log|\mathcal V(\alpha)|
    =\widetilde O\!\left(k^{3/2}\alpha^{-2}(1+\ell)^3\right),
\]
as required.
\end{proof}

\section{Necessity examples for assumptions}\label{sec:parameter-lower-bounds}

\subsection{Necessity of the comparison condition~\eqref{eq:comparison_condition}}\label{appendix:comparison-necessity}
\begin{lemma}[Necessity of the comparison condition]\label{lem:comparison-condition-necessary}
There is an absolute constant $c>0$ such that, for every sufficiently large $k$, the equal-weight mixture of $k$ identical components $\mathcal N(0,1)$ on $\mathbb R$ has $\pi_{\min}=1/k$ and $\zeta_\mu=1$, but an i.i.d.\ sample of any integer size
\[
    1\le N\le \frac{ck^2}{\log^2 k}
\]
fails to be a $1/2$-coreset with probability at least $3/4$. Moreover, its latent cost is $V_\mu=1$, while some $k$-center queries have population cost tending to zero as $k\to\infty$. Thus no fixed comparison constant is available for this family.
\end{lemma}

\begin{proof}
Set $\pi_j=1/k$ and $\mu_j=\mathcal N(0,1)$ for every $j\in[k]$. Then $a_j=0$ and $r_j^2=1$, giving the stated parameters. If $N\le k$, a query containing all sample points, padded by repeated centers if necessary, has empirical cost zero and positive population cost. Hence assume $N>k$.

Let $g\sim\mathcal N(0,1)$, and choose $b>1$ so that $\Pr[|g|>b]=1/(4N)$. Use the fixed grid
\[
    C=\left\{-b+\frac{2b(i-1)}{k-1}:i\in[k]\right\}.
\]
For $c\le1$, we have $N\le k^2$, and the Gaussian tail bound gives
\[
    b^2\le2\log(8N)\le6\log k.
\]
With probability at least $1-N\Pr[|g|>b]=3/4$, every sample lies in $[-b,b]$. On this event,
\[
    \cost(X,C)\le\frac{b^2}{(k-1)^2}.
\]
Outside $[-b,b]$, the nearest center is an endpoint. Since $b\ge1$,
\[
\begin{aligned}
    \cost(\mu,C)
    &\ge\frac{2}{\sqrt{2\pi}}\int_{b+1/b}^{b+2/b}(t-b)^2e^{-t^2/2}\,\mathrm dt\\
    &\ge\frac{2e^{-4}e^{-b^2/2}}{\sqrt{2\pi}\,b^3}
    \ge\frac{e^{-4}}{4Nb^2}.
\end{aligned}
\]
The last inequality uses $1/(8N)=\Pr[g>b]\le e^{-b^2/2}/(\sqrt{2\pi}\,b)$. Consequently, on the same event,
\[
    \frac{\cost(X,C)}{\cost(\mu,C)}
    \lesssim\frac{Nb^4}{(k-1)^2}\lesssim c.
\]
Choosing the absolute constant $c$ sufficiently small makes the ratio at most $1/4$, proving coreset failure.

Finally, take $N=\lfloor ck^2/\log^2 k\rfloor$. Then $b\to\infty$, and the same grid satisfies
\[
    \cost(\mu,C)
    \le\frac{b^2}{(k-1)^2}
       +\mathbb E\!\left[g^2\mathbf1_{\{|g|>b\}}\right]
    \longrightarrow0,
\]
whereas $V_\mu=1$. The components remain isotropic Gaussians with balanced weights and identical radii, so the example isolates the need for the comparison condition.
\end{proof}

\subsection{Necessity of a tail assumption}\label{appendix:tail-necessity}
We next show the necessity of quantitative tail control. The two Gaussian families that follow separately match the $\pi_{\min}^{-1}\varepsilon^{-2}$ and $k\zeta_\mu\varepsilon^{-2}$ terms, up to logarithmic factors. Every observation has weight $1/N$. Queries with fewer than $k$ distinct centers are padded by repeated centers, as permitted by $C\in(\mathbb R^d)^k$. Throughout, $V_\mu=\sum_{j\in[k]}\pi_jr_j^2$ denotes the latent-component cost.

\begin{theorem}[Necessity of tail control]\label{thm:tail-lower-bound}
For every $d\ge1$, $0<\pi\le1/2$, and $\varepsilon\in(0,1)$, there exists a mean-zero distribution $\mu$ supported in the unit ball, with covariance $(\pi/d)I_d$, for which an i.i.d.\ sample of any integer size $1\le N\le(2\pi)^{-1}$ fails to be an $\varepsilon$-coreset for $k=1$ with probability at least $1/2$. Moreover, $\pi_{\min}=\zeta_\mu=1$ and $V_\mu\le\cost(\mu,C)$ for every one-center query $C$.
\end{theorem}

\begin{proof}
Let $\mu$ put mass $1-\pi$ at the origin and mass $\pi$ uniformly on the unit sphere in $\mathbb R^d$. By symmetry, its mean is zero and its covariance is $(\pi/d)I_d$. Regard $\mu$ as a single component, so $A=\{0\}$ and $\pi_{\min}=\zeta_\mu=1$. For every $c\in\mathbb R^d$,
\[
    \cost(\mu,\{c\})=\pi+\|c\|^2\ge\pi=V_\mu.
\]
For the fixed query $C=\{0\}$, the population cost is $\pi>0$, whereas the empirical cost is zero whenever every sample is the origin. This event has probability
\[
    (1-\pi)^N\ge1-N\pi\ge\frac12,
\]
and gives relative error one. Since $\pi$ can be arbitrarily small, no finite sample bound depending only on $k,d,\varepsilon$ is possible. The obstruction concerns tails relative to the second moment: although the support is bounded and all moments are finite,
\[
    \frac{\mathbb E_{p\sim\mu}\|p\|^4}{\bigl(\mathbb E_{p\sim\mu}\|p\|^2\bigr)^2}=\frac1\pi.
\]
\end{proof}

\subsection{Dependence on the minimum mixture weight}\label{appendix:pi-min-necessity}

\begin{theorem}[Dependence on the minimum mixture weight]\label{thm:pi-min-lower-bound}\label{lemma:unweighted_worst_case}
There are absolute constants $c,c_0>0$ such that, for every $k\ge2$, $0<\varepsilon\le c_0$, and prescribed $\pi_{\min}\in(0,1/k]$, there exists a componentwise isotropic Gaussian mixture in some dimension $d\ge\log k$ with minimum weight $\pi_{\min}$ and $\zeta_\mu=1$ for which an i.i.d.\ sample of any integer size $1\le N\le c\pi_{\min}^{-1}\varepsilon^{-2}$ fails to be an $\varepsilon$-coreset with probability at least $1/8$.
\end{theorem}

\begin{proof}
Choose $d\ge\max\{k,1000\varepsilon^{-2}\}$, and let $e_1,\ldots,e_d$ be the standard basis. Set
\[
    \pi_1=\pi_{\min},\qquad \pi_j=\frac{1-\pi_{\min}}{k-1}\quad(j\ge2).
\]
Since $\pi_{\min}\le1/k$, every $\pi_j$ with $j\ge2$ is at least $\pi_{\min}$. For the Gaussian components, set
\[
    a_1=e_1,\qquad a_j=\sqrt{\pi_{\min}}e_j\quad(j\ge2),\qquad \sigma_j=\frac{\pi_{\min}}d\quad(j\in[k]).
\]
Thus the minimum mixture weight is $\pi_{\min}$. Moreover, every component has squared radius $r_j^2=d\sigma_j=\pi_{\min}$, so $\zeta_\mu=1$.

 A sample point $p$ can be written as
\[
    p=a(p)+g,\qquad g\sim\mathcal N\!\left(0,\frac{\pi_{\min}}d I_d\right).
\]
Let $I$ be the indicator that the latent component of $p$ is the first component $\mu_{1}$, so $I\sim\operatorname{Bernoulli}(\pi_{\min})$. If $I=1$ then $\|a(p)\|^2=1$, whereas if $I=0$ then $\|a(p)\|^2=\pi_{\min}$. Hence
\[
    \|a(p)\|^2=\pi_{\min}+(1-\pi_{\min})I, \quad \mathbb E\|a(p)\|^2=2\pi_{\min}-\pi_{\min}^2, \quad \quad \mathbb E\|g\|^2=\pi_{\min}
\]

We use the fixed query $C=\{0\}$. Therefore

\[
    \cost(\mu,C)=\mathbb E\|p\|^2=3\pi_{\min}-\pi_{\min}^2\le3\pi_{\min}.
\]

Now let $p_1,\ldots,p_N$ be the sample, write $p_i=a(p_i)+g_i$, and let $I_i$ indicate that $p_i$ was generated by the first component. Then $|X_1|=\sum_{i=1}^N I_i$, and expanding the squared norm pointwise gives
\[
    \|p_i\|^2=\|a(p_{i})+g_{i}\|^2=\pi_{\min}+(1-\pi_{\min})I_i+2\langle a(p_i),g_i\rangle+\|g_i\|^2.
\]
Consequently,
\[
\begin{aligned}
    \cost(X,C)-\cost(\mu,C)
    &=(1-\pi_{\min})\left(\frac{|X_1|}{N}-\pi_{\min}\right)\\
    &\quad+\frac1N\sum_{i=1}^N
       \left(2\langle a(p_i),g_i\rangle+\|g_i\|^2-\pi_{\min}\right).
\end{aligned}
\]
We first bound the second term, then use the component-count fluctuation to
obtain relative error larger than $\varepsilon$.
Since $g\sim\mathcal N(0,(\pi_{\min}/d)I_d)$ and
$\mathbb E\|a(p)\|^2\le2\pi_{\min}$, each centered summand in the second term
has variance at most $10\pi_{\min}^2/d$; the cross term vanishes by
the symmetry $g\mapsto-g$. Independence and Chebyshev's inequality give
\[
    \Pr\!\left[
       \left|\frac1N\sum_{i=1}^N
       \left(2\langle a(p_i),g_i\rangle+\|g_i\|^2-\pi_{\min}\right)\right|
       >\varepsilon\pi_{\min}\right]
    \le\frac{10}{dN\varepsilon^2}
    \le\frac1{100}.
\]

It remains to show that the component-count fluctuation is large. Since $|X_1|\sim\operatorname{Bin}(N,\pi_{\min})$, a standard binomial anti-concentration bound gives absolute constants $c,c_0>0$ such that, whenever $0<\varepsilon\le c_0$ and $N\le c\pi_{\min}^{-1}\varepsilon^{-2}$,
\[
    \Pr\!\left[\left|\frac{|X_1|}{N}-\pi_{\min}\right|>8\varepsilon\pi_{\min}\right]\ge e^{-2}.
\]
For completeness, take $c\le1/1024$ and $c_0\le1/16$. If $N\pi_{\min}\le1$, the event $|X_1|=0$ has probability $(1-\pi_{\min})^N\ge e^{-2N\pi_{\min}}\ge e^{-2}$ and gives the required deviation. If $N\pi_{\min}>1$, then
\[
    \mathbb E\bigl(|X_1|-N\pi_{\min}\bigr)^2=N\pi_{\min}(1-\pi_{\min})\ge\tfrac12,\qquad
    \mathbb E\bigl(|X_1|-N\pi_{\min}\bigr)^4
    \le5\bigl(N\pi_{\min}(1-\pi_{\min})\bigr)^2.
\]
Paley--Zygmund applied to $\bigl(|X_1|-N\pi_{\min}\bigr)^2$ gives
\[
    \Pr\!\left[\bigl||X_1|-N\pi_{\min}\bigr|>
    \sqrt{N\pi_{\min}(1-\pi_{\min})/8}\right]
    \ge49/320>e^{-2}.
\]
The sample-size bound and $\pi_{\min}\le1/2$ imply $(8\varepsilon N\pi_{\min})^2\le N\pi_{\min}(1-\pi_{\min})/8$, proving the claim in this case as well.
Intersect this event with the event that the second term in the error decomposition has absolute value at most $\varepsilon\pi_{\min}$, whose complement was bounded above by Chebyshev's inequality. Since $1-\pi_{\min}\ge1/2$, the error decomposition gives
\[
    |\cost(X,C)-\cost(\mu,C)|>4\varepsilon\pi_{\min}-\varepsilon\pi_{\min}=3\varepsilon\pi_{\min}>\varepsilon\cost(\mu,C).
\]
The intersection has probability at least $e^{-2}-1/100>1/8$, proving the lower bound. Since $d\ge\log k$, Corollary~\ref{cor:mixture_quantization_lower_bound} gives
\[
    \cost(\mu,C)\ge k^{-2/d}V_\mu
    \qquad\text{for every query }C.
\]
Thus this mixture satisfies the global comparison condition with a universal constant.
\end{proof}

\subsection{Dependence on the scale imbalance}\label{appendix:zeta-necessity}

\begin{theorem}[Dependence on the scale imbalance]\label{thm:zeta-lower-bound}
There is an absolute constant $c>0$ such that, for every $k\ge4$, every prescribed $\zeta\in[1,k]$, and every $0<\varepsilon\le1/128$, there exists an equal-weight, componentwise isotropic Gaussian mixture $\mu$ in some dimension $d\ge\log k$ whose scale imbalance satisfies $\zeta/2\le\zeta_\mu\le2\zeta$, and for which an i.i.d.\ sample of any integer size $1\le N\le ck\zeta\varepsilon^{-2}$ fails to be an $\varepsilon$-coreset with probability at least $0.98$.
\end{theorem}

\begin{proof}
Set $m=\lfloor k/\zeta\rfloor$. Then
\[
    \zeta\le\frac{k}{m}\le2\zeta.
\]
Choose $d$ to be a large enough absolute multiple of $\varepsilon^{-2}\log(k/\varepsilon)$. Give every component weight $1/k$, and set
\[
    \mu_j=\begin{cases}\mathcal N(0,I_d), & j\le m,\\ \mathcal N\!\left(10\sqrt d\,e_1,\frac{\varepsilon m}{1000k}I_d\right), & j>m.\end{cases}
\]
The first $m$ components have squared radius $d$, while the remaining components  have squared radius $\varepsilon md/(1000k)$. 

Recall the definition of $\zeta_{\mu}$ and that $r_j^2=\sigma_j d$, hence the scale imbalance of the constructed mixture is
\[
    \frac{\max_{j\in[k]}\sigma_j}{\sum_{j\in[k]}\pi_j\sigma_j}=\frac{\max_{j\in[k]}r_j^2}{\sum_{j\in[k]}\pi_j r_j^2}
    =\frac{d}{m/k\cdot d+(1-m/k)\varepsilon md/(1000k)}=\frac{k/m}{1+(1-m/k)\varepsilon/1000},
\]

which lies between $k/(2m)$ and $k/m$, and therefore between $\zeta/2$ and $2\zeta$.

Thus the first $m$ components receive only an $m/k$ fraction of the observations but contribute essentially all of the latent cost. Also $N\le k^2\varepsilon^{-2}$ for $c\le1$, so the choice of $d$ is independent of $N$ throughout the stated range.

Let
\[
    n=\sum_{j=1}^m|X_j|
\]
be the number of sampled points from the first $m$ components. When $n>0$, partition these points in their sampling order into $\min\{k-1,n\}$ nonempty groups whose sizes differ by at most one; below, $B$ denotes such a group. With probability at least $0.98$, the following bounds hold simultaneously:
\[
    \bigl|\|p\|^2-d\bigr|\le\frac{\varepsilon d}{10},\qquad |\langle p,e_1\rangle|\le\frac{\varepsilon\sqrt d}{10}\qquad(p\in X_1\cup\cdots\cup X_m),
\]
\[
    \left\|\sum_{p\in B}p\right\|\ge\frac12\sqrt{|B|d}\qquad\text{for every group }B, \quad \text{and} \quad
    \frac1N\sum_{j=m+1}^k\sum_{p\in X_j}\|p-a_j\|^2\le\frac{\varepsilon md}{k}.
\]
Conditional on the component labels, every group sum is $\mathcal N(0,|B|I_d)$. Gaussian and chi-square tail bounds, followed by a union bound, give failure probability at most $4Ne^{-\varepsilon^2d/800}+ke^{-d/8}$ for the first two displays. The expectation in the last display is at most $\varepsilon md/(1000k)$, so Markov's inequality bounds its failure probability by $1/1000$. Our choice of $d$ makes the total failure probability smaller than $0.02$. Work on this event below.

Consider a query containing $10\sqrt d\,e_1$ and between one and $k-1$ centers of norm $\sqrt d$. For a fresh $p\sim\mathcal N(0,I_d)$, the expected maximum inner product with the latter centers is at most $\sqrt{2d\log k}$. Adding the distant center gives no improvement when $\|p\|\le2\sqrt d$. Its expected improvement is therefore at most
\[
    \mathbb E\!\left[(\|p\|+\sqrt d)^2\mathbf 1_{\{\|p\|>2\sqrt d\}}\right]\le\bigl(\mathbb E(\|p\|+\sqrt d)^4\bigr)^{1/2}\Pr[\|p\|>2\sqrt d]^{1/2}\le8de^{-d/4}.
\]
Here $\mathbb E\|p\|^4=d(d+2)$, and the Gaussian radius bound gives $\Pr[\|p\|>2\sqrt d]\le e^{-d/2}$. Thus every such query satisfies
\begin{equation}\label{eq:scale-lower-bound-population}
        \cost(\mu,C)\ge\frac{m}{k}\left(2d-2\sqrt{2d\log k}-8de^{-d/4}\right)\ge(2-\varepsilon)\frac{md}{k}.
\end{equation}

First consider the fixed query $C=\{\sqrt d\,e_1,10\sqrt d\,e_1\}$. On our event,
\[
    (2-\varepsilon)\frac{nd}{N}\le\cost(X,C)\le(2+\varepsilon)\frac{nd}{N}+\frac{\varepsilon md}{k}.
\]
Its population cost lies between $(2-\varepsilon)md/k$ and $(2+\varepsilon)md/k$, where the upper bound follows by assigning each block to its designated center. Unless this fixed query already violates the coreset condition, these bounds imply, for $\varepsilon\le1/6$,
\begin{equation}\label{eq:scale-lower-bound-count}
    (1-3\varepsilon)\frac{m}{k}\le\frac nN\le(1+3\varepsilon)\frac{m}{k}.
\end{equation}
In particular, $n>0$. Thus small sample sizes are already handled by the fixed query; we never need a separate concentration bound for $n$.

Now use the group directions to construct the sample-dependent query
\[
    C=\{10\sqrt d\,e_1\}\cup\left\{\sqrt d\,\frac{\sum_{p\in B}p}{\left\|\sum_{p\in B}p\right\|}:B\text{ is a group}\right\}.
\]
It has at most $k$ centers. The latent labels are used only to exhibit a query in the supremum. Since the groups have balanced sizes and $k-1\ge k/2$,
\[
    \sum_B\left\|\sum_{p\in B}p\right\|\ge\frac{\sqrt d}{2}\sum_B\sqrt{|B|}\ge\frac{\sqrt d}{8}\min\{n,\sqrt{kn}\}.
\]
Assign every group to its corresponding center and the remaining observations to the distant center. Expanding the squared distances gives
\[
    \cost(X,C)\le(2+\varepsilon)\frac{nd}{N}+\frac{\varepsilon md}{k}-\frac{d}{4N}\min\{n,\sqrt{kn}\}.
\]
Using \eqref{eq:scale-lower-bound-count},
\[
    \cost(X,C)\le\frac{md}{k}\left(2+9\varepsilon-\frac18\min\left\{1,\sqrt{\frac{k^2}{Nm}}\right\}\right).
\]
Since $\zeta\le k/m$ and $N\le ck\zeta\varepsilon^{-2}$, we have $N\le ck^2m^{-1}\varepsilon^{-2}$. Hence the minimum above is at least $128\varepsilon$ when $c\le128^{-2}$ and $\varepsilon\le1/128$. Together with \eqref{eq:scale-lower-bound-population},
\[
    \cost(X,C)\le(2-7\varepsilon)\frac{md}{k}<(1-\varepsilon)\cost(\mu,C).
\]
Thus, on the event of probability at least $0.98$, either the fixed query or the sample-dependent query certifies failure. Moreover, $d\ge\log k$ and Corollary~\ref{cor:mixture_quantization_lower_bound} imply $\cost(\mu,C)\ge k^{-2/d}V_\mu$ for every query $C$. Hence the mixture satisfies the global comparison condition, in addition to the bounds proved for the two witnesses.
\end{proof}

\end{document}